\documentclass[twocolumn, aps, prd, amscd, amsmath, amssymb,verbatim,nofootinbib,usenatbib]{revtex4}

\usepackage{graphicx}
\usepackage{verbatim}

\def\nobf{}
\def\hb{\hat{\boldsymbol \beta}}
\def\bb{{\boldsymbol \beta}}

\def\bbR{{\boldsymbol \beta}_R}
\def\bR{\beta_R}
\def\gR{\gamma_R}
\def\g{\gamma}
\def\Res{\mathcal{R}}
\def\Sh{\mathcal{\tilde{S}}}
\def\Ch{\mathcal{\tilde{C}}}
\def\ChT{\mathcal{\tilde{C}}2}

\def\BM{{\bf B}_M}
\def\BR{{\bf B}_R}
\def\EM{{\bf E}_M}
\def\ER{{\bf E}_R}

\def\hr{\hat {\bf r}}
\def\rr{{\bf r}}

\def\vb{v_{S, x}}

\begin{document}
\title[Rindler Dipoles]{Big Black Hole, Little Neutron Star:\par  Magnetic Dipole Fields in the Rindler Spacetime}
  

\author{Daniel J. D'Orazio$^1$\thanks{dorazio@astro.columbia.edu} and Janna Levin${}^{2,3}$}
\affiliation{${}^1$ Department of Astronomy, Columbia University, New York, NY 10027}
\affiliation{${}^2$ Department of Physics and Astronomy,
Barnard College, Columbia University, NY, NY 10027}
\affiliation{${}^3$ Division of Physics, Mathematics, and Astronomy, California Institute of Technology, Pasadena, California 91125, USA}

\begin{abstract}

As a black hole and neutron star approach during inspiral,
the field lines of a magnetized neutron star eventually thread the
black hole event horizon and a
short-lived electromagnetic
circuit is established.
The black hole acts as a battery that provides power to the circuit, thereby lighting up the pair just before merger. 
Although originally suggested as an electromagnetic counterpart to gravitational-wave detection, a black hole battery is of more general interest as a novel luminous astrophysical source. To aid in the theoretical understanding,
we present analytic solutions for the electromagnetic
fields of a magnetic dipole in the presence
of an event horizon. In the limit that the neutron star is very close to a
Schwarzschild horizon, the Rindler limit, we can solve Maxwell's
equations exactly for a magnetic dipole on an arbitrary worldline. We
present these solutions here and investigate a proxy for a
small segment of the neutron star orbit around a big black hole. We find
that the voltage the black hole battery can provide is in the range
$\sim 10^{16}$ statvolts with a projected luminosity of $10^{42}$ ergs/s
for an $M=10M_\odot$ black hole, a neutron star with a B-field of $10^{12} G$, and an
orbital velocity $\sim 0.5 c$ at a distance of $3M$ from the horizon. Larger black holes provide less power for binary separations at a fixed number of gravitational radii. 
The black hole/neutron star system therefore has a significant power supply to
light up various elements in the circuit possibly powering bursts, jets, beamed radiation, or
even a hot spot on the neutron star crust.

\end{abstract}

\maketitle

\section{Introduction}
Although intrinsically dark, a black hole (BH) can
potentially act as a battery in an electromagnetic
circuit -- a battery that can power great luminosities when connected
to other elements in the circuit \citep{MP2_MTBZ:1982}. 

\begin{figure}
\begin{center}
\includegraphics[scale=0.35]{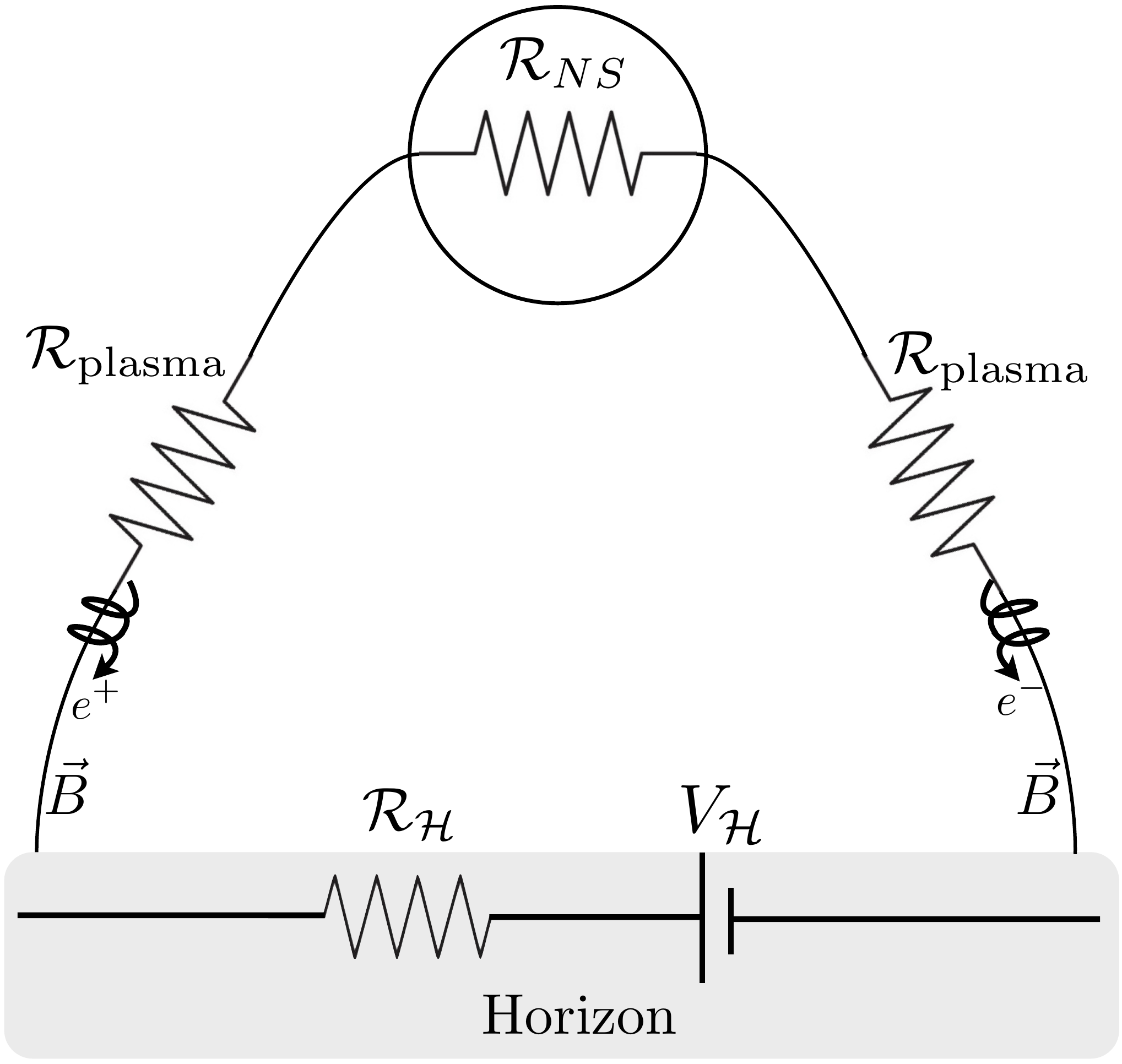} 
\end{center}
\caption{Neutron star - Rindler horizon effective circuit
  diagram. Magnetic field lines act as wires connecting the neutron
  star to the horizon. Current flows in (out) of the horizon via 
  positively (negatively) charged particles spiraling in tight Larmor radii around magnetic field lines into the horizon. }
\label{CircuitDiagram}
\end{figure}

Blandford-Znajek famously proposed a
BH battery as the power source for quasar jets \cite{BZ:1977}. In their
well-known model, a spinning BH twists a strong magnetic field
anchored in an accretion disk to create an 
emf that powers an energetic jet. The BH spins down as energy is lost
to the luminosity of the jet.
In a related yet novel scenario, it was recently proposed
\cite{McL:2011} that a
magnetized neutron star (NS) in orbit with a BH
could light up. When the BH orbits within the magnetosphere 
of the NS, the relative motion of
the BH through the NS dipole field could generate
an emf. The BH acts as a battery, the field lines as
wires, the charged particles of the NS magnetosphere as current
carriers, and the NS itself behaves as a resistor. In
principle, the orbit would wind down as angular momentum is lost to the circuit, 
although in practice gravitational radiation drains angular momentum by far the
faster. The circuit is illustrated schematically in
Figure \ref{CircuitDiagram}. (See also
Refs.\ \cite{Piro:2012,DLai:2012,Lyut:2012,Palenzuela:2010,Palenzuela:2013}
for related systems.)

BH-NS pairs may generate gamma-ray bursts
during merger via the Blandford-Znajek mechansim; 
when the NS is tidally disrupted, accreting
material tows a magnetic field into the ringing BH \citep{NarayanPacz:1992, LeeRuiz:2005, FaberBaumgarte:2006, ShibataUryu:2007, ShibataTaniguchi:2008, Etienne:2009, Rezzolla:2011, Etienne:2012, EastPret:2012, Giacomazzo:2013}. 
If the BH is big enough, however, the NS will not be tidally
disrupted prior to merger but instead will be swallowed whole,
prohibiting the post-merger gamma-ray burst. 
Since AdLIGO will be most sensitive to binaries with larger BH's
\citep{AdLIGO:2012}, it is important to note that the electromagnetic circuit of 
\citep{McL:2011} may be the only electromagnetic counterpart to
the gravitational-wave signal.

In this paper, we describe the BH-NS circuit in an
analytic calculation valid for large BHs. Very near a large
BH, the event horizon looks like a flat wall and in this limit
the Schwarzschild metric can be approximated by a Rindler metric --
the metric of a
flat spacetime as measured by observers with uniform proper-accelerations. 
In the Rindler limit, calculations are simplified while
some of the key physics is retained.
Due to acceleration, the Rindler observer also sees a flat-wall 
event horizon and so the relevant interaction of the EM field with 
an horizon is present. Also, since the Rindler observer is just  on a special
worldline in flat spacetime, calculations can be carried out in the
Minkowski spacetime and transformed to Rindler, a significant
calculational advantage.
(The Rindler limit is also used by \citep{MP3_MS:1985} to investigate the fields of a point charge interacting with an horizon.)

We consider a magnetic dipole on an arbitrary worldline 
near the flat-wall event horizon and derive analytic expressions for the electromagnetic fields.
We find that 
a battery is established when the worldline of the source
incorporates motion parallel to the horizon and the pair have
approached within the light cylinder of the NS.\footnote{Because we
  do not capture effects from spatial curvature, an actual BH-NS pair
  may establish a battery even with head-on motion.} As the pair draws closer
under the effects of gravitational radiation, the power of the battery
and luminosity of the circuit hits a maximum, just prior to merger. We evaluate the maximum
power the black-hole battery would provide to a completed circuit, and thus the maximum
luminosity generated. In addition, we estimate the maximum
energy to which plasma particles
could be accelerated, and thus the type of emission the circuit is capable 
of producing.  
As a preview of the conclusions, we quote here the rough scaling of the voltage
and luminosity:
\begin{align}
V^{\rm{max}}_\mathcal{H} &\sim 3.3 \times 10^{16} \left( \frac{B_p}{10^{12}\ \rm{G}} \right) \left( \frac{M}{10 M_{\odot} } \right)^{-2} \rm{statvolts}  \nonumber\\
 \mathcal{L}^{\rm{max}} &\sim 1.3 \times 10^{42} \left(
 \frac{B_p}{10^{12}\ \rm{G}} \right)^2 \left( \frac{M}{10 M_{\odot} }
 \right)^{-4} \frac{\rm{erg}}{\rm{s}} 
\label{Eq:conc}
\end{align}
where $M$ is the mass of the BH and $B_p$ is the magnetic field strength at the poles of the
NS.
(Readers who prefer to skip the derivations in favor of the
conclusions can fast-forward to the results of \S \ref{Consequences for the BH-NS Binary}.)
These scalings only apply at a fixed height $3M$ above the horizon and are dependent on the unknown resistivities of the plasma
and of the NS. Eqs.\ (\ref{Eq:conc}) should therefore be taken as a guide only. Still, even with these
caveats, the conclusion is that a BH-NS circuit could power high-energy bursts of radiation 
visible to current missions, especially for the special case of magnetar-strength NS fields. This intriguing possibility calls for
more detailed predictions of the timescales and spectra of emission, a topic for future explorations.
We hope that, in addition to the above estimates, the 
electro-vacuum example this paper provides will be a resource for further analytic studies and numerical experiments.

\section{Set-up and Limits}
\subsection{Rindler Spacetime}
\label{Set-up and Approximations}

Consider the line element in Minkowski spacetime
\begin{equation}
ds^2=-dT^2+dX^2+dY^2+dZ^2 \ \ 
\end{equation}
The following coordinate transformation,
\begin{align}  
T&= z \hbox{sinh}(g_H t)         \hbox{  } \hbox{  }   \hbox{  } \hbox{  }   \hbox{  } \hbox{  } \hbox{  }     X=x     \nonumber \\
Z&= z \hbox{cosh}(g_H t)     \hbox{  } \hbox{  }  \hbox{  } \hbox{  }   \hbox{  }    \hbox{  }   Y=y,
\label{MtoR}
\end{align}
leads to the Rindler line element,
\begin{align}
ds^2 &= -\alpha^2  dt^2 + dx^2 + dy^2 + dz^2 \nonumber \\
\alpha &= g_H z
\label{alphaDef}
\end{align}
where the lapse function $\alpha$ measures the difference in Rindler observer proper time $\tau_R$ and Rindler coordinate time $t$.
For reference, the inverse transformation is given by
\begin{align}  
t&=   \frac{1}{g_H}  \tanh^{-1}\left[ \frac{T}{Z} \right]   \nonumber \\
z&= \sqrt{Z^2 - T^2} ,
\label{RtoM}
\end{align}
and we may write the non-inertial, uniformly accelerated trajectory in Minkowski coordinates as,
\begin{align}
X^\mu_R & =(T,0,0,Z) =(z_R\sinh(g_Ht),0,0,z_R\cosh(g_Ht))\nonumber \\
u_R^\mu & = \frac{dX_R}{d\tau_R}=(\gR,0,0,\gR \bR)\nonumber \\
a_R^\mu & = \frac{du_R}{d\tau_R}=z_R^{-1}(\gR\bR, 0, 0, \gR)
\label{RindWL}
\end{align}
where
\begin{align}
\bR &= \frac{dZ}{dT} =\frac{T}{Z}\nonumber \\
\gR & = (1-\bR^2)^{-1/2} \ .
\end{align}
It is also useful to express these in Rindler coordinates
\begin{align}
\gR & =\cosh(g_Ht) \nonumber \\
\gR \bR & = \sinh(g_Ht) \ .
\end{align} 
The 4-acceleration has constant magnitude:
\begin{equation}
a_R^\mu a^R_\mu = z_R^{-2}
\end{equation}
and so observers of constant Rindler coordinate $z_R$ have a
$4$-acceleration of constant magnitude according to a Minkowski
observer. 

\begin{figure}
\begin{center}
\includegraphics[scale=0.35]{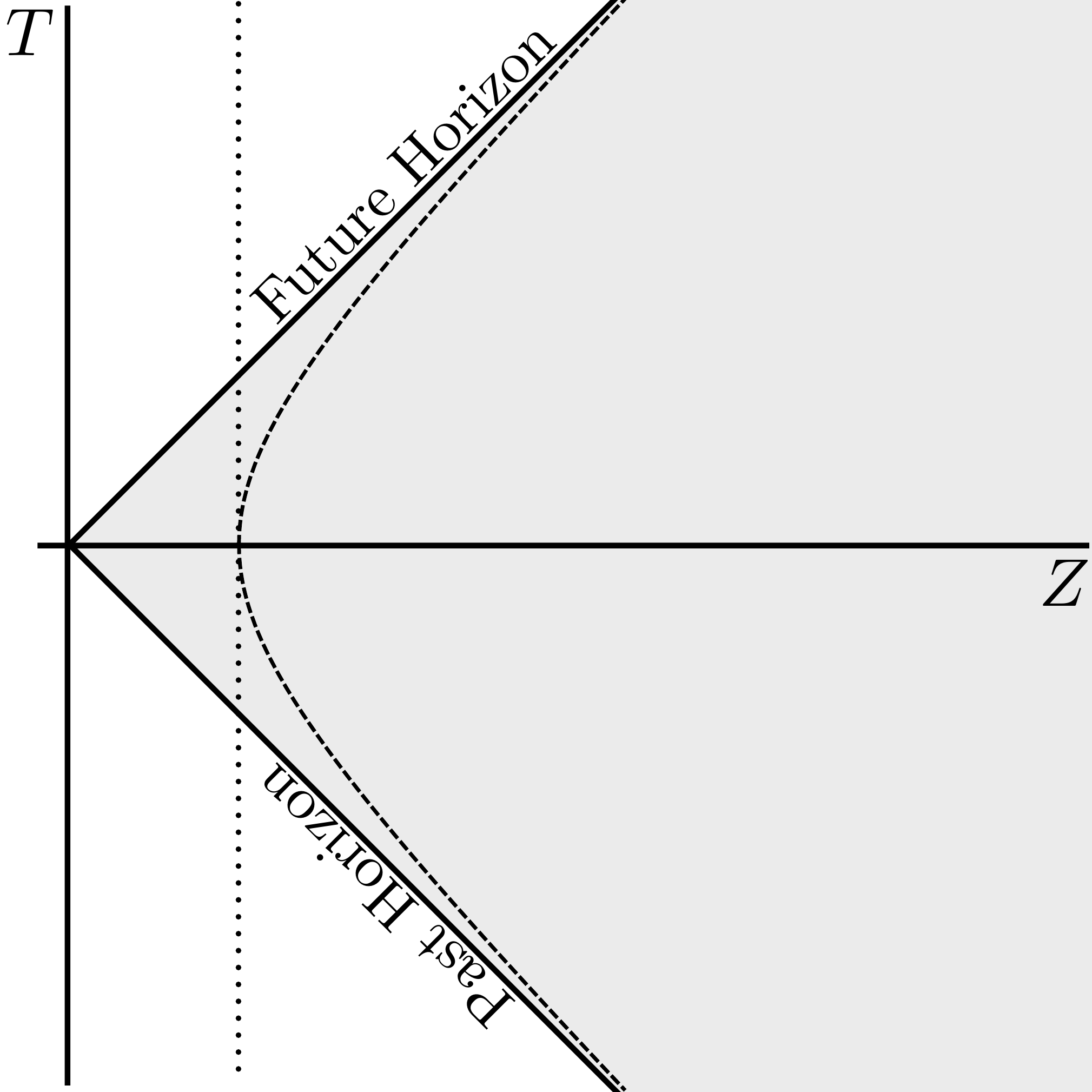}
\end{center}
\caption{Rindler space is the shaded wedge  given by $T > \pm Z$,
  $Z>0$ on the Minkowski spacetime diagram. The dotted vertical line
  is the trajectory of a Minkowski observer and the dashed hyperbolic
  line is that of a Rindler observer as viewed by a Minkowski
  observer. In the frame of the Rindler observer an event horizon
  exists at $T=\pm Z$.}
\label{Fig:MinkRind}
\end{figure}

Figure \ref{Fig:MinkRind} is a Minkowski spacetime diagram
demonstrating the wedge occupied by the Rindler spacetime (shaded
region). The worldline of a stationary Minkowski observer is denoted
by the vertical dotted line while the worldline of a Rindler observer
is denoted by the dashed hyperbolic trajectory (Eqs.\
(\ref{RindWL})). Due to their accelerations, Rindler
observers are causally disconnected from the non-shaded region of
Minkowski space in Figure \ref{Fig:MinkRind} and thus experience an
event horizon at $z=0$ ($T=\pm Z$, $Z>0$).

With the choice of $g_H=1/(4M)$ and the transformations,
\begin{align}
x = 2M \phi \quad  y = 2M\left( \theta -\pi/2\right) \quad z = 4M\left( 1-2M/r\right)^{1/2} 
\label{RindtoSch}
\end{align}
the Rindler line element approximates the
Schwarzschild line element around the point $(r, \phi, \theta)=( 2M, 0, \pi/2)$. Errors of order unity in
the approximation to the Schwarzschild spacetime occur when $z
\rightarrow 4M$ ($\alpha \rightarrow 1$) and $y \rightarrow 2M$ (see \textit{e.g.}
\citep{MP3_MS:1985}). The Rindler limit retains some key
features of the spacetime, including gravitational red-shifting and time
dilation as well as the event horizon, although it necessarily misses
elements of spatial curvature.

\subsection{Electrodynamical Properties of an Event Horizon and the Horizon Battery}
\label{Electrodynamical Properties  of an Horizon}

To understand and interpret power generation by the BH-NS circuit, we
first review some key features of horizon electrodynamics \cite{MPbook}. 
We consult observers who are at a fixed
location {\it relative to the event horizon}. These fiducial observers
can tell us if the event horizon has established charge separation and
therefore a battery. 
Around a BH, these observers must
accelerate to maintain a fixed location and avoid plunging into the BH. Similarly, in Rindler space,
our fiducial observers accelerate to maintain a fixed location $z$
from the event horizon. So while stationary relative to the horizon,
our fiducial observers are not stationary in an absolute sense -- they
are non-inertial and so
must burn fuel to stay at their Rindler-coordinate location.

We are therefore after the electric $\ER$ and magnetic $\BR$ fields
measured by a Rindler observer -- fields due to a magnetic dipole
source on an arbitrary worldline -- and we want to determine these fields
everywhere outside the event horizon. Problematically, the fields
of a Rindler observer will necessarily experience divergences at the
event horizon due to infinite time dilation.
Following the membrane paradigm \cite{MPbook}, we construct a
timelike hyper-surface stretched over the true, null-horizon. On the
stretched horizon, fields will be finite. We then
apply electromagnetic boundary conditions on this fictitious surface. Since
electric field lines can only terminate or originate on sources, the stretched horizon is assigned hypothetical surface charge to
satisfy the boundary conditions of any normal $\ER$
component. Similarly, the stretched horizon is assigned hypothetical surface current to
satisfy the boundary conditions of any tangential $\BR$
component. As can be derived from local versions of Gauss's law and Ampere's law, the fictitious 
charge density and surface current are given by
\begin{align}
 \mathbf{E_R} \cdot \mathbf{n} \big{|}_{\mathcal{H}} = 4 \pi
 \sigma_{\mathcal{H}}   \qquad \alpha \mathbf{B_R}\big{|}_{\mathcal{H}} = 4 \pi \mathbf{\mathcal{J}_{\mathcal{H}}} \times \mathbf{n}
 \label{EBBCs}
\end{align}
where $\mathbf{n}$ is the unit normal to the horizon and $\mathcal{H}$
denotes evaluation at the stretched horizon. 
The interpretation then is that electric fields terminate on charges
in the stretched horizon, and magnetic fields parallel to the horizon
are sourced by surface currents. 

Combining (\ref{EBBCs})  along with the horizon normal component of the
differential form of Ampere's law gives charge conservation on the horizon 
\begin{align}
\frac{\partial \sigma_{\mathcal{H}}}{\partial t} + \nabla \cdot \mathbf{\mathcal{J}_{\mathcal{H}}} = -\left( \alpha j_n \right)_{\mathcal{H}}
\label{HChargeCons}
\end{align}
where $\left( \alpha j_n \right)_{\mathcal{H}}$ is the normal component of 
currents entering (positive charges flowing in) and leaving (negative 
charges flowing in) the horizon in units of universal time. The divergence 
is the two-dimensional divergence computed on the horizon. 

The lesson of the membrane paradigm: when electromagnetic field
boundary conditions are
applied, the horizon behaves as if it were a conductor with
the resistivity of free space\footnote{This follows from (\ref{EBBCs}) as well as using
stationary observers to measure the fields. See Ch. 2 of \cite{MPbook}.}. On this (hypothetical) conductor may
exist (hypothetical) surface
charges and currents. Eq.\ (\ref{HChargeCons}) tells us that
charge is conserved as current flows.
In the vacuum calculations presented here, the right hand side of
(\ref{HChargeCons}) will always be zero (to fractional errors of order
$\alpha_{\mathcal{H}}$ arising from stretching the horizon) as there
is no plasma to carry current off the horizon.

We will be particularly interested in the case where the motion of a
magnetic field relative to the Rindler observers induces an electric
field that has normal components to the horizon. These normal
components source a surface charge density on the horizon that must, when
integrated over the black hole area, amount to zero net charge for an
initially uncharged black hole. Therefore, charge separation is
induced on the horizon and that gradient
can be interpreted as creating a battery. 
If an external circuit is connected to the horizon then
the horizon emf associated with the charge
separation will drive a current in the circuit. The instantaneous emf of such a
horizon battery is given by
\begin{align}
V_{\mathcal{H}} = \int{\left[ \alpha \mathbf{E_R}\right]_{\mathcal{H}} \cdot ds} \ ,
\label{VH}
\end{align}
remembering that an electric potential is only well-defined for
electric fields that originate and terminate on source charges, albeit
hypothetical source charges in this case.
The situation is analogous to a conventional chemical battery. 
In the
horizon battery, energy of motion of the magnetic field source
replaces the chemical energy. In the
specific case of a NS orbiting a Schwarzschild BH,
the energy source is the spin and orbital energy of the binary. 

Figure \ref{CircuitDiagram}  shows the equivalent electrical circuit
of such a system. The horizon battery drives current in the form of
charged magnetosphere particles spiraling along the NS
magnetic field lines. Current enters the horizon via positive-charge
carriers (positrons) riding magnetic field lines into the horizon and
leaves the horizon via negative-charge carriers (electrons) flowing
into the horizon. The current flows through three resistors comprised of
the NS, the plasma, and the BH. If we know the electric field
induced from the orbital motion of the magnetic dipole we can compute
an horizon battery voltage. We may calculate the power, as observed at infinity, dissipated
by the $i^{th}$ resistive component of the system, 
\begin{align}
\label{CircuitPower}
\mathcal{P} =  \frac{  V^2_{\mathcal{H}}  }{( \Res_{\mathcal{H}} + \Res_{NS} + 2\Res_{\rm{plasma}})^2 } \Res_{i} ,
\end{align}
to approximate the luminosity generated by that component.
While the resistance of the BH horizon is set by the resistivity of
free space \cite{MPbook}, 
the resistances of the NS and plasma are interesting unknowns.
Although the primary calculations done here are all in vacuum, in \S
\ref{Consequences for the BH-NS Binary} we 
use Eq.\ (\ref{CircuitPower}) to estimate the power and
find that there is potential for significant bursts of energy from
black hole batteries.

First, we find exact closed form solutions for the
electromagnetic fields of a magnetic dipole on an arbitrary
worldline. We then implement those
solutions for specific dipole trajectories.

\section{A magnetic Dipole in arbitrary motion}
\label{Text:Solutions}

\subsection{The Electromagnetic Four-Potential}
In Minkowski spacetime, Maxwell's equations for the 4-potential $A^{\alpha}$ are,
\begin{equation}
\Box A^{\alpha}(\nobf{x}) - \partial^{\alpha} \left( \partial_{\beta} A^{\beta} \right)= \frac{4 \pi}{c}  J^{\alpha}(\nobf{x})
\label{MxPotMink1}
\end{equation}
where $J^{\alpha}(\nobf{x})$ is the 4-current as a function of the
coordinates.\footnote{We use Gaussian units to write Maxwell's equations. In writing Maxwell's equations we have included the proper factors of $c$. However, everywhere else, in writing the Rindler metric and the 4-velocities etc. we have set $G=c=1$.}
We choose to work in the Lorentz Gauge $  \partial_{\beta} A^{\beta}
= 0$. Then Maxwell's equations for the 4-potential become sourced
wave-equations,
\begin{equation}
\Box A^{\alpha}(\nobf{x}) = \frac{4 \pi}{c} J^{\alpha}(\nobf{x}).
\label{MxPotMinkL}
\end{equation}

We choose the 4-current for a point dipole source
\begin{align}
J^{\alpha}(\nobf{x}) =  \hbox{  } \nabla_{\mu} \int { Q^{\alpha \mu}(\tau) \delta^{(4)}\left[ \nobf{x} -\nobf{x_S}(\tau) \right]   d \tau}
\label{eq:4JDipole}
\end{align}
where $\tau$ is the proper time of the dipole source, not to be confused with the proper time of the Rindler observers $\tau_R$, and the antisymmetric dipole tensor,
\begin{align}
Q^{\alpha \mu}(\tau) &= V^{\alpha}p^{\mu} - p^{\alpha}V^{\mu} + \epsilon^{\alpha \mu}_{ \hbox{  } \hbox{  } \hbox{  }\rho \sigma} V^{\rho} m^{\sigma},
\label{DipoleTensor}
\end{align}
is the decomposition of electric $p$ and magnetic $m$ parts
\citep{RS:1995, Rowe2:1986}. (See appendix \S \ref{DipoleMoments} for more detail.) Notice that $V$ is the
instantaneous 4-velocity of the {\it source}. 
Also, hereafter $X$ will denote observer coordinates and $X_S$ will denote the
coordinates along the trajectory of the dipole source. 
The antisymmetric tensor is fixed by $\epsilon^{0123}=1$.

\begin{figure}
\begin{center}
\includegraphics[scale=0.35]{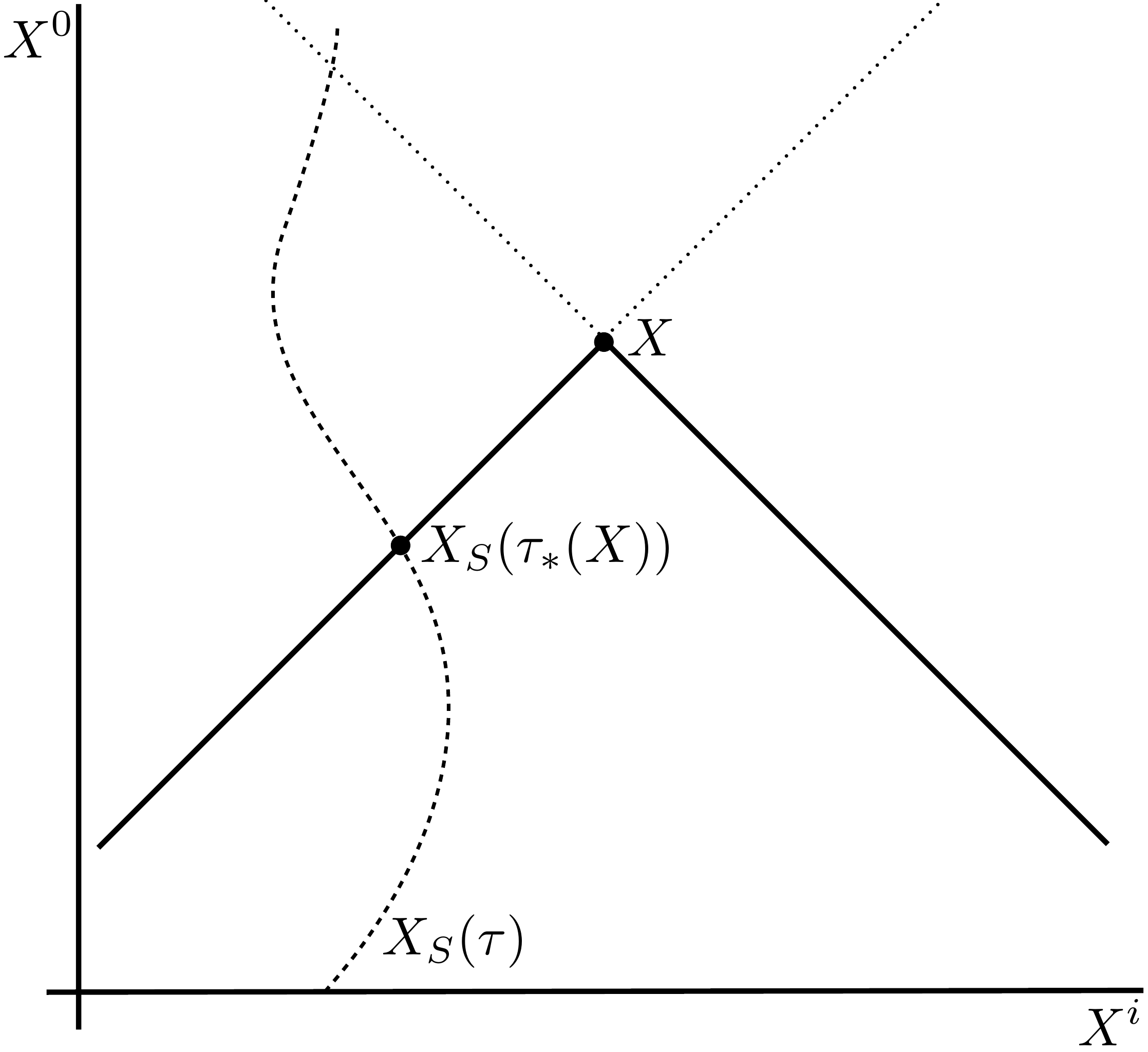} 
\end{center}
\caption{Diagram demonstrating the relationship between source trajectory coordinates
  $X_S(\tau)$, observer coordinates $X$, and the retarded proper
  time $\tau_*(X)$ at the intersection of the past light cone of an
  observer at $X$ and the source trajectory.}
\label{RetardedDiagram}
\end{figure}

The solution for $A^{\alpha}$ is derived in Appendix \ref{AppendixA} and can be written in the Minkowski frame, off of the worldline of the source, as
\begin{align}
 A^{\alpha} &=  
\nabla_{\mu}  \left[   \frac{Q^{\alpha \mu} }{
    r \cdot  V  }    \right]_{*}   \ \ .
\label{DipoleSoln1}
\end{align} 
By $r\cdot V $ between 4-vectors we mean the inner product
$g_{\mu\nu}r^\mu V^\nu$.
Since the source may be moving, we must account for the fact that an
observer at $X$ will observe fields due to the source in the past, it
taking the speed of light for the source information to get to the
observer. Therefore, 
the 4-potential is always evaluated at
the retarded time $T_*$ as represented graphically in Figure \ref{RetardedDiagram}.
The retarded time is found as a function of observer coordinates by
imposing the null condition.
We define the relative distance between an observer and a point on the
source trajectory in Minkowski coordinates as
\begin{equation}
r^\mu =
\begin{pmatrix}r^0 \\
{\bf{r}}
\end{pmatrix}
\equiv 
\begin{pmatrix}
T-T_{S} \\
{\bf{X}}-{\bf{X}}_{S}(T_S)
\end{pmatrix} \quad .
\end{equation}
The null condition is
$(r_\mu r^\mu )_* =0$,
with subscript $*$ denoting evaluation at $T_*$. Then 
\begin{equation}
r^0=\sqrt{\rr \cdot \rr}=r
\end{equation} 
and
\begin{equation}
({\bf{X}}-{\bf{X}}_S (T_*))^2= (T-T_*)^2 \quad .
\label{Eq:Tcond}
\end{equation}

It should be noted that the covariant derivative in
(\ref{DipoleSoln1}) is taken with respect to the coordinates
$X^\mu$. Since the solution to the null condition for $T_*$ is
dependent on the position of the observer, $T_* = T_*(X^{\mu})$,
$T_*$ is acted on by the covariant derivative.

For completeness, we
expand the 4-potential further.
Since $(r_\mu r^\mu)_* =0$ we may write,
\begin{equation}
r_\mu \nabla_\nu r^\mu |_* =0=
r_\mu \left (\delta^\mu_\nu-\nabla_\nu \tau \ V^\mu \right
)|_*=r_\nu- (r\cdot V)\nabla_\nu\tau |_* \quad .
\nonumber 
\end{equation}
Given this extremely useful relation, we can compile a list
of gradients that we will need in order to evaluate Eq.\ (\ref{DipoleSoln1}) and construct the field tensor:
\begin{align}
\nabla_\mu \tau & = \frac{r_\mu}{(r\cdot V)} \nonumber \\
\nabla_\mu r^\nu &= \delta_{\mu}^{ \nu}-\frac{r_\mu \ V^\nu}{(r\cdot V)}
\nonumber \\
\nabla_\mu V^\nu &= \frac{r_\mu \ a^\nu}{(r\cdot V)} \nonumber \\
\nabla_\mu a^\nu &=  \frac{r_\mu}{(r\cdot V)}\dot{a}^{\nu}  \nonumber \\
\nabla_\mu m^\nu &=  \frac{r_\mu}{(r\cdot V)}\dot{m}^{\nu} 
\label{Grads}
\end{align}
where an overdot denotes a $\tau$ derivative. Evaluation at $*$ is
implied in the relations (\ref{Grads}).

When there is only a magnetic dipole moment ${\bf m}$, then $Q^{\alpha \mu}=\epsilon^{\alpha \mu}_{\quad \ \ \rho \sigma}V^\rho m^\sigma$ and we can
expand the 4-potential as:
\begin{widetext}
\begin{align}
A^\alpha (x) &= \epsilon^{\alpha}_{\quad \mu \rho \sigma} r^\mu \left [\frac{a^\rho
  m^\sigma  + V^\rho \dot  m^\sigma }{(r\cdot V)^2} 
 -\frac{V^\rho m^\sigma (1+r\cdot a)}{(r\cdot V)^3}\right ]\Bigg{|}_{*}   \ .
\label{DipleSolnLong}
\end{align}
Note that the RHS of Eq.\ (\ref{DipleSolnLong})
reduces to the usual stationary dipole solution for a constant dipole at rest ($V^{\alpha}=(1,0,0,0)$ and $\dot m=0$), as it must.

Given the 4-potential, the electromagnetic field tensor is
\begin{align}
F_{\alpha \beta} = \nabla_{[\alpha} A_{\beta]} = \nabla_{\alpha} A_{\beta} - \nabla_{\beta} A_{\alpha}
\label{Fmunu}
\end{align}
Using (\ref{Grads}) and (\ref{DipleSolnLong}) to evaluate (\ref{Fmunu}) we find
\begin{align}
F_{\alpha \beta} (x) &= - \epsilon_{  [\alpha \beta]  \rho \sigma} \frac{ a^\rho m^\sigma  + V^{\rho} \dot{m}^{\sigma}  }{(r \cdot V)^2} \nonumber \\
&+     \frac{r_{[\alpha} \epsilon_{ \beta] \mu  \rho \sigma}}{(r \cdot V)^3} \Big{\{} r^{\mu} \left( 2a^{\rho} \dot{m}^{\sigma} + V^{\rho} \ddot{m}^{\sigma} + \dot{a}^{\rho} m^{\sigma}   \right)   -  V^{\mu} \left( a^{\rho} m^{\sigma} + V^{\rho} \dot{m}^{\sigma} \right) \Big{\}}
\nonumber \\
&- 2   \frac{ V_{[\alpha} \epsilon_{ \beta] \mu \rho \sigma}}{(r \cdot V)^3} \Big{\{}r^{\mu} \left( a^{\rho} m^{\sigma} + V^{\rho} \dot{m}^{\sigma} \right)  \Big{\}}  
 -  \frac{ a_{[\alpha} \epsilon_{\beta] \mu \rho \sigma}  }{(r \cdot V)^3} r^{\mu} V^{\rho} m^{\sigma} 
+   \epsilon_{ [\alpha \beta]  \rho \sigma} \frac{ V^\rho m^\sigma}{(r \cdot V)^3} (1+ r \cdot a) 
\nonumber \\
&-    \frac{r_{[\alpha} \epsilon_{\beta] \mu \rho \sigma}}{(r \cdot V)^4} \Big{\{} 3 r^{\mu} \left( a^{\rho} m^{\sigma} +  V^{\rho} \dot{m}^{\sigma}   \right) (1+ r \cdot a) + r^{\mu} V^{\rho} m^{\sigma}   (r \cdot \dot{a}) -  V^{\mu} V^{\rho} m^{\sigma}  (1+ r \cdot a) \Big{\}}
\nonumber \\
&+ 3     \frac{ V_{[\alpha} \epsilon_{\beta] \mu \rho \sigma}}{(r \cdot V)^4}  r^{\mu} V^{\rho} m^{\sigma} (1+ r \cdot a)
\nonumber \\
&+ 3   \frac{ r_{[\alpha} \epsilon_{\beta] \mu \rho \sigma}}{(r \cdot V)^5}  r^{\mu} V^{\rho} m^{\sigma} (1+ r \cdot a)^2.
\label{FmunuLong}
\end{align}
Again, evaluation at $*$ is implied in the expression (\ref{FmunuLong})\footnote{Note that the $(1 + r \cdot a)$ term, in (\ref{DipleSolnLong}) and  (\ref{FmunuLong}), becomes $(c^2 + r \cdot a)$ upon restoring units.}.

The electromagnetic fields for an observer with 4-velocity
$u^\beta$ are
\begin{align}
E^{\alpha} = F^{\alpha \mu} u_{\mu}    \qquad  B^{\alpha} =  \frac{1}{2} \epsilon^{\alpha \mu \gamma \delta}F_{\gamma \delta} u_{\mu}
\label{EBtoF}
\end{align}
A stationary Minkowski observer has 4-velocity $u^\mu=(1,0,0,0)$
and the Minkowski fields drop out:
\begin{align}
\EM & = -{\mathbf \nabla} A^{0} - \nabla_0 {\bf A} \nonumber \\
\BM & = {\mathbf \nabla} \times {\bf A} \quad .
\label{Eq:EMBM}
\end{align}
In vector notation, 
\begin{align}
\label{DipleSolnVec}
A^{0} = &   \frac{  \left(\mathbf{a}\times \mathbf{m} \right) \cdot
  \mathbf{r}  +  \left(\mathbf{V}\times \mathbf{\dot{m}} \right) \cdot
  \mathbf{r}}{(r\cdot V)^2}    - \frac{ \left(\mathbf{V}\times\mathbf{m}\right) \cdot
  \mathbf{r}} {(r\cdot V)^3}  \left( 1 + r\cdot a\right) \quad \bigg{|}_*   \nonumber \\
\mathbf{A}  = &
 \frac{ 
r^0\left(\mathbf{a}\times\mathbf{m}\right)  
+ 
r^0\left(\mathbf{V}\times\mathbf{\dot{m}} \right)
-  
a^0\left(\mathbf{r}\times\mathbf{m}\right) 
-
V^0  \left(\mathbf{r}\times\mathbf{\dot{m}}\right) 
+
m^0   \left(\mathbf{r}\times\mathbf{a}\right)  
+  
\dot{m}^0   \left(\mathbf{r}\times\mathbf{V}\right) }{(r\cdot V)^2}\Bigg{|}_{*} 
 \nonumber  \\ 
  -&  \frac{   
r^0( \mathbf{V}\times\mathbf{m} ) 
-
V^{0}  (\mathbf{m}\times\mathbf{r}) 
+ 
m^0 \left(\mathbf{r}\times\mathbf{V}\right) }{(r\cdot V)^3}  \left( 1 + r\cdot a\right) \Bigg{|}_{*}  
\end{align}
\end{widetext}
where the source kinematics can be expressed as
\begin{align}
\label{Eq:kinematics}
V^{\alpha} &= (V^0, \bf{V}) = (\gamma_S,  \gamma_S \boldsymbol{\beta}_S )   \\
\bb_S &= \frac{d \mathbf{X}_S}{dT} \nonumber \\
a^{\alpha} &=  \left(\gamma_S^4 \left(\bb_S \cdot
\frac{d \bb_S}{dT} \right), \gamma_S^2  \frac{d \bb_S}{dT} +  \gamma_S^4
\left(\bb_S \cdot  \frac{d \bb_S}{dT} \right) \bb_S \right)  \nonumber
\end{align}
where $\gamma_S$, $\bb_S$ are the instantaneous Lorentz factor and Lorentz
boost of the {\it source},  not to be confused with $\gR$, $\bR$
of the Rindler {\it observer}.
A dipole with only a magnetic rest frame moment $\bf{m_S}$
moving at $\bb$ relative to our Minkowski observer has moments
\begin{align}
\label{Eq:onlydipole}
p^{\alpha} &=  \left ( 0, \vec 0\right ) \\
m^{\alpha} &=  \left ( \gamma_S \bb_S \cdot {\bf{m_S}}, {\bf{m_S}}
  +(\gamma_S -1)(\hb_S \cdot \mathbf{m_S}) \hb_S   \right) \nonumber
\end{align}
as explained in more detail in appendix \ref{DipoleMoments}.
With the values in Eqs.\ (\ref{Eq:kinematics})-(\ref{Eq:onlydipole}), the
Minkowski fields can be
computed. Rindler fields are then transformed from the Minkowski
fields.

We find the Rindler fields ${\bf E}_R$ and ${\bf B}_R$ (primed)
expressed in terms of Minkowski fields $\EM,\BM$ (unprimed) and the Rindler 4-velocity (\ref{RindWL}) via the transformations,
\begin{align}
E^{\alpha'} = \frac{\partial x^{\alpha'}}{\partial x^{\alpha}}
F^{\alpha \mu} u^R_{\mu}     \hbox{  } \hbox{  }  \hbox{  } \hbox{  }
\hbox{  } \hbox{  }  B^{\alpha'} = \frac{1}{2} \frac{\partial
  x^{\alpha'}}{\partial x^{\alpha}} \epsilon^{\alpha \mu \gamma
  \delta}F_{\gamma \delta} u^R_{\mu}.
\label{EBtoFRind}
\end{align}
Here $u_R^\mu$ is the Rindler velocity according to a Minkowski observer and
the coordinate transformation expresses the components of the fields in the Rindler
basis.

A compact way to expand Eqs.\ (\ref{EBtoFRind}) exploits the fact that any vector can be decomposed as
\begin{align}
\ER & = \hb_R(\hb_R\cdot \ER) +\hb_R\times (\ER\times\hb_R) \nonumber \\
& = \ER^\perp +\ER^\parallel
\label{Eq:defperp}
\end{align}
where $\perp$ and $\parallel$ refer to components perpendicular to the Rindler
horizon and parallel to the Rindler horizon respectively. (So $\perp$
is parallel to $\bb_R$ and $\parallel$ is perpendicular to $\bb_R$.) 
The fields as measured by a Rindler observer are then expressed
conveniently in terms of
the fields as measured by a Minkowski observer as
\begin{align}
\ER & =\EM^{\perp} +\gR\EM^{\parallel} +\gR(\bb_R\times \BM^{\parallel}) \nonumber \\ 
\BR & =\BM^{\perp} +\gR\BM^{\parallel} -\gR(\bb_R\times
\EM^{\parallel})  \quad .
\label{Eq:trans}
\end{align}

Although we will focus on computing a charge gradient on the horizon
to gauge the power output of the BH-circuit in the following examples,
it is also instructive to
consider the Poynting flux driven by the Rindler dipole. Given the
electromagnetic fields as measured by the Rindler observer, we
may compute the Poynting vector in Rindler space as seen by an
observer at infinity,
\begin{align}
\mathbf{S} = \frac{\alpha^2}{4 \pi} \mathbf{E}_R \times \mathbf{B}_R,
\label{RPoynt}
\end{align}
where one factor of $\alpha$ converts from locally measured energy to energy at infinity and the second factor converts from proper time measured by the local stationary observer to the universal time of the 3+1 split (see \S \ref{Electrodynamical Properties  of an Horizon}).

To understand the meaning of the Poynting flux in this case, we integrate Poynting's theorem over the entire Rindler 3-volume, bounded at infinity and the horizon,
 \begin{align}
 \frac{d U}{dt} = -\int{ \mathbf{S}_{\infty} \cdot dA} - \int{ \mathbf{E}_H \cdot  \mathbf{\mathcal{J}_H}  \hbox{ }  dA}.
 \label{PoyntThm}
 \end{align}
The last term on the right is evaluated over the stretched horizon
since this is the only location in the volume where there are non-zero
currents (we could of course add a plasma and get more currents).  In
the absence of radiation at infinity, we see that any change in EM energy 
$U$ must be due to ohmic dissipation from horizon surface currents. 

Generally, the Poynting flux perceived by a Rindler observer can
be expressed in terms of Minkowski fields as 
\begin{align}
4\pi\alpha^{-2} {\bf S}= & \gR E_M^\perp \left [ (\hb_R \times
  \BM^\parallel) +\bR \EM^\parallel \right ]  \nonumber \\
& + \gR B_M^\perp \left [- (\hb_R \times
  \EM^\parallel) +\bR \BM^\parallel \right ]  \nonumber \\
& - \gR^2 \bb_R \left [ (B_M^{\parallel})^2+(E_M^{\parallel})^2 \right] \nonumber \\
& +\gR^2 \left(1+ \bbR^2 \right) \left [\EM^\parallel \times \BM^\parallel\right ]
\label{Eq:Poynt}
\end{align}
The first two terms represent flux parallel to the horizon.
The third term is always into the horizon and is due solely to the
Rindler motion. The final term
can be in or out of the horizon and is proportional to the in or out ($\pm Z$) Poynting flux that would be
observed by a Minkowski observer. This final term is the only term that could contribute to power coming 
out of the dipole-horizon system. However, it can be negated by the inward flux do to Rindler observer motion.

In our vacuum calculations, the
Poynting flux can only tell us about radiation from the moving
dipole fields, since we have not included a plasma. 
Instead, we look for the existence of
a battery to ascertain if there is a power source. When
a magnetosphere is added, the black-hole battery will power an outward
Poynting flux at infinity delivering radiation to a distant observer.

\section{A Freely Falling Dipole Solution}
\label{A Freely Falling Dipole Solution}
\begin{figure}
\begin{center}$
\begin{array}{c}
\includegraphics[scale=0.33]{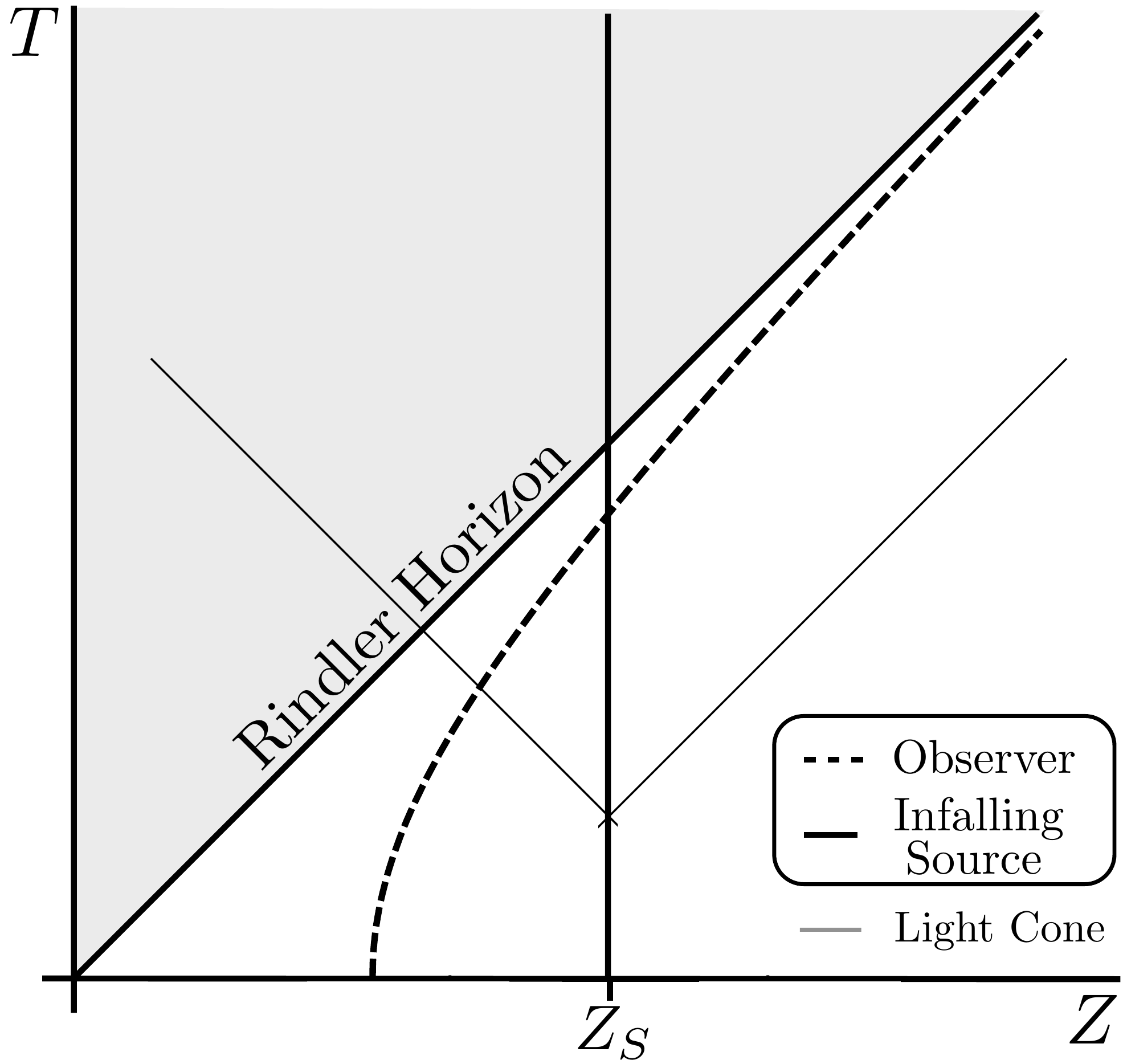} \\
\includegraphics[scale=0.33]{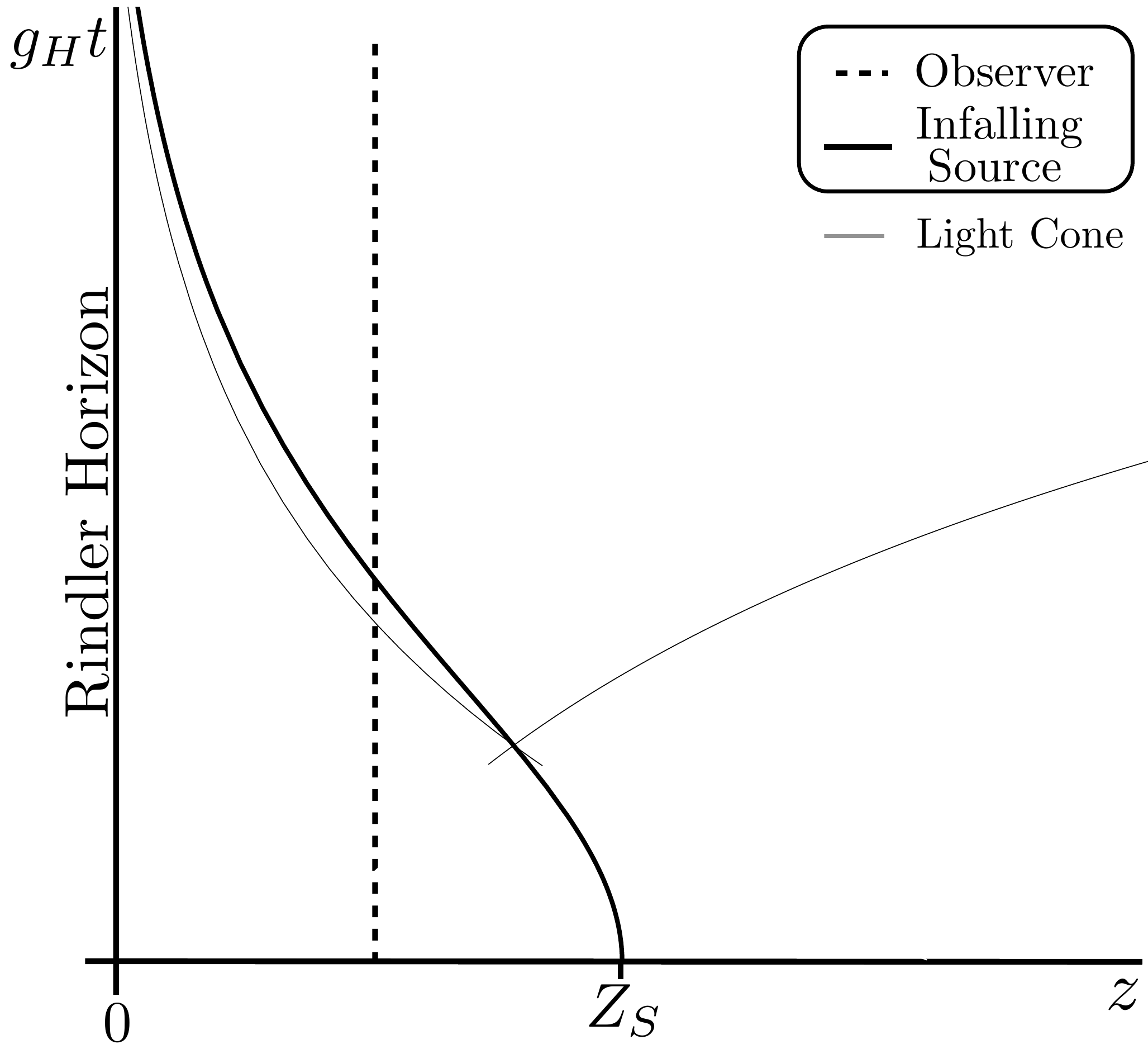}  
\end{array}$
\end{center}
\caption{Spacetime diagrams for the infalling Rindler dipole of
  section \S\ref{A Freely Falling Dipole Solution}. Also shown is the
  worldline of a Rindler observer.
 The top panel is drawn by Minkowski observers, the bottom
  panel is drawn by Rindler observers. Note that
  $Z_S = z_S$ at $T=t=0$, hence the labeling of the initial source
  position.}
\label{ST_INF}
\end{figure}

As a check of the above dipole solutions, we consider a dipole source
that is stationary in Minkowski space at the location
$(X_S=0,Y_S=0,Z_S=$constant).
According to the Rindler observer, the magnetic dipole appears to fall
straight into the event horizon.
In Minkowski coordinates, the world line is characterized by
\begin{align}
r^{\alpha} &= \nonumber
\begin{pmatrix}
 T-T_S  \\
X  \\
Y  \\
Z-Z_S
\end{pmatrix}   \\
V^{\alpha}  &=
\begin{pmatrix}
 1 \\
0  \\
0 \\
0
\end{pmatrix}\qquad
a^{\alpha}  =
\begin{pmatrix}
0  \\
0  \\
0 \\
0 \label{Eq:InF_rva}
\end{pmatrix} .
\end{align}
The trajectory is plotted as the dotted worldline in Figure \ref{ST_INF}
 as seen by Minkowski observers (top panel) and by Rindler 
 observers (bottom panel). The retarded time can be found in closed form:
\begin{equation}
T_*=T-\sqrt{X^2+Y^2+(Z-Z_S)^2}\nonumber \ .
\end{equation}
Because our source is stationary in Minkowski spacetime,
there is no dependence on $T$ in the field solutions and 
\begin{equation}
\left(r \cdot V\right)_*=  - r_*
\end{equation}

Then our 4-potential becomes simply
\begin{align}
A^{\alpha}(x) = \frac{ \epsilon^{\alpha }_{ \quad \mu 0 \sigma}r^{\mu} m^{\sigma} }{r^3} \bigg{|}_*
\end{align}
which, written more familiarly, is the potential of a stationary magnetic dipole
\begin{align}
A^0  = 0, \quad  {\bf{A}} (x) = \frac {\bf{m_S} \times \hat{\bf{r}}}{r^2}.
\end{align}
where $\hat{\bf{r}}=\textbf{r}/r$ and ${\bf{m_S}}$ is the source 3-dipole moment.

The nonzero field components as viewed by the
Minkowski observer in the rest frame of the dipole are 
\begin{align}
\EM &=0  \nonumber \\
\BM &=\frac{\left[3(\bf{m_S} \cdot \hat{\bf{r}})\hat{\bf{r}} - \bf{m_S} \right]}{r^3}.
\label{BDipMinkGen}
\end{align}

The fields as measured by our Rindler
observer are related to the Minkowski observer fields according to
the transformation law Eq.\ (\ref{Eq:trans}), which in this case simplifies to
\begin{align}
\ER &= \gR(\bbR \times \BM^{\parallel}) = \gR(\bbR \times \BM) \nonumber \\
\BR &= \BM^{\perp} +\gR \BM^{\parallel}.
\label{Eq:ERBR1}
\end{align}

We can plot the fields observed by a Rindler observer
in Rindler coordinates if we express 
$\rr$ in Rindler coordinates
\begin{equation}
\rr =\left (x,y,z\gR -Z_S \right ),
\label{Eq:rRindler}
\end{equation}
with $Z_S$ just a number for this example.

A slightly different path to the same answer is to 
transform the $4$-potential directly into Rindler coordinates and build
the Rindler observer's electromagnetic field tensor. Both approaches give the
same result, as they must.

Eq.\  (\ref{EBBCs}) gives the horizon current and charge density,
\begin{align}
\sigma_{\mathcal{H}} &\equiv \frac{\bf{E_R^{\perp}}}{4 \pi} \bigg{|}_{\mathcal{H}} =0   \nonumber \\
\mathbf{\mathcal{J}_\mathcal{H}} &\equiv \left[ \frac{1}{4 \pi}    \hat{\bb}_R \times  \alpha \mathbf{B_R^{||}}  \right]_{\mathcal{H}}  \nonumber \\
&= \frac{g_H z_\mathcal{H}}{4 \pi}  \left[    \hat{\bb}_R \times  \mathbf{B_M^{||}} \gamma_R   \right]_{z=z_\mathcal{H}}  
\label{SHJH_INF}
\end{align}
where $  \hat{\bb}_R$ is the unit normal to the Rindler horizon
and $z_\mathcal{H}$ is the position of the stretched Rindler horizon.
Since there is no charge on the horizon, there is no potential drop on
the horizon -- that is, no battery has been established. A freely
falling dipole does not generate a power supply in the Rindler limit.

\begin{figure}
\begin{center}
\includegraphics[scale=0.45]{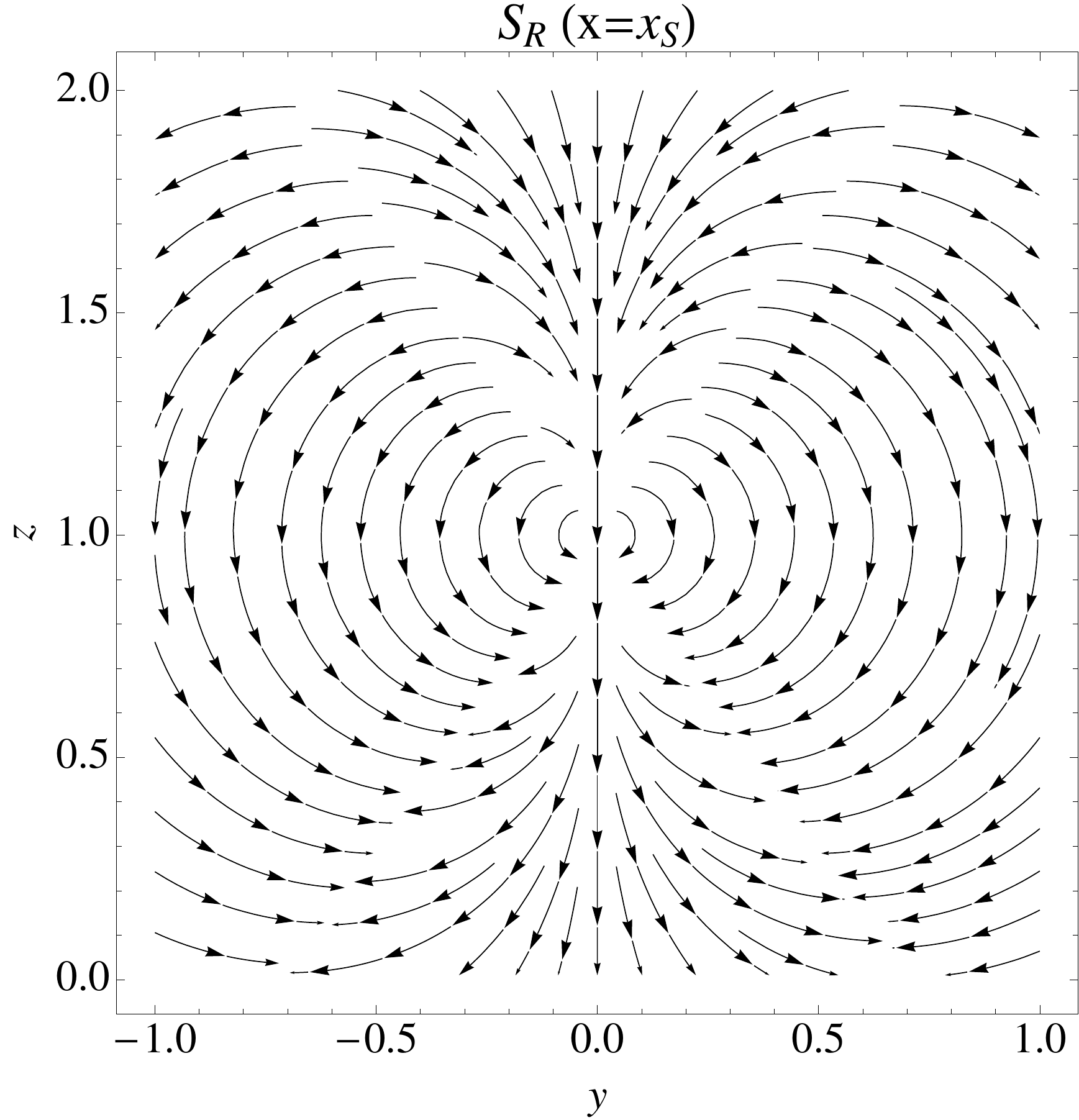} 
\end{center}
\caption{The $x=x_S=0$ slice (plane containing the dipole) of the Poynting flux for the infalling dipole
 as viewed by Rindler observers. The axes are in units of $Z_S$. }
\label{SInf}
\end{figure}

Since $\EM=0$, we already know from Eq.\ (\ref{Eq:Poynt}) that there is no outward
directed Poynting flux anywhere. Neither the Rindler observer nor the Minkowski
observer sees any radiation. For completeness,
we write the Rindler Poynting vector explicitly
\begin{align}
\mathbf{S} = \frac{g_H^2 z^2}{4 \pi}   \left[\gR \bR B_M^{\perp} \BM^{\parallel} -\gR^2 (B_M^{\parallel})^2 \bbR \right ]
\label{RPoynt_InfD}
\end{align}
and plot streamlines of $\mathbf{S}$ in Figure \ref{SInf}. 
Notice there is a component of the Poynting flux parallel to the
horizon and there is a component of
the Poynting flux into the horizon, both due to the observer's motion
outward.

\begin{figure*}
\begin{center}$
\begin{array}{cc}
\includegraphics[scale=0.33]{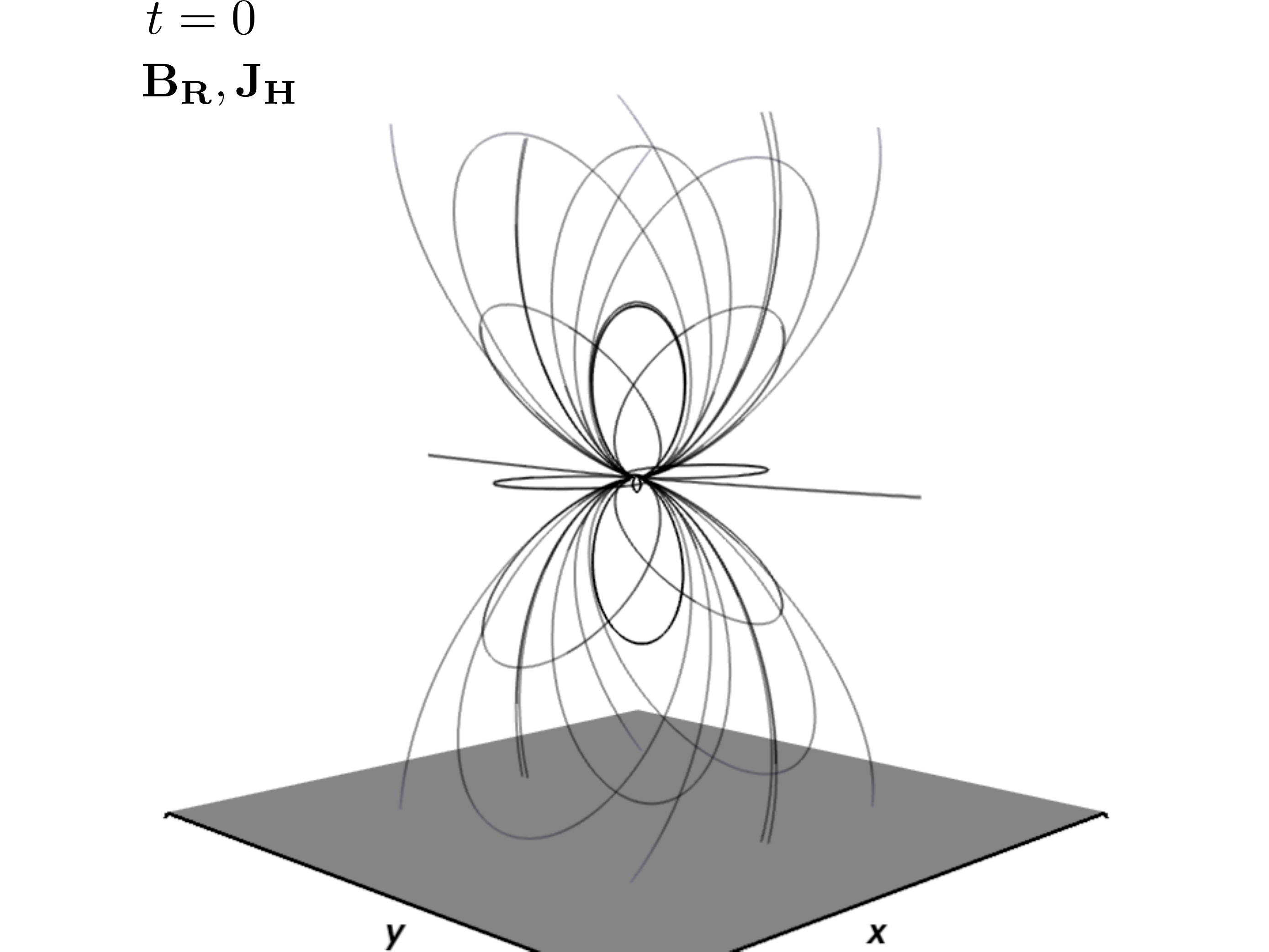} &
\includegraphics[scale=0.33]{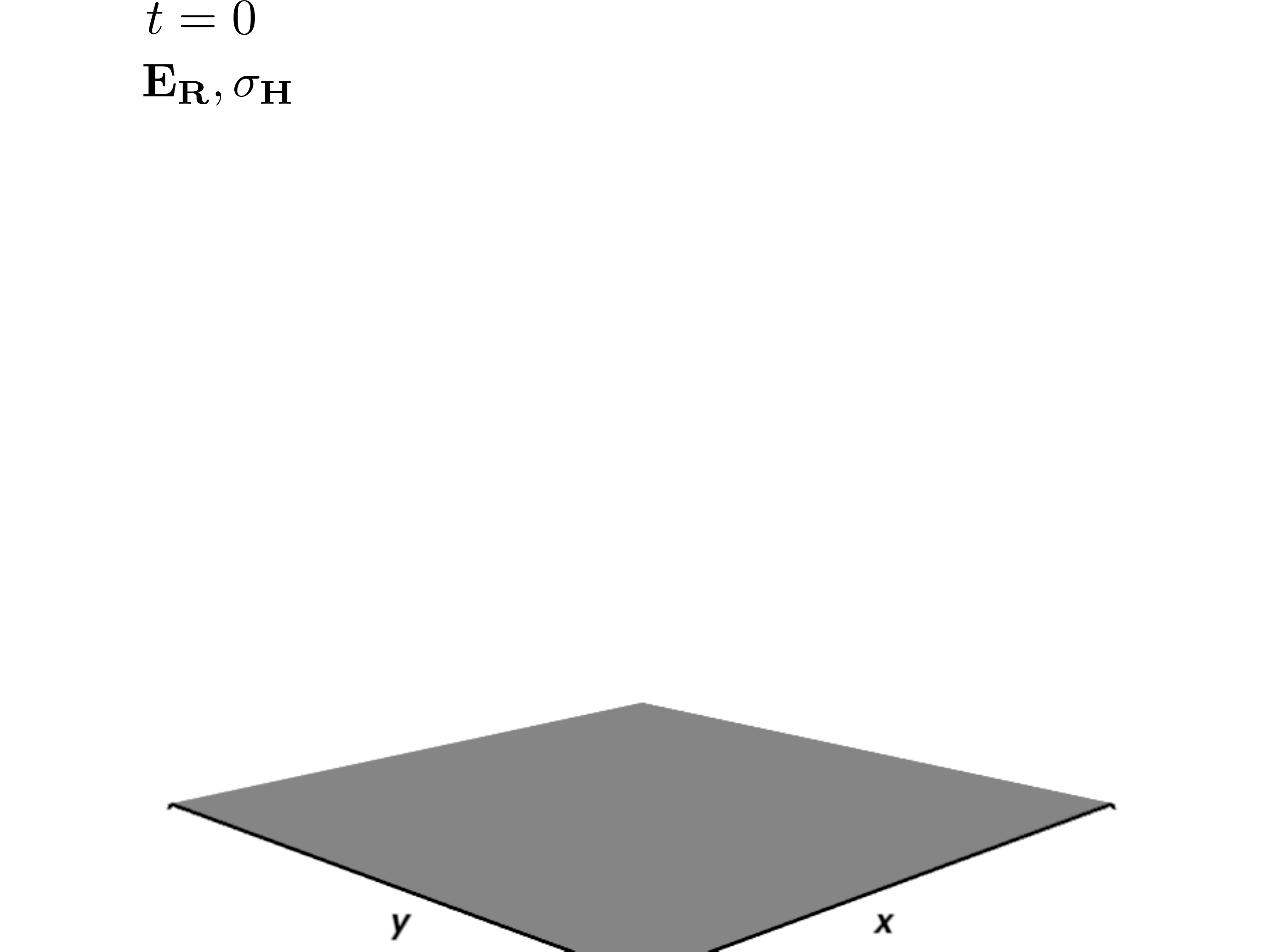}  \\ \\
\includegraphics[scale=0.33]{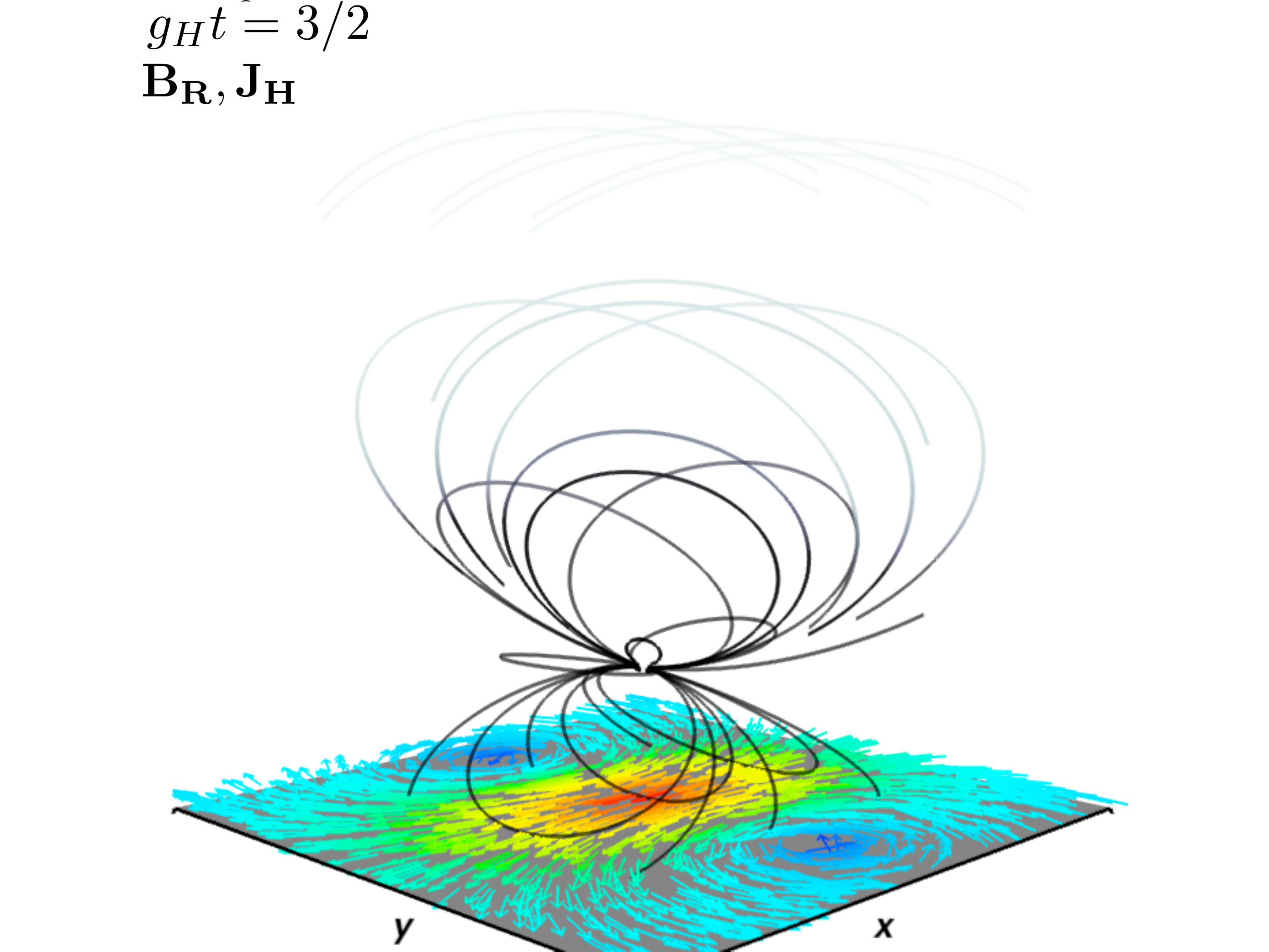} &
\includegraphics[scale=0.33]{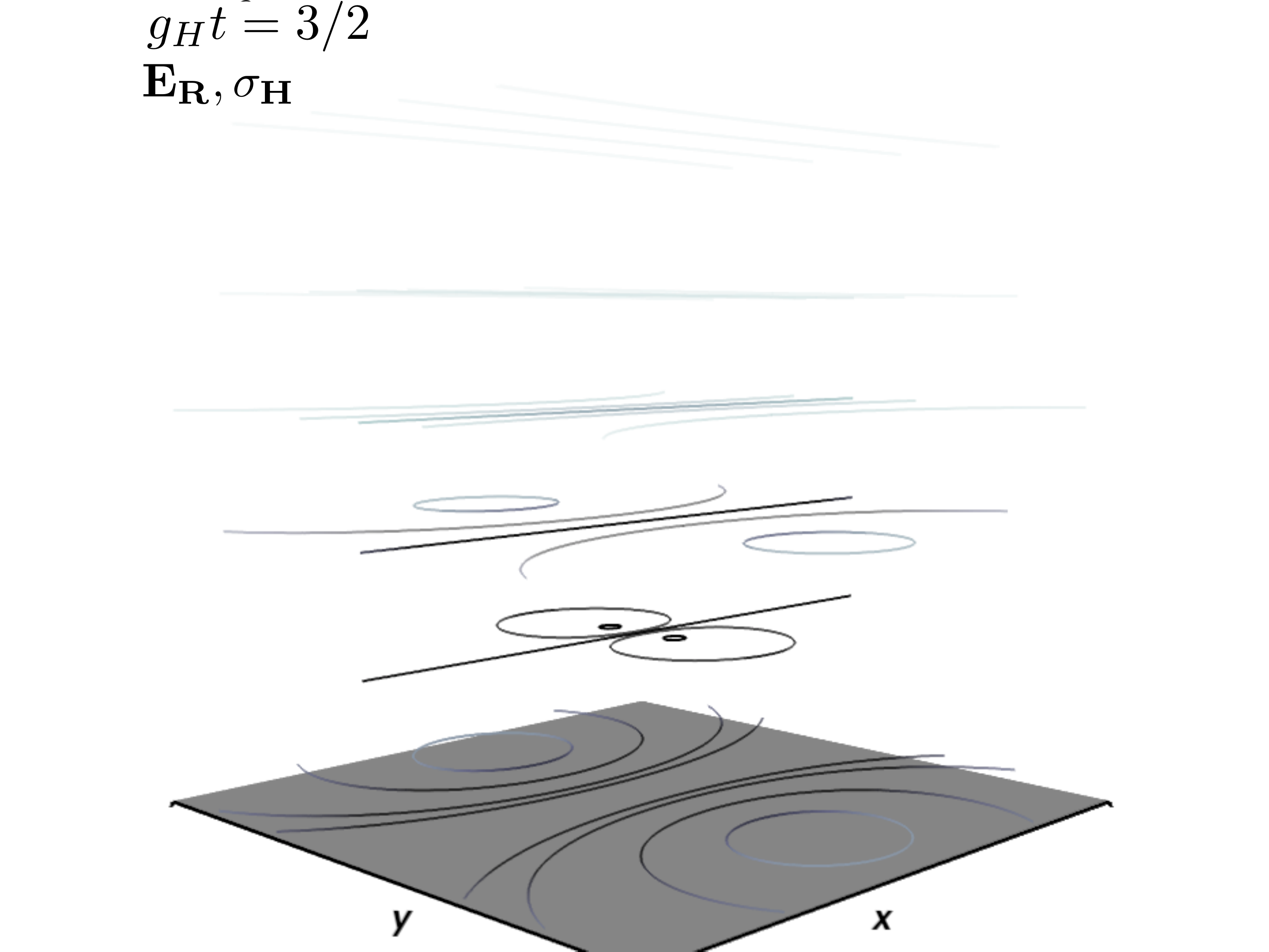} \\ \\
\includegraphics[scale=0.33]{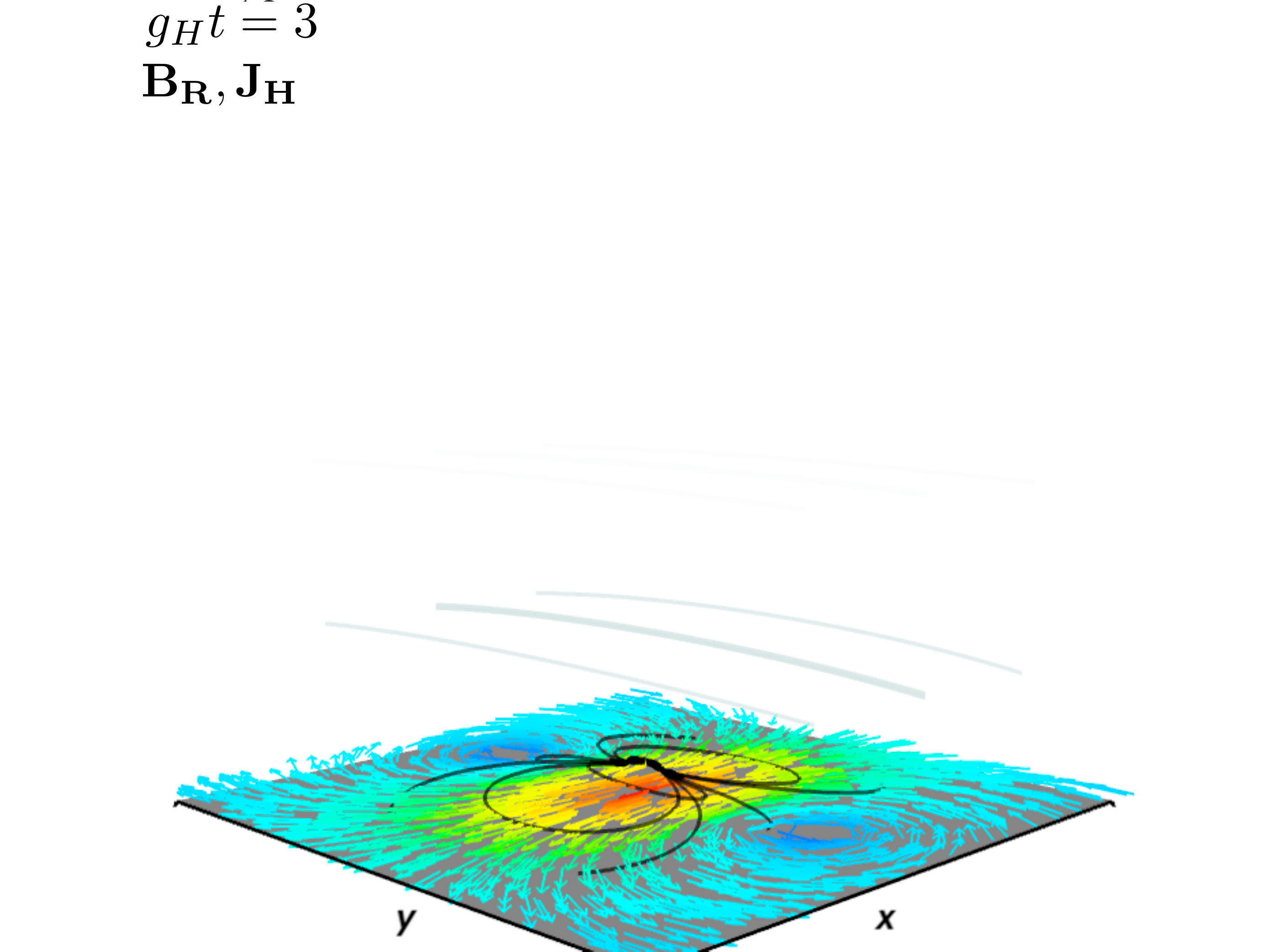} &
\includegraphics[scale=0.33]{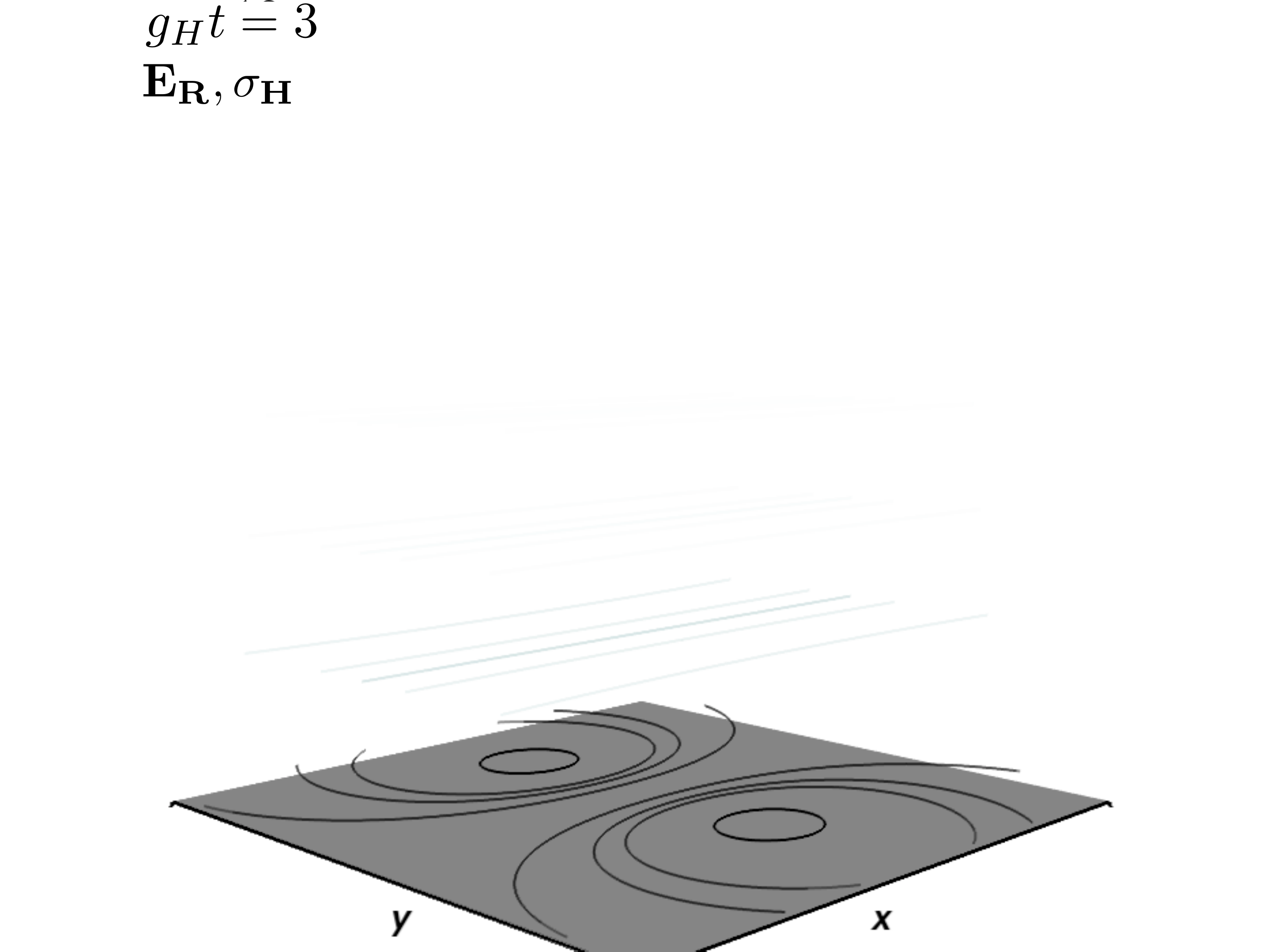} 
\end{array}$
\end{center}
\caption{3D visualization of the magnetic dipole field lines of a
  dipole falling from initial height $z_S(t=0)=Z_S$ above the Rindler
  horizon, denoted by the gray plane at $z=0$. The visualization
  region is a cube with side length $2 Z_S$. On the left, magnetic
  field lines and the corresponding horizon current densities
  $\mathbf{J_{\mathcal{H}}}$ are plotted. On the right, electric field
  lines and corresponding charge densities $\sigma_{\mathcal{H}}$ (0
  here) are plotted on the stretched horizon located at $z_H = 0.01
  Z_S$.}
\label{InfRD}
\end{figure*}

For the sake of illustration, we write out the 
components of $\ER$ and $\BR$ from Eq.\ (\ref{Eq:ERBR1}) for the infalling dipole explicitly for
the case
$\mathbf{m_S} = m \mathbf{\hat{e}_y}$.
Using $\gR=\cosh(g_Ht)$ and $\gR\bR=\sinh(g_Ht)$ and the
magnitude of $r$ from Eq.\ (\ref{Eq:rRindler}):
\begin{align}
\label{Inf_Rind_Comps}
B_R^{x} &= \frac{3 m x y}{r^5} \hbox{cosh}[g_H t] \nonumber\\
B_R^{y} &= -\frac{ m \left(r^2-3y^2 \right)}{r^5} \hbox{cosh}[g_H t] \nonumber\\
B_R^{z} &=  \frac{3 m y\left(z \hbox{cosh}[g_H t] - Z_S\right)}{r^5}   \nonumber\\
E_R^{x} &=  \frac{ m \left(r^2-3y^2 \right)}{r^5} \hbox{sinh}[g_H t]  \nonumber\\
E_R^{y} &= \frac{3 m x y}{r^5} \hbox{sinh}[g_H t] \nonumber\\
E_R^{z} &=0  
\end{align}
Using (\ref{Inf_Rind_Comps}), we plot the fields, and horizon charge
densities and currents at three different times during the infall in
Figure \ref{InfRD}.  As we have already seen from Eqs.\
(\ref{SHJH_INF}), there are no charges set up on the horizon and thus no
battery. However, there are currents moving in circles along the
horizon. These are the currents implied in the discussion surrounding
(\ref{PoyntThm}) which are responsible for dissipating the energy in
the EM fields as they pass through the horizon. Note that the
divergence of $\mathbf{\mathcal{J}_\mathcal{H}}$ in this case is 0, as can be
seen from the purely rotational nature in Figure
\ref{InfRD}. Recalling Eq.\ (\ref{HChargeCons}), we see that this
must be the case for charge conservation to hold in vacuum where
currents normal to the horizon, $j_{n}$, must be zero.

The freely-falling worldline provides a helpful test of our solutions,
but no power for an electromagnetic circuit. This system would
remain dark, unlike the orbit we explore in the next
section. (For an actual BH-NS system at separations that probe spatial curvature, a 
  battery may be established even with pure infall. The effect is not captured here in a
  flat-wall limit.)

\section{A Boosted, Freely Falling Dipole Solution}
\label{A Boosted, Freely Falling Dipole Solution}

As a second test of the solutions, consider a source that
stays at constant $Z_S$ in Minkowski but is boosted in 
the $X$, $Y$ plane. Relative to our Rindler observer, 
the dipole will appear to fall through the horizon but on an arc.

\begin{figure}
\begin{center}
\includegraphics[scale=0.4]{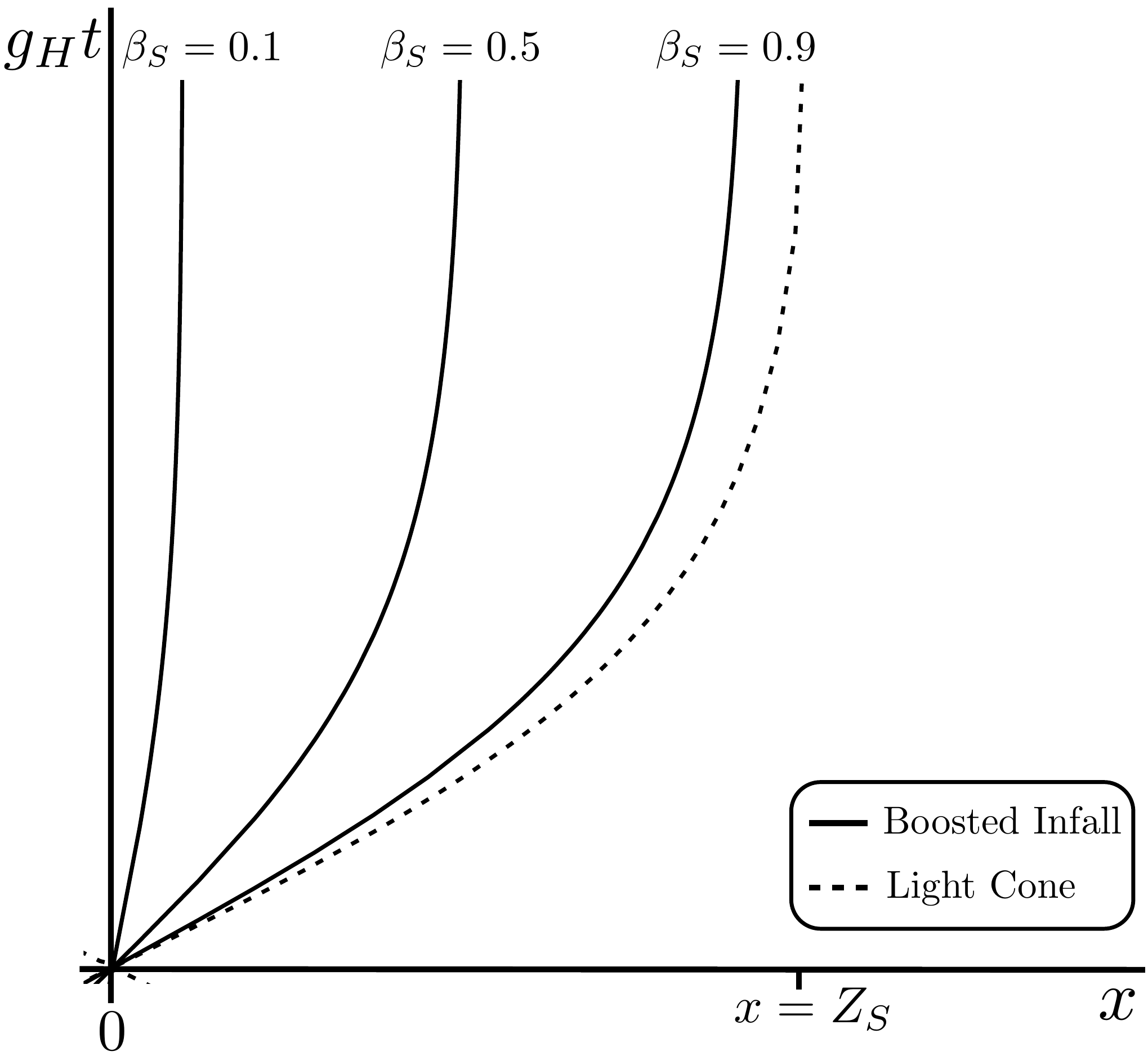} 
\end{center}
\caption{Spacetime diagram depicting the $x$-component of the
  infalling, boosted dipole worldline of \S\ref{A Boosted, Freely
    Falling Dipole Solution} from the Rindler observer's perspective,
  for three different values of $\beta_S$. The z-component of the
  worldline is identical to that portrayed for the infalling dipole in
  the bottom panel of Figure \ref{ST_INF} except the light cone
  structure is altered. Because the infalling boosted dipole
  approaches the speed of light in the z-direction, the
 motion in the x-direction must go to zero ($dx/dt\rightarrow 0$). This is evident from the
  worldlines in this figure which asymptote to vertical lines.}
\label{ST_INFBoost}
\end{figure}

Taking the boost to be at constant velocity in the $X$-direction, as seen by the Minkowski observer, we have
\begin{align}
\label{Eq:BoostedFF}
r^{\alpha}  &=
\begin{pmatrix}
T-T_S   \\
X-\beta_S T_S \\
Y  \\
Z -Z_S
\end{pmatrix} \\
V^{\alpha} &=
\begin{pmatrix}
 \gamma_S \\
\gamma_S \beta_S \\
0 \\
0
\end{pmatrix} \qquad
a^{\alpha} =
\begin{pmatrix}
0  \\
0  \\
0  \\
0
\end{pmatrix} \nonumber
\end{align}
The coordinate $Z_S$ is again simply a number, $\beta_S$ and $\gamma_S =
\left( 1 - \beta^2_S\right)^{-1/2}$ are also constant. It is important
to note that although the Minkowski observer sees the dipole boosted
in the $X$-direction at a constant velocity, the Rindler observer sees the
dipole slow down in the $x$-direction as it speeds up in the $z$. This is 
illustrated in Figure \ref{ST_INFBoost}.

The retarded time can be found in closed form:
\begin{align}
T_*=  & \gamma^2_S{(T-\beta_S X)}-    \\
& \gamma_S \sqrt{ \gamma^2_S \left({T-\beta_S X}\right)^2  {-T^2+X^2+Y^2+(Z-Z_S)^2} } \nonumber 
\end{align}
so that 
\begin{align}
\left(r \cdot V\right)_*= - \left[ \gamma^2_S (X-\beta_S T)^2+Y^2+(Z-Z_S)^2 \right]^{1/2}  
\label{rV_InfB}
\end{align}
Eq.\ (\ref{DipleSolnLong}) with $a = 0$ and $\mathbf{\dot{m}} = 0$ then gives the 4-potential for a boosted Minkowski dipole. 
From the 4-potential or the field tensor, Minkowski $\EM$ and $\BM$
can be derived from Eq.\ (\ref{Eq:EMBM}) or (\ref{EBtoF}).

As a check, we may also derive the electromagnetic fields by writing the field tensor for
a dipole in the rest frame of a Minkowski observer, transform to a
boosted frame, and then transform to the accelerated Rindler
frame. 

Let the reference frame of the Minkowski observer at rest with
respect to the dipole be denoted by a double prime, the frame of the 
Minkowski observer boosted relative to the source by a single prime, and the Rindler frame by no
prime. Then the field tensor $F^{\alpha'' \beta''} (X^{\mu''})$ is
constructed from Eq.\ (\ref{BDipMinkGen}). The field tensor in the
Minkowski boosted frame is given by
\begin{equation}
 F^{\alpha' \beta'} (\mathbf{X'}) = \Lambda^{\alpha'}_{\sigma''} \Lambda^{\beta'}_{\rho''} F^{\sigma'' \rho''} (\mathbf{X''})
 \label{LLFboost}
\end{equation}
where $\Lambda^{\alpha'}_{\sigma''}$  is the Lorentz transformation for a boost in the $X$ direction. The boosted coordinates $X^{\mu'}$ are given in terms of the rest frame coordinates $X^{\mu''}$ via an inverse Lorentz transformation. The Rindler $\bf{E_R}$ and $\bf{B_R}$ fields are then found via Eq.\  (\ref{RindWL}) and (\ref{EBtoFRind}),
\begin{align}
E_R^{\alpha} (\bf{x}) &= \frac{\partial x^{\alpha}}{\partial x^{\alpha'}} F^{\alpha' \mu'} {(\bf{X'})} u^R_{\mu'}   \nonumber \\  
B_R^{\alpha} (\bf{x}) &= \frac{1}{2} \frac{\partial x^{\alpha}}{\partial x^{\alpha'}} \epsilon^{\alpha' \mu' \gamma' \delta'}F_{\gamma' \delta'}(\mathbf{X'}) u^R_{\mu'}.
\label{EBRindfromFB}
\end{align}
where $u^R_{\mu'}$ are the components of the Rindler observer's
4-velocity as viewed by the boosted Minkowski observer. 
The $X^{\mu'}(x^{\mu})$ are given by Eqs.\  (\ref{MtoR}) and we have
again kept Rindler coordinates lower case while Minkowski coordinates
are upper case.

Carrying out the above procedure, we start with an observer co-moving with the dipole. This observer sees fields,
\begin{align}
{\bf E_{M^{\prime \prime}}} & = 0 \nonumber \\
{\bf B_{M^{\prime \prime}}} & = \frac{3 \hr^{\prime\prime} ({\bf
    m_S}\cdot \hr^{\prime\prime})-{\bf m_S}}{\rr^{\prime\prime \ 3}} 
\label{Eq:MinkD}
\end{align}
where ${\bf m_S}$ is the constant rest-frame value of the dipole's
magnetic moment and 
$\rr^{\prime\prime}$ is a radial coordinate in the rest-frame of the dipole. 
There exists another Minkowski observer boosted by $-\bb_S$ relative to
the source who measures the fields
\begin{align}
{\bf E_{M^\prime}} & =  -\g_S(\bb_S\times {\bf B_{M^{\prime \prime}}})  \nonumber \\
{\bf B_{M^\prime}} & ={ \bf B_{M^{\prime \prime}}} +(\g_S-1) \hat {\bb}_S\times ({\bf B_{M^{\prime \prime}}}\times \hat {\bb}_S)
\label{Eq:transstep}
\end{align}
We obtain the Rindler fields by an application of Eq.\  (\ref{Eq:trans}):
\begin{align}
\ER & =\g_S \gR \left( \bb_R \times \mathbf{ B_{M^{\prime \prime}}} \right)  \nonumber \\
-&  \left[ \g_S \hat{\bb}_R \cdot \left(\bb_S \times \mathbf{ B_{M^{\prime \prime}}}   \right) \right] \hat{\bb}_R  \nonumber \\
-& \left[   \g_R  \mathbf{ B_{M^{\prime \prime}}} \cdot \left(  \left( \g_S -1\right) |\bb_R| \hat{\bb}_S - \g_S |\bb_S|  \hat{\bb}_R \right) \right] \left( \hat{\bb}_R \times \hat{\bb}_S  \right)
 \nonumber \\ 
\BR & = \g_S \gR  \mathbf{ B_{M^{\prime \prime}}} \nonumber \\
-& \left[ \g_S \left( \g_R -1 \right) \left(  \mathbf{ B_{M^{\prime \prime}}}  \cdot \hat{\bb}_R\right) \right] \hat{\bb}_R \nonumber \\
-& \left[\g_R   \mathbf{ B_{M^{\prime \prime}}}  \cdot  \left( \left(\g_S -1 \right) \hat{\bb}_S  - \g_S |\bb_S| \bb_R\right)   \right] \hat{\bb}_S
\label{Eq:ERBR_BINF}
\end{align}

The horizon charge and current densities are
\begin{align}
\sigma_{\mathcal{H}} &\equiv \frac{\bf{E^{\perp}_R}}{4 \pi} \bigg{|}_{\mathcal{H}} = -\frac{ \hat{\bb}_R \cdot (\gamma_S \bb_S \times  \mathbf{B}^\parallel_{M^{\prime \prime}}  )_{z=z_\mathcal{H}}  }{4 \pi}   \nonumber \\
\mathbf{\mathcal{J}_\mathcal{H}} &\equiv \left[ \frac{1}{4 \pi}    \hat{\bb}_R \times  \alpha \mathbf{B_R^{||}}  \right]_{\mathcal{H}}  \nonumber \\
&= \g_R  \frac{g_H z_\mathcal{H}}{4 \pi}  \left[ \g_S  \left(    \hat{\bb}_R \times  \mathbf{B}^\parallel_{M^{\prime \prime}}    \right)\right. \nonumber \\
-& \left.     \mathbf{ B_{M^{\prime \prime}}}  \cdot  \left( \left(\g_S -1 \right) \hat{\bb}_S  - \g_S |\bb_S| \bb_R\right)      \left(\hat{\bb}_R \times \hat{\bb}_S \right) \right]_{z=z_\mathcal{H}} 
\label{SHJH_BINF}
\end{align}
This example manifests charge separation and therefore a voltage drop
across the event horizon. We have established a BH battery. 

To express these Rindler fields in Rindler coordinates, we perform a 
Lorentz transformation on the Minkowski 4-vector $r^{\prime \prime}$ 
for a boost in the x-direction and use Eqs.\ (\ref{MtoR}) to write,
\begin{equation}
\rr^{\prime \prime} = (\g_S \left(x-\beta_S z \hbox{sinh}[g_H t] \right), y, z \hbox{cosh}[g_H t] -Z_S).
\label{r_BInf}
\end{equation}
Eqs.\ (\ref{r_BInf}), (\ref{Eq:MinkD}), and (\ref{Eq:ERBR_BINF}) then give the Rindler fields in Rindler coordinates.

The Rindler fields derived in this manner agree with the fields
derived from inserting (\ref{Eq:BoostedFF}) into the 4-potential as
they must.

\begin{figure*}
\begin{center}$
\begin{array}{cc}
\includegraphics[scale=0.33]{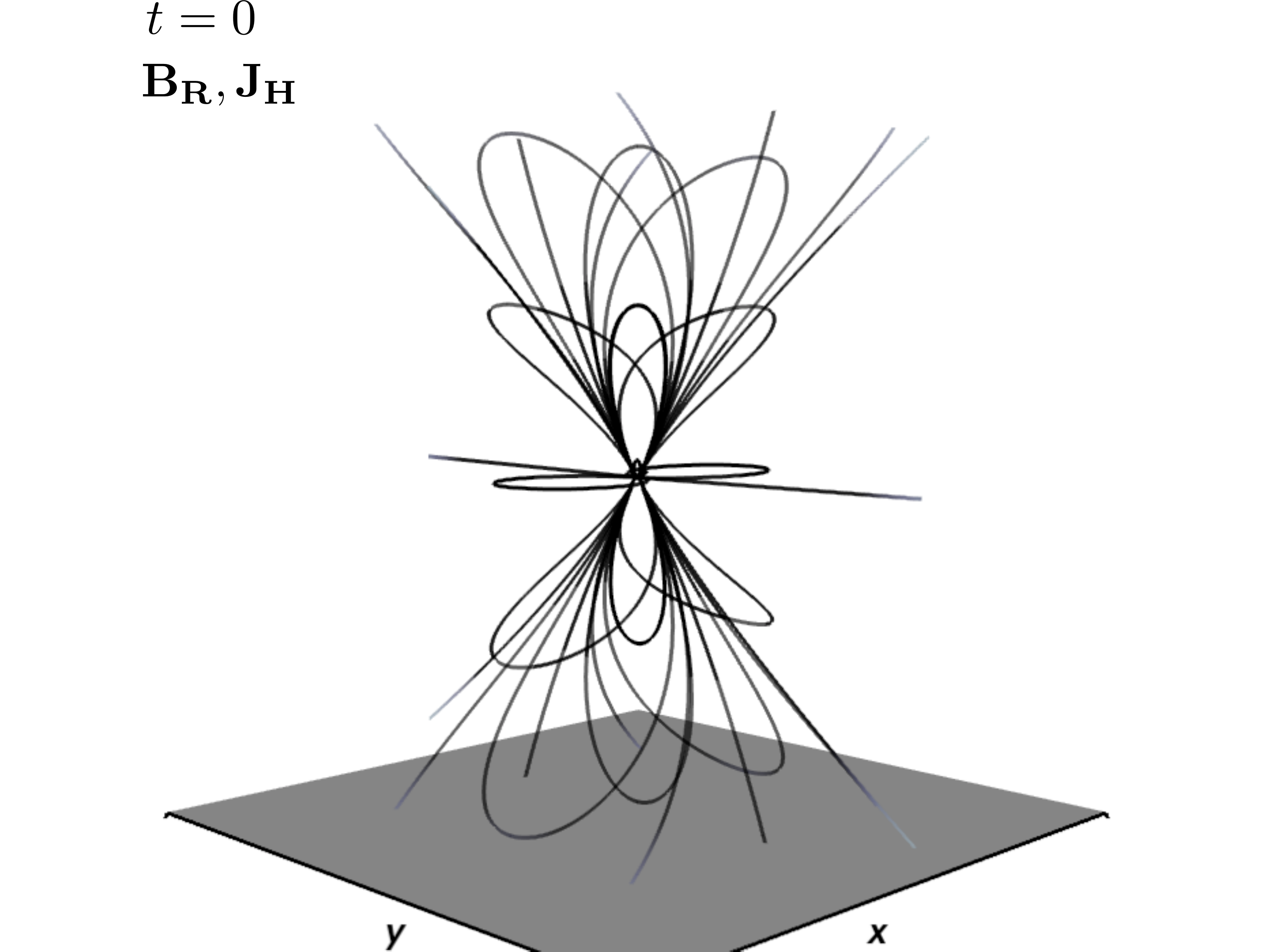} &
\includegraphics[scale=0.33]{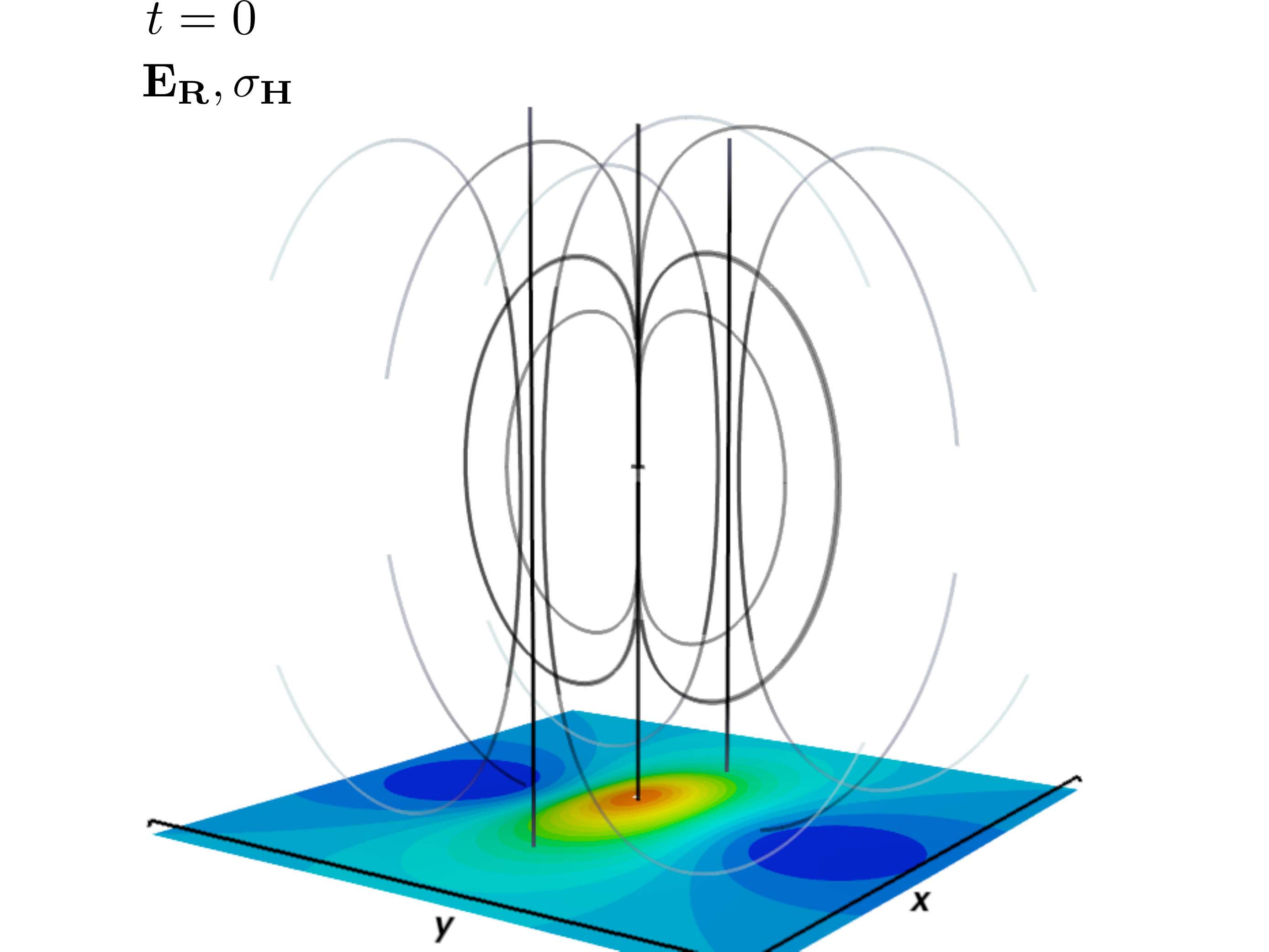}  \\ \\
\includegraphics[scale=0.33]{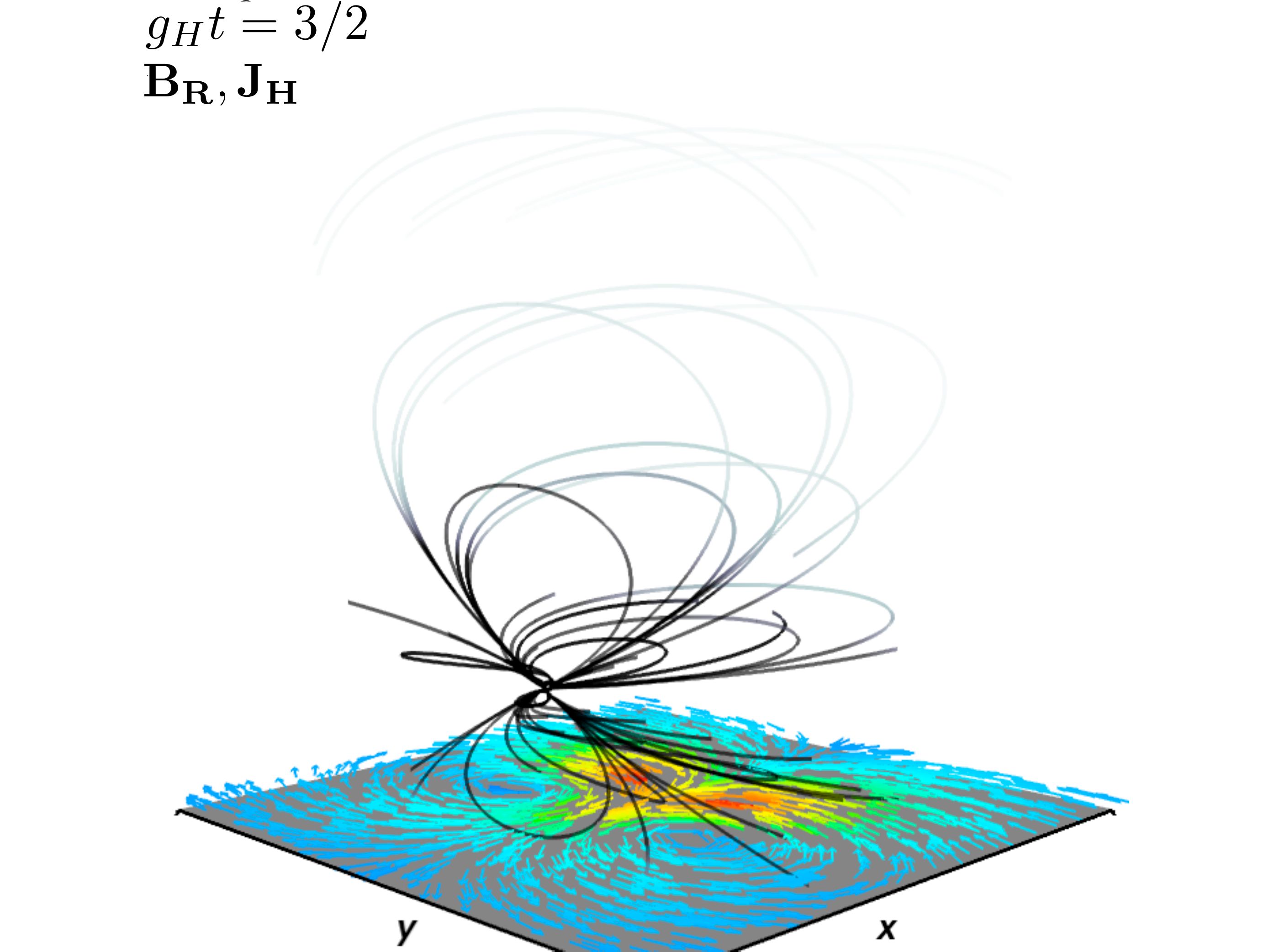} &
\includegraphics[scale=0.33]{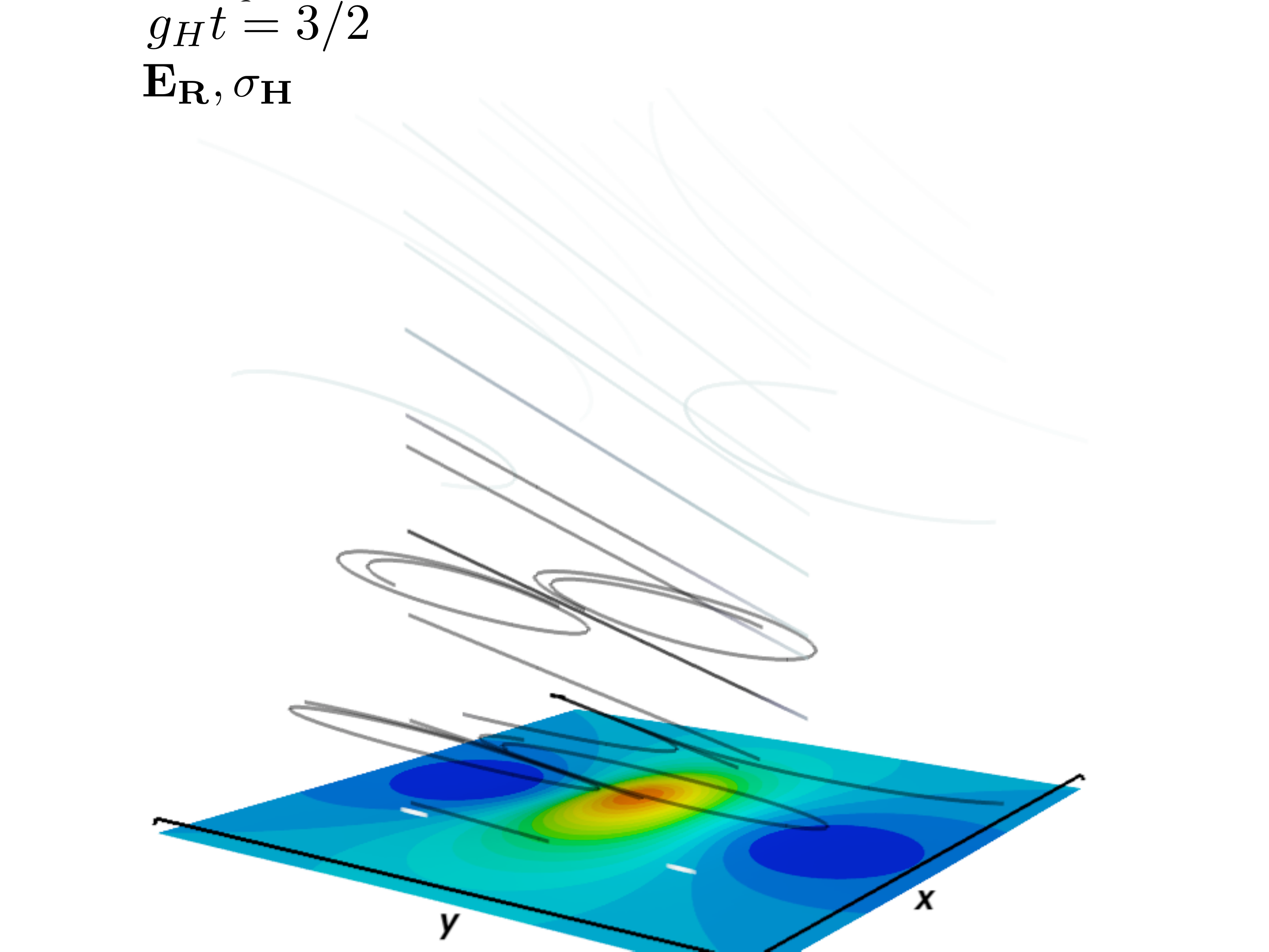} \\ \\
\includegraphics[scale=0.33]{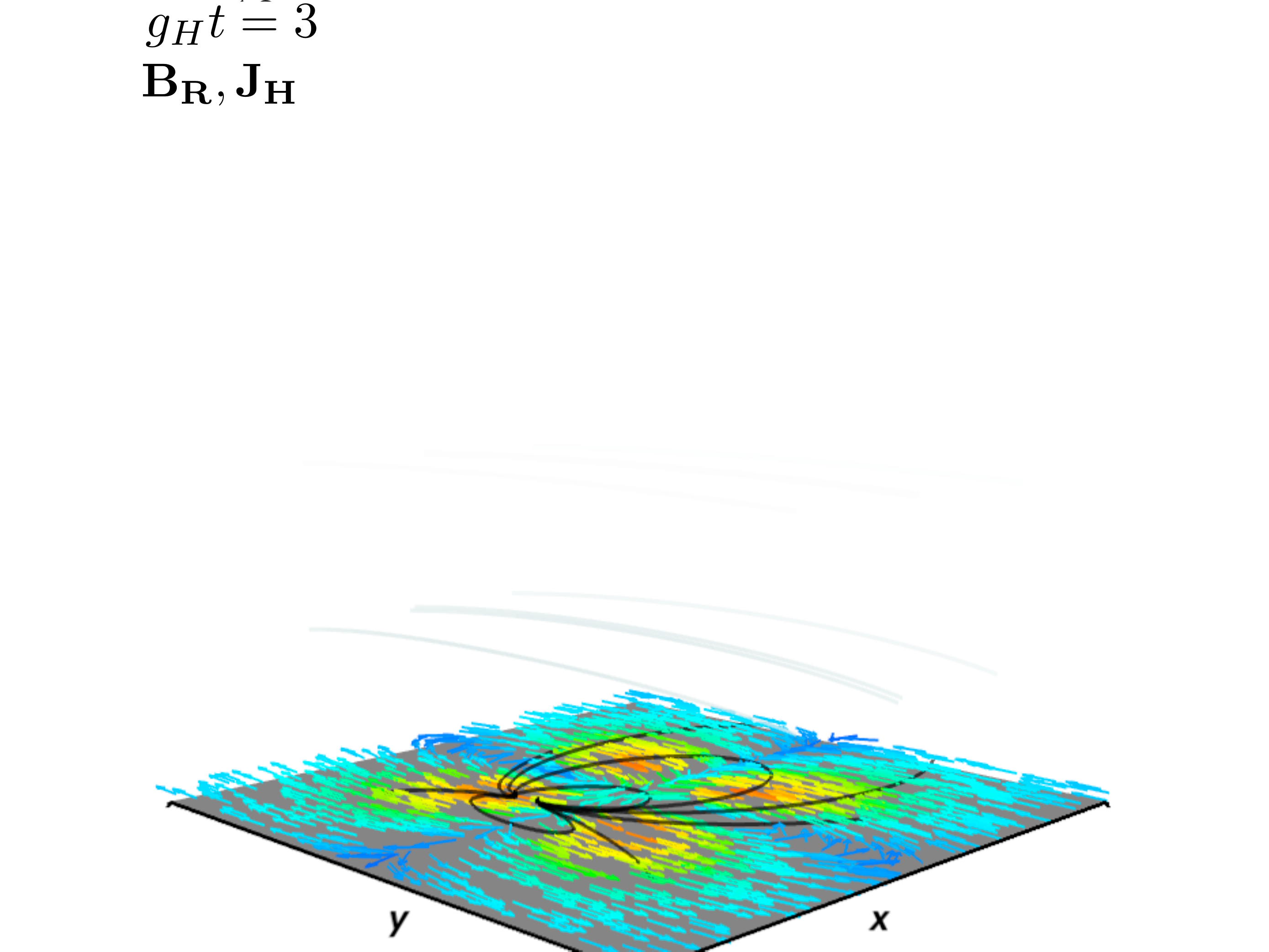} &
\includegraphics[scale=0.33]{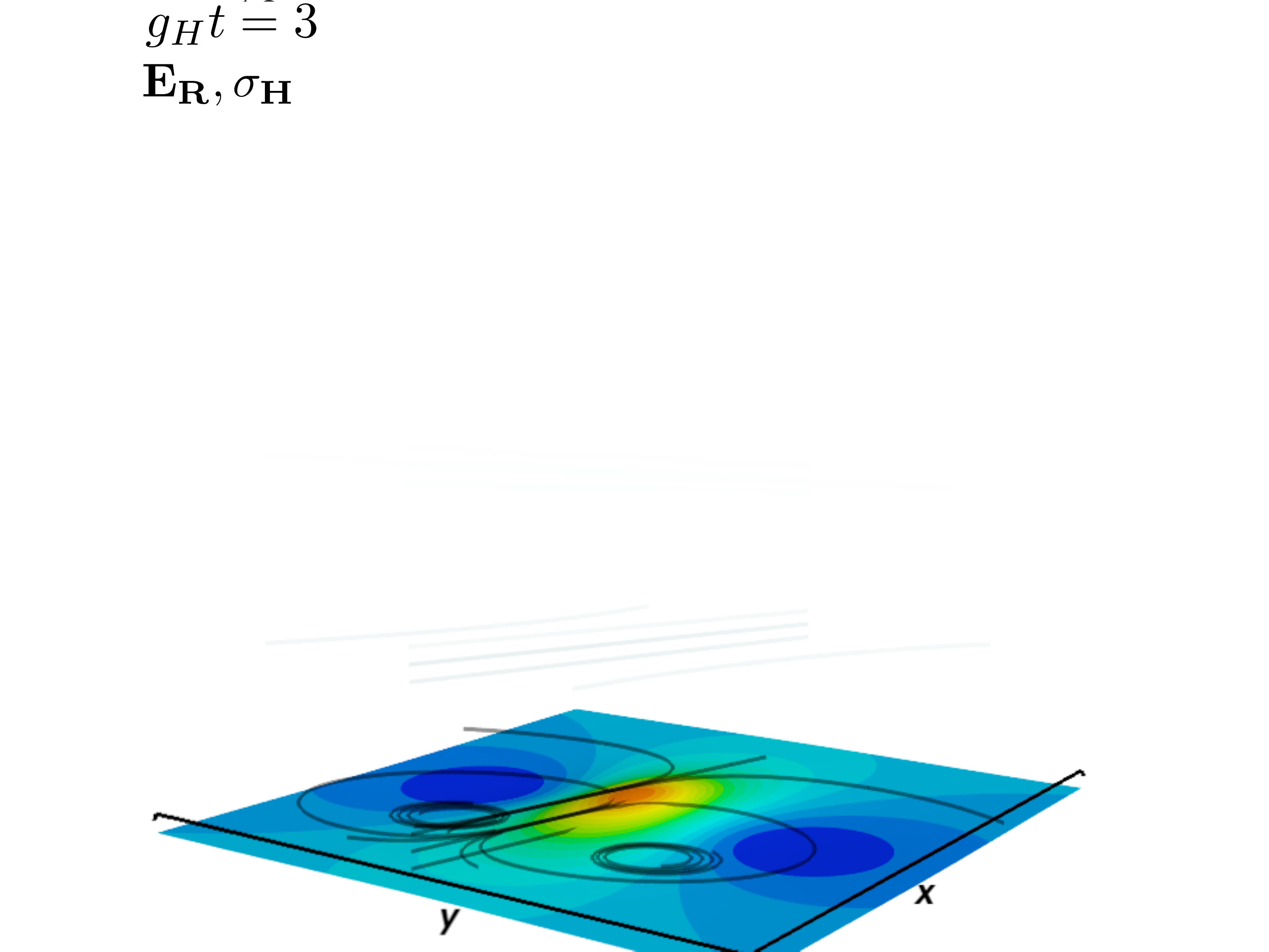} 
\end{array}$
\end{center}
\caption{3D visualization of the field lines of a dipole spiraling
  into the Rindler horizon from initial height $z_S(t=0)=Z_S$ with an
  initial boost of  $\beta_S=0.9$ in the $x$-direction.  The
  visualization region spans from $-2Z_S$ to $2Z_S$ in the x- and
  y-directions and extends $2 Z_S$ above the Rindler
  horizon. Surface currents $\mathbf{J_{\mathcal{H}}}$ and surface
  charge densities $\sigma_{\mathcal{H}}$ are plotted on the stretched
  horizon located at $z_H = 0.01 Z_S$.}
\label{BInfRD}
\end{figure*}

Choosing $\mathbf{m_S} = m \mathbf{\hat{e}_y}$, given that we boost in
the x-direction, leads to the simplest form for the observed 4-dipole moment,
\begin{equation}
m_R^\mu= m_{M^{\prime \prime}}^\mu= 
m_{M^{\prime}}^\mu=
 (0,{\bf m_S}).
\end{equation}
We write out the components of $\ER$ and $\BR$ for the boosted, infalling dipole explicitly:
\begin{align}
B_R^{x} &= \gamma_S \frac{ 3m y \left(  x \hbox{cosh}[g_H t] -\beta_S Z_S \hbox{sinh}[g_H t] \right)}{r^5} \nonumber\\
B_R^{y} &= - \gamma_S \frac{ m \left(r^2 -3 y^2 \right)}{r^5}  \hbox{cosh}[g_H t]  \nonumber\\
B_R^{z} &=   \gamma_S \frac{3 m y\left(z \hbox{cosh}[g_H t] - Z_S \right)}{r^5}   \nonumber\\
E_R^{x} &= \gamma_S \frac{ m \left( r^2 - 3y^2 \right)}{r^5} \hbox{sinh}[g_H t]  \nonumber\\
E_R^{y} &=  \gamma_S \frac{3 m y \left\{ x \hbox{sinh}[g_H t] + \beta_S \left(z - Z_S \hbox{cosh}[g_H t]  \right)  \right\} }{r^5} \nonumber\\
E_R^{z} &=  \gamma_S \frac{ m \left(  r^2 - 3y^2 \right)}{r^5} \beta_S 
\label{InfBoostEB}
\end{align}
where $r$ is the RHS of (\ref{rV_InfB}) in Rindler coordinates. Using the above, we plot the fields, and horizon charge and current densities, given by Eqs.\  (\ref{SHJH_BINF}), at three different times during the inspiral in Figure \ref{BInfRD}.

\begin{figure}
\begin{center}$
\begin{array}{c}
\includegraphics[scale=0.38]{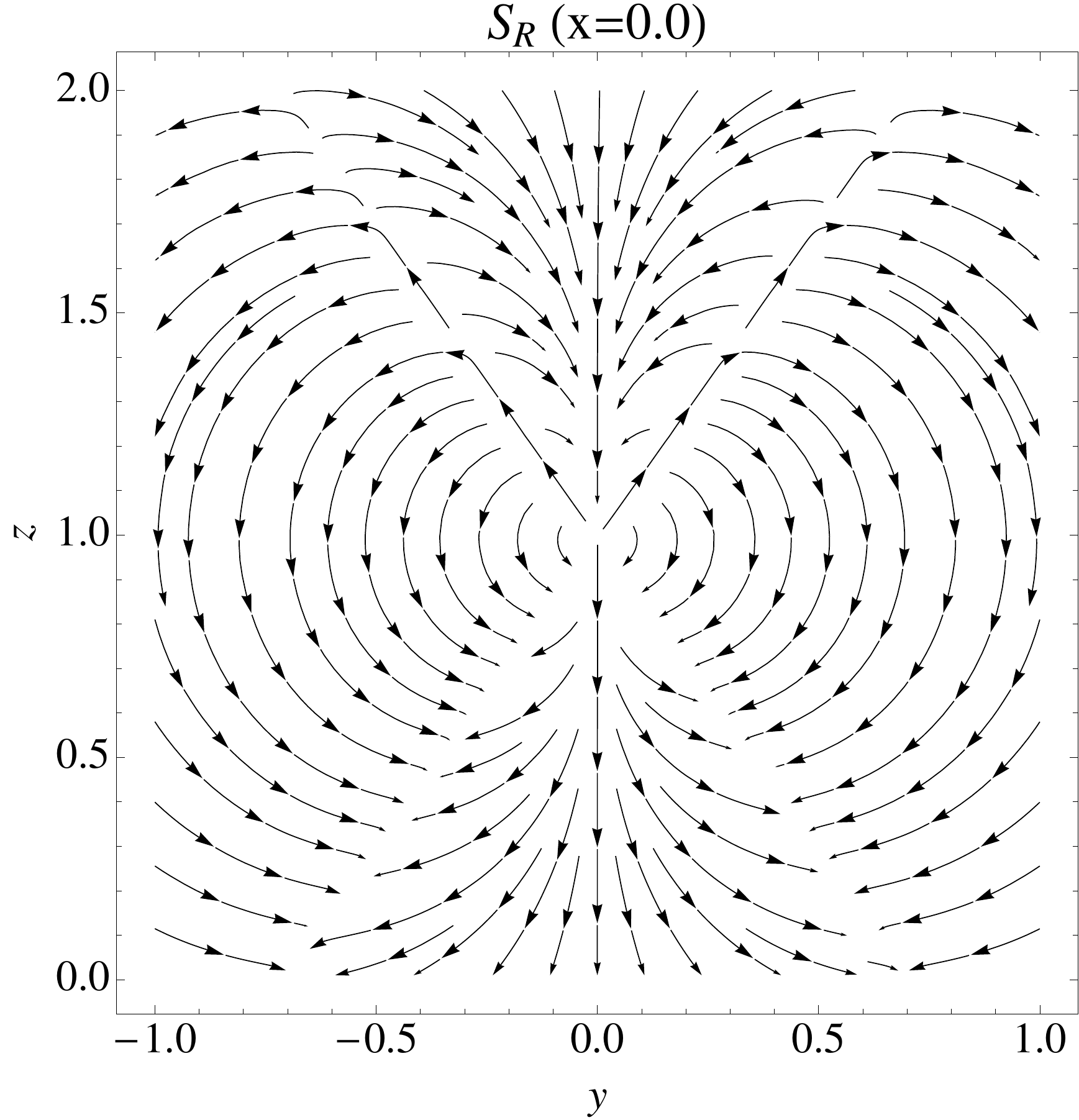} \\
\includegraphics[scale=0.38]{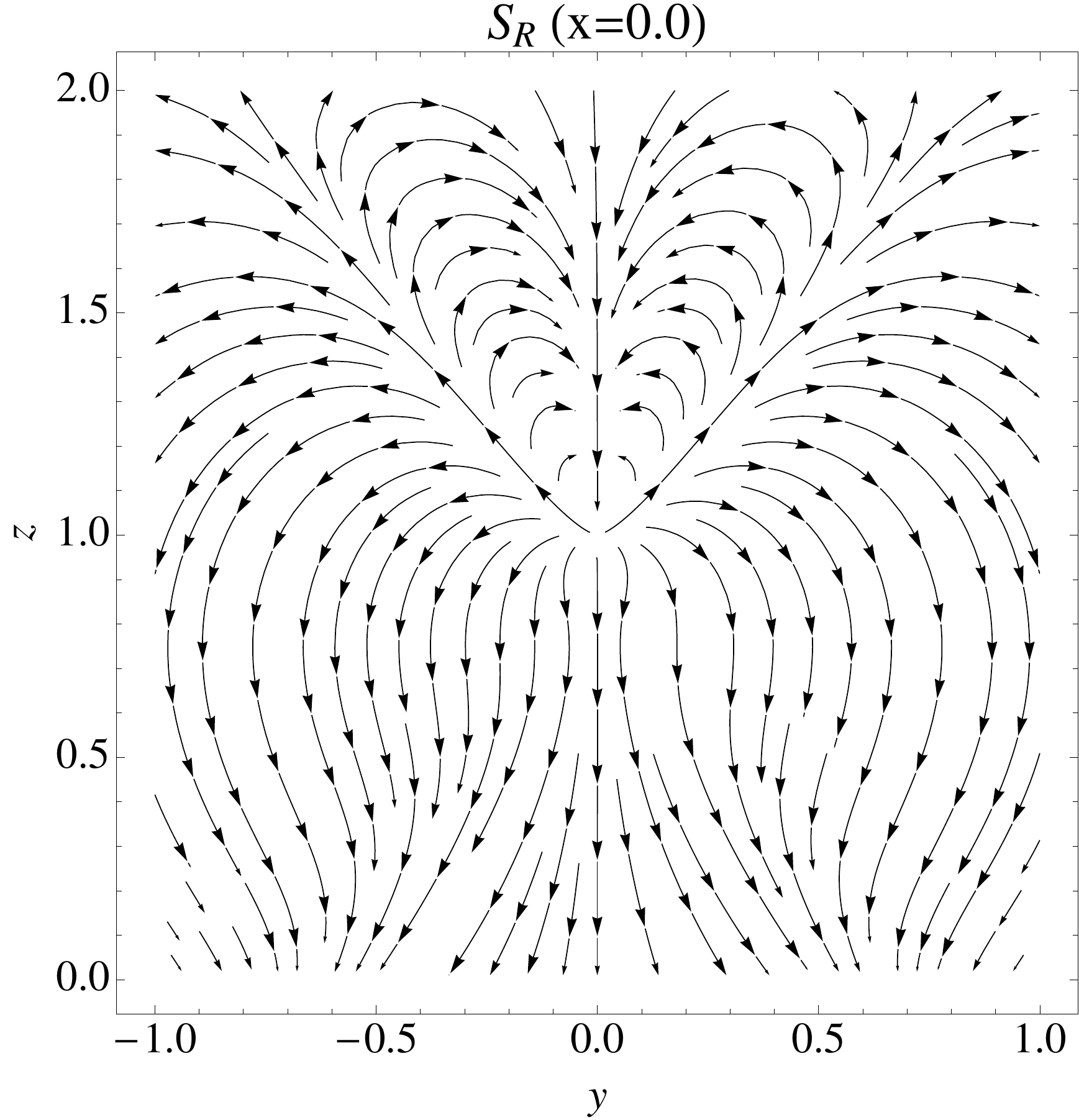}  \\
\includegraphics[scale=0.38]{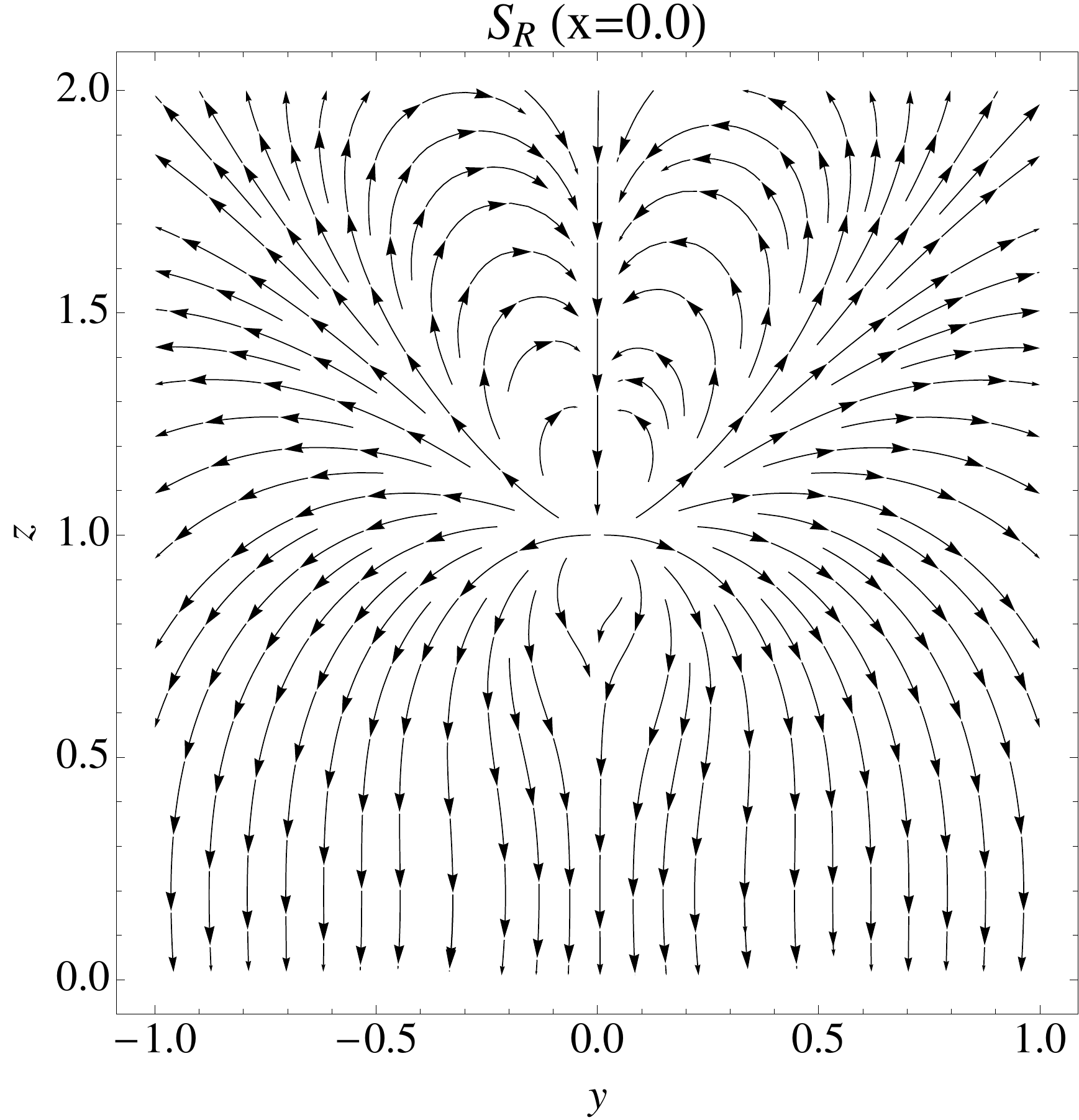} 
\end{array}$
\end{center}
\caption{An $x=x_S=0$ slice of the Poynting flux for the infalling boosted dipole of
  \S \ref{A Boosted, Freely Falling Dipole Solution} as viewed by Rindler observers for three different
  boost magnitudes in the $x$-direction, $\beta_S=0.1, 0.5, 0.9$. The Poynting flux is 0 at infinity despite outward components of the field in the region plotted here. The axes are in units of $Z_S$.}
\label{BInf_Poynt}
\end{figure}

Figure \ref{BInf_Poynt} shows the Poynting flux generated by the above
fields for $\beta_S = 0.1, 0.5, 0.9$. The Poynting flux is directed
into the horizon below the dipole signifying the dissipation of the
field energy into the horizon (via ohmic dissipation from horizon
currents). The increasingly uniform $z$ component of the Poynting flux
for increasing $\beta_S$ is due to the increasing disparity between
$t$ and $t_*$ (observers see further into the relative past of the dipole) for larger $\beta_S$ and smaller $z$. There is no
observed Poynting flux at infinity in this case and hence no radiation
from the moving dipole in vacuum. We elaborate on the above points further in  \S\ref{c}. 

\begin{figure}
\begin{center}$
\begin{array}{c}
\includegraphics[scale=0.38]{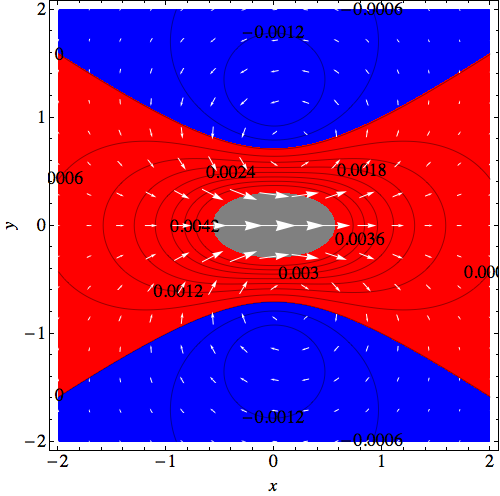} \\
\includegraphics[scale=0.38]{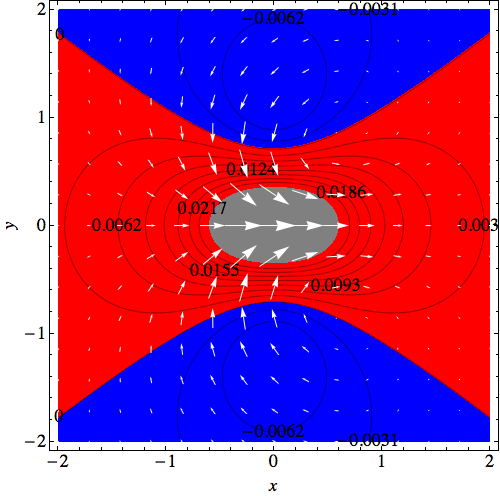}  \\
\includegraphics[scale=0.38]{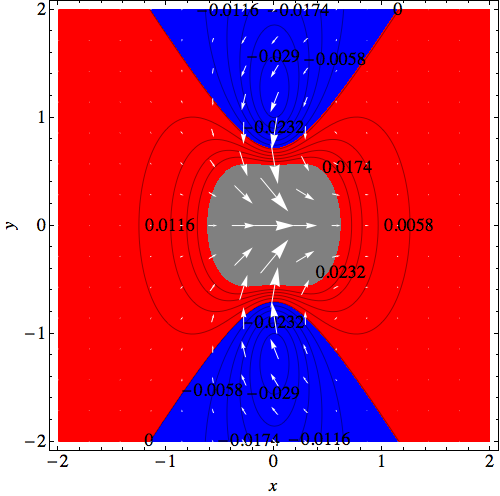} 
\end{array}$
\end{center}
\caption{Current density vectors (white) overlaid on contours of
  charge density on the stretched horizon ($\alpha_\mathcal{H}=10^{-4}$) of the infalling boosted
  dipole with rest frame magnetic moment in the $y$-direction. From
  top to bottom, the magnitude of the boost in the $x$-direction
  increases from $\beta_S=0.1, 0.5, 0.9$.  As inferred from the last of
  Eqs.\  (\ref{InfBoostEB}), the magnitude of the charge density
  increases with $\beta_S$. Also the shape of the charge separation is
  squeezed in the direction of source boost as indicated by
  Eq.\  (\ref{HCharge_IB}). All of the snapshots are taken at $g_H t= 1$
  and the contour labels are arbitrarily scaled. The gray regions are
  regions of steeply increasing $\sigma_\mathcal{H}$ which have 
  been removed to more clearly view the contour structure. The axes are in units of $Z_S$. }
\label{BInf_SigvBet}
\end{figure}

Figure \ref{BInfRD}, as well as the expression for $E_R^{z}$ and
Eq.\  (\ref{EBBCs}) for the horizon charge density, indeed confirm
that charge separation occurs on the stretched horizon. Figure \ref{BInf_SigvBet} explores the horizon
charge density further. As we will elaborate in \S\ref{c}, this charge
separation can be considered a result of tangential components of the
dipole magnetic field sourcing horizon currents via
(\ref{EBBCs}). Because no currents are entering or leaving the horizon
in the vacuum case, these horizon currents pile up charge on the
horizon, that is the divergence of the horizon current in (\ref{HChargeCons}) is not 0, 
but cancels a time changing charge density. This is why the currents 
seem to flow towards regions of positive charge in Figure \ref{BInf_SigvBet}. We see also that, as might
have been expected, the magnitude of the charge separation grows with
the speed of the boost, i.e. the more energy given to boost 
the dipole along the horizon, the higher the voltage of the horizon battery. 

 The line of zero charge density in the plane of the stretched horizon is given by,
\begin{align}
x\big{|}_{\sigma_{\mathcal{H}}=0} &= \beta_S z_{\mathcal{H}}
\hbox{sinh}[g_H t] \nonumber \\
&\pm \gamma^{-1}_S \sqrt{2 y^2 - (z_{\mathcal{H}}
  \hbox{cosh}[g_H t] - Z_S)^2} \ .
\label{HCharge_IB}
\end{align} 
On the true horizon,
\begin{align}
x\big{|}_{\sigma_{H}=0} =  \pm \gamma^{-1}_S \sqrt{ 2 y^2 - Z^2_S } \ .
\label{0Charge_IB}
\end{align} 
On the stretched horizon the shape of the charge separation boosts along the horizon at late times, when $z_{\mathcal{H}} \hbox{sinh}[g_Ht]$ becomes large. On the true horizon however the charge separation is stationary reflecting the freezing in of fields on the horizon. 

As can be seen in Figure \ref{BInf_SigvBet},  the $\gamma^{-1}_S$ pre-factor in equation
(\ref{0Charge_IB}) morphs the geometry of the charge separation from
that of roughly equal parts positive and negative charge at low
$\beta_S$, to that of smaller regions of larger negative charge density
squeezed to the sides of the dipole in the direction of its motion for
larger $\beta_S$. 

The charge separation and corresponding battery emf is a direct
consequence of the boosted motion parallel to the horizon. We will see
this feature again in the final example (\S \ref{c}).
In the penultimate section \S \ref{Consequences for the BH-NS Binary},
we estimate the power produced by a black-hole battery, the luminosities
attained in the circuit, and the energy scale of the emission.

\section{Rindler Dipole}
\label{Stationary Rindler Dipole}
Now suppose there is a magnetic dipole that is uniformly accelerated
so that it lives at constant Rindler coordinate $z_S$. While the
Minkowski observers see this dipole accelerate and asymptote to a
null trajectory, the Rindler observers
see a source dipole at fixed coordinate distance above the horizon.

This is the first case for which we no longer have a check of our
solutions. Nor do we have an obvious alternative method of
calculation.
We must
compute fields from our exact solution for the 4-potential from \S \ref{Text:Solutions}. The kinematics
of the accelerated source are characterized by 
\begin{align}
r^{\mu} &=
\begin{pmatrix}
T-T_{S}\nonumber \\
X \nonumber \\
Y \nonumber \\
Z- Z_S(T_S)
\end{pmatrix}  \qquad
V^{\mu} = \gamma_S
\begin{pmatrix}
1\nonumber \\
0 \nonumber \\
0 \nonumber \\
\beta_S
\end{pmatrix} \nonumber \\
a^{\mu} &=  \frac{\gamma^2_S}{Z_S }
\begin{pmatrix}
\beta_S \nonumber \\
0 \nonumber \\
0 \nonumber \\
1
\end{pmatrix} \quad 
\dot{a}^{\mu} = \frac{\gamma^3_S}{Z^2_S }
\begin{pmatrix}
1 \nonumber \\
0 \nonumber \\
0 \nonumber \\
\beta_S
\end{pmatrix} 
\end{align}
where, in this case, $\beta_S=\tanh(g_Ht_S)=T_S/Z_S$, $\gamma_S=\cosh(g_Ht_S)=(1-(T_S/Z_S)^2)^{-1/2}$, and 
$Z_S= \sqrt{z^2_S + T^2_S}=z_S \hbox{cosh}[g_H t_S]$, where $z_S$ is the constant height of the Rindler dipole above the horizon.

Here, the light cone condition is easier to solve in Rindler coordinates.
Evaluating the source at the retarded time, $t_S=t_*$, the light cone condition is
\begin{equation}
x^2 +y^2 + z^2 + z^2_S -2z z_S \hbox{cosh}\left[g_H(t-t_*)\right] = 0.
\label{Eq:nullrepeat}
\end{equation}
For $z_S=$constant, we find
\begin{equation}
t_*(x) =t - g_H^{-1}\cosh^{-1}\left ( \frac{x^2 +y^2 + z^2 + z^2_S
}{2z z_S }\right )
\label{Eq:nullrepeat}
\end{equation}
and $T_*=z_S \sinh(g_H t_*)$.

\begin{figure*}
\begin{center}$
\begin{array}{cc}
\includegraphics[scale=0.3]{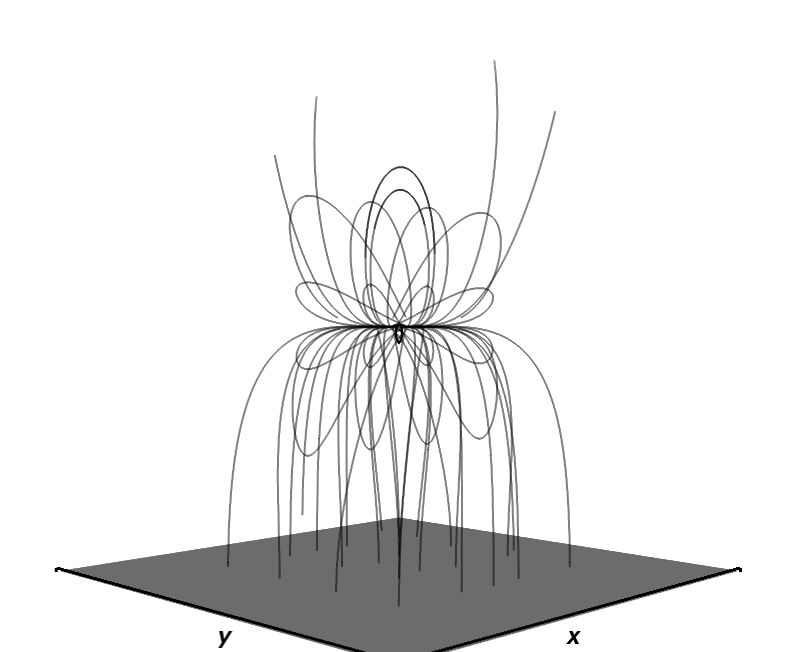} &
\includegraphics[scale=0.3]{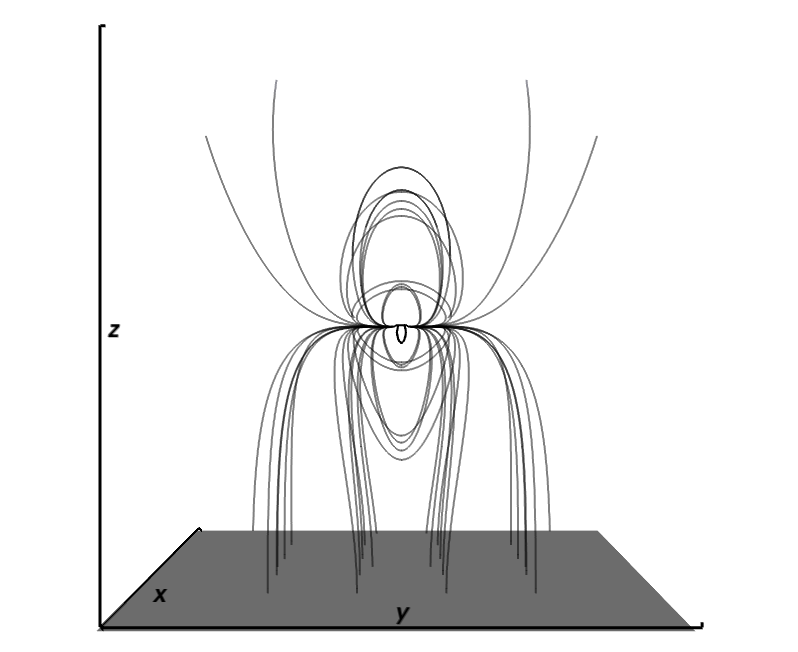} \\
\includegraphics[scale=0.3]{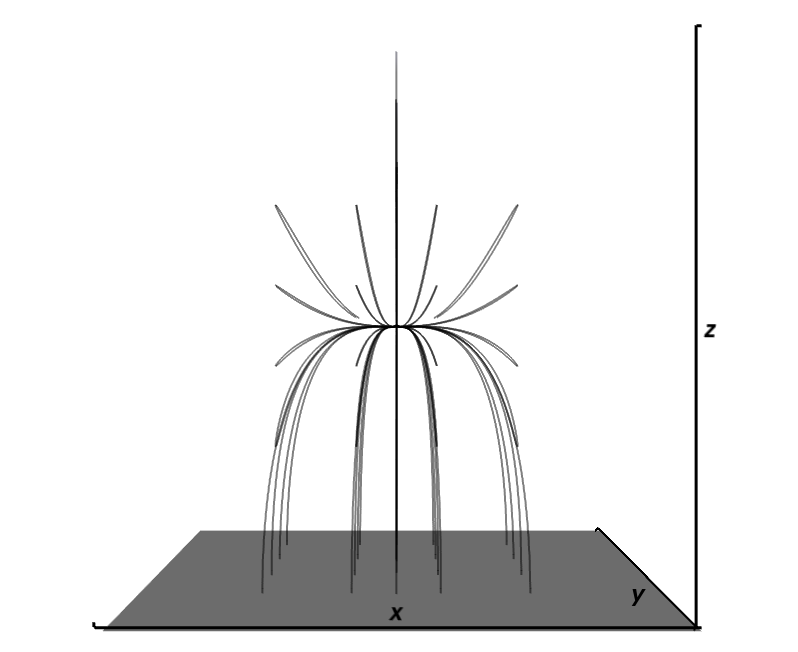} &
\includegraphics[scale=0.3]{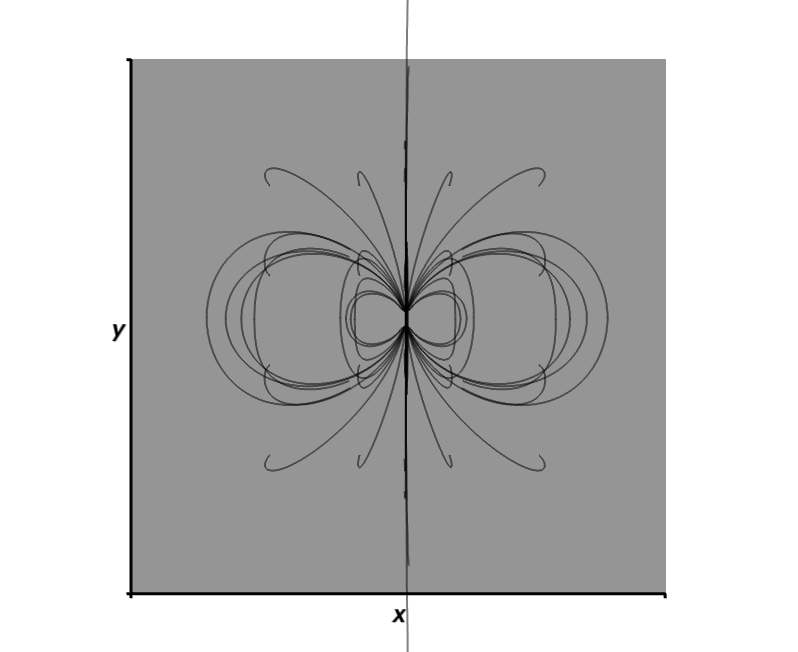} 
\end{array}$
\end{center}
\caption{3D visualization of the magnetic dipole field lines hovering
  at constant height $z_S$ above the Rindler horizon, denoted by the
  gray plane at $z=0$. The visualization region is a cube with side
  length $2 z_S$.}
\label{SRD}
\end{figure*}

Now that we have an expression for the retarded time, we can find the
fields for our Rindler observer following the prescription of \S \ref{Text:Solutions}.
For the sake of illustration, we write the Rindler fields for the
specific case where $\mathbf{m_S} = m \mathbf{\hat{e}_y}$:
\begin{align}
B_R^x =& -\frac{48 m x y z z^2_S}{ \left(r_- r_+ \right)^5 } \left[ r^2_+ - 2z z_S  \right]  \nonumber \\
B_R^y =& \frac{8 m z z^2_S}{\left(r_- r_+\right)^3} + B_R^x \frac{y}{x}
 \nonumber \\
B_R^z =&  \frac{16 m y z^2_S}{\left(r_- r_+\right)^5} \left[ (r_- r_+)^2 + 6 z^2z^2_S\right] + B_R^x \frac{z}{x} \nonumber \\
\mathbf{E_R} =& 0 \\
r_{\pm} &= \sqrt{x^2 + y^2 + (z \pm z_S )^2} \nonumber
\end{align}
We plot the above fields in Figure \ref{SRD} from 4 different
points of view; looking down each coordinate axis and looking
from a position half way between the $x$ and $y$ axes. From the above
expressions we see that even though the Minkowski dipole is
accelerated, Rindler
observers see no radiation field, nor do they see any electric field
at all. This is surprising since the Minkowski observers see a Poynting flux as well as radiation\footnote{Note that the field tensor (\ref{FmunuLong}) has terms which fall off as $1/|r|$ and hence generate a radiation field, as long as there is a non-zero dipole acceleration (See also \citep{Heras:1998})}. However, from the expression for the Rindler Poynting flux in terms of Minkowski fields (\ref{Eq:Poynt}), we see that the purely Minkowski term $\left [\EM^\parallel \times \BM^\parallel\right ]$ is exactly balanced by terms due to accelerations of the Rindler observers, which account for field energy moving past them as they accelerate. Note that this is also consistent with our choice of the rest frame moments ${\bf{m_S}}$ and ${\bf{p_S}}$.

As can be seen in Figure \ref{SRD}, at the horizon, the fields align themselves
perpendicular to the horizon, i.e ${\bf B}_{\mathcal{H}}=0$. This is a consequence of
the conductor-like properties of the horizon, Eqs.\ (\ref{EBBCs}),
along with the ingoing wave boundary conditions, ${\bf B}_{\mathcal{H}} = -\hat{e}_z
\times {\bf E}_{\mathcal{H}}$ and ${\bf E}_{\mathcal{H}} = \hat{e}_z
\times {\bf B}_{\mathcal{H}}$ which are a result of
choosing stationary observers to measure the fields.

Since $\ER=0$, no battery is established for the Rindler dipole. But then, no
battery would be expected from this configuration given that the dipole is
fixed relative to the horizon. When we introduce relative motion, as
we do in the next section, we will once again see a power source
generated in the form of an event-horizon battery.

\section{Rindler Dipole Boosted Parallel to the Horizon}
\label{c}
We would like to imagine a worldline for the source dipole that
mimics a magnetized NS in orbit around a BH. The
physical motion we want to represent
is best imitated by a source dipole at some fixed Rindler height above
the horizon $z_S$, but 
moving
parallel to the horizon with some fixed Rindler velocity, $v_{S,x}=$constant, so that
\begin{equation}
x_S=v_{S,x} \alpha_S t_S
\ .
\label{vsx}
\end{equation}
The kinematic
ingredients are then expressed in Minkowski coordinates as
\begin{align}
r^{\mu} &=
\begin{pmatrix}
T-T_{S}\nonumber \\
X - X_S (T_S) \nonumber \\
Y \nonumber \\
Z- Z_S(T_S)
\end{pmatrix}  \nonumber \qquad
V^{\mu}  = \gamma_S
\begin{pmatrix}
1   \\
\beta_{S,X}  \\
0 \\
\beta_{S,Z}
\end{pmatrix} \nonumber \\
a^{\mu} &= \frac{\gamma^2_S}{Z_S}
\begin{pmatrix}
\beta_{S,Z} \\
0 \\
0  \\
1
\end{pmatrix} \qquad
\dot{a}^{\mu} = \frac{\gamma^3_S}{Z^2_S }
\begin{pmatrix}
1 \nonumber \\
0 \nonumber \\
0 \nonumber \\
\beta_{S,Z}
\end{pmatrix} 
\end{align}
Here $\gamma_S$ is the total Lorentz factor computed with
$\bb_S=\beta_{S,X} {\bf{e_X}} + \beta_{S,Z} {\bf{e_Z}}$ and
\begin{align}
\beta_{S,X}&={v_{S,x}}\frac{z_S}{Z_S} \nonumber \\
\beta_{S,Z}&=\frac{T_S}{Z_S}
\end{align}
and again
$Z_S= \sqrt{z^2_S + T^2_S}=z_S \hbox{cosh}[g_H t_S]$, with $z_S$ 
the constant height of the Rindler dipole above the horizon. 
Notice that as long as $|\vb|\le 1$, the source will
travel slower than the speed of light at all times, $\beta_S \le 1$.

\begin{figure}
\begin{center}
\includegraphics[scale=0.4]{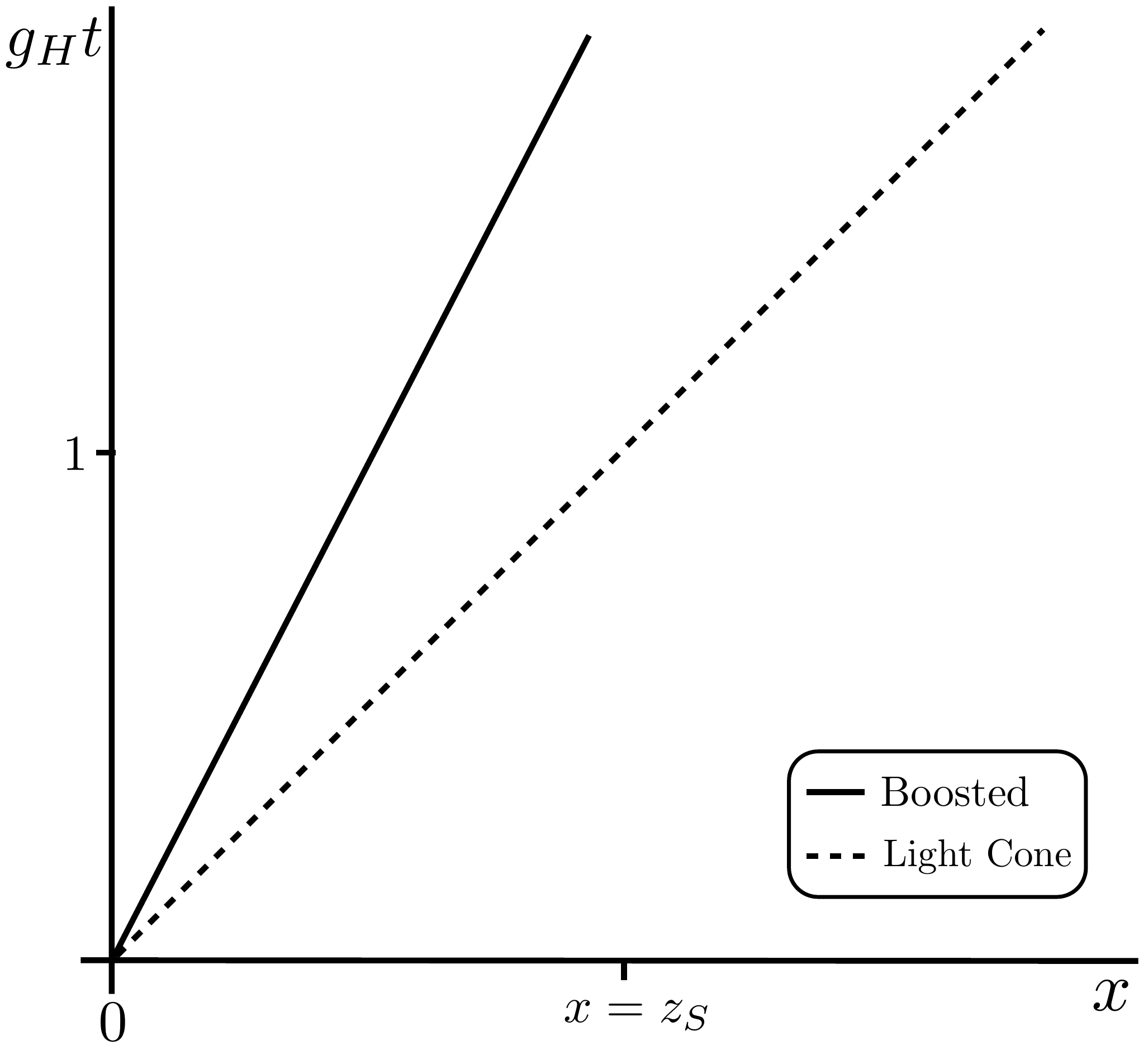} 
\end{center}
\caption{Spacetime diagram in the Rindler frame depicting the worldline of a source boosted parallel to the Rindler horizon (\S\ref{c}). The light ray (dotted line) has slope $dt/dz = (g_Hz_S)^{-1}$ and the worldline has slope $(v_{S,x} g_Hz_S)^{-1}$, where $z_S$ is the constant position of the source above the Rindler horizon.}
\label{ST_Boost}
\end{figure}

The light-cone condition in Rindler coordinates can no longer be
found in closed form for $t_S=t_*$. We can however
write $A^{\alpha}$ or $F^{\alpha \beta}$ in terms of $t_*$ (or $T_*$)
and solve numerically for the retarded time.  
It is extremely helpful that we never have to take explicit
derivatives of $t_*$ since the first relation in Eq.\ (\ref{Grads}) allows us to
re-express derivatives in terms of more transparent variables.

\begin{figure*}
\begin{center}$
\begin{array}{c}
\includegraphics[scale=0.33]{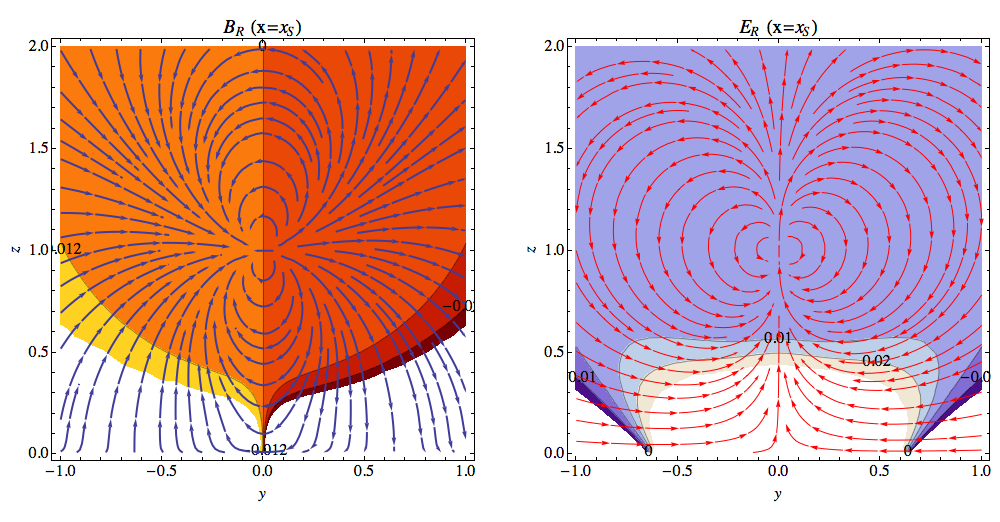} \\
\includegraphics[scale=0.33]{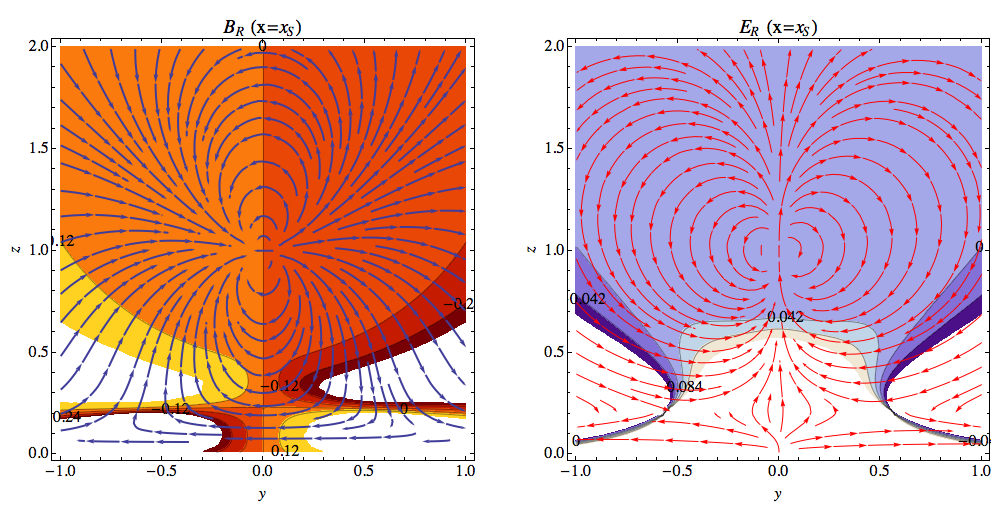} \\
\includegraphics[scale=0.33]{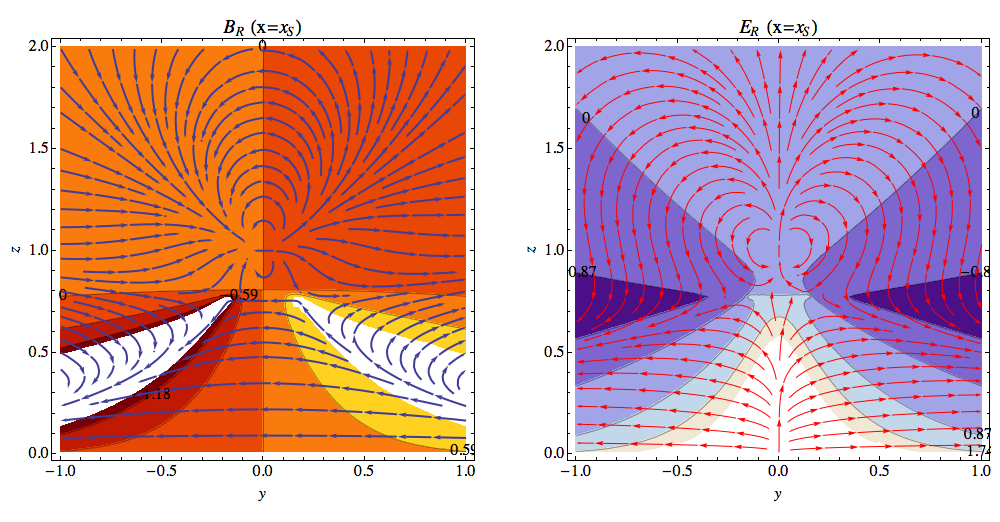}
\end{array}$
\end{center}
\caption{Streamlines of the magnetic (left) and electric (right) fields in the plane $x=0$ for a magnetic dipole with dipole
  moment ${\bf m}\propto \hat {\bf e}_y$ and with three different boost velocities
  increasing from top to bottom $\vb=0.1, 0.5, 0.9$ in the $x$-direction. Plotted over the streamlines are contours of 
  the x-components of the fields relative to the $y-z$
  magnitude. Darker regions represent negative values and lighter
  regions represent positive values. The
  white regions are clipped to better view the contour structure. The
  snapshots here are taken at $g_Ht = 1/4$, however the fields retain
  the same structure for all time except for their motion in the
 x-direction (out of the page). The axes are in units of $z_S$. 
 Since the source is boosted,
 observers near the horizon see the fields as they would have been when 
 the dipole was further away in the negative x-direction. The result is an
 observed dragging of the fields along the horizon in the 
 negative x-direction. The observed larger radius of curvature of the
 dipole lobes manifests itself as the flattening of the field lines.  As can be gathered
 from Figure \ref{ST_Boost}, this effect is intensified 
 for larger boost factor $\vb$.  In the $\vb=0.9$ case, plotted at the
 bottom of the figure, the 2D slice of the magnetic field
 loses its dipolar structure in most of the region 
 below the source.  As $\vb$ approaches $1$, the slope of the source
 worldline approaches the light cone
 slope and an observer at a given $z$ will see further and further
 into the relative past of the dipole. 
  Note also that the contours in the left panels show that the circulation direction of the dipole lobes 
 changes sign at a value of $z$ which gets larger for larger $\vb$. This change in sign 
 results since observers near the horizon see fields from further in the past 
 when the fields were pointing in a different x-direction. The increase in z-location 
 of this turning point for larger $\vb$ can again be understood from Figure \ref{ST_Boost}.
}
\label{Fig:EByz_Boost}
\end{figure*}

Figure \ref{Fig:EByz_Boost} plots the fields of a parallel-boosted dipole for
the choice of a magnetic dipole moment in the $y$-direction and a boost in
the $x$-direction. Each panel plots streamlines of the magnetic (blue)
or electric (red) fields in the $y-z$ plane containing the source. Also 
plotted are contours of $B^x_R/ \sqrt{(B^y_R)^2 + (B^z_R)^2}$
or $E^x_R/\sqrt{(E^y_R)^2 + (E^z_R)^2}$ to give a sense of 
the 3D nature of the fields. Successive rows correspond to increases in $\vb$.  The fields 
do not evolve in time except for their constant (universal-time) velocity motion in the $x$-direction. 

The dipolar magnetic field structure
flattens near the horizon due to time dilation.
Observers below the dipole source 
 see the dipole as it was further in the past, when the dipole was
 further away in the negative $x$-direction, than do observers the same 
 distance above the source. 
This leads to an overall dragging of field lines along the horizon as explained
 in more detail in the figure captions.

\begin{figure}
\begin{center}$
\begin{array}{c}
\includegraphics[scale=0.38]{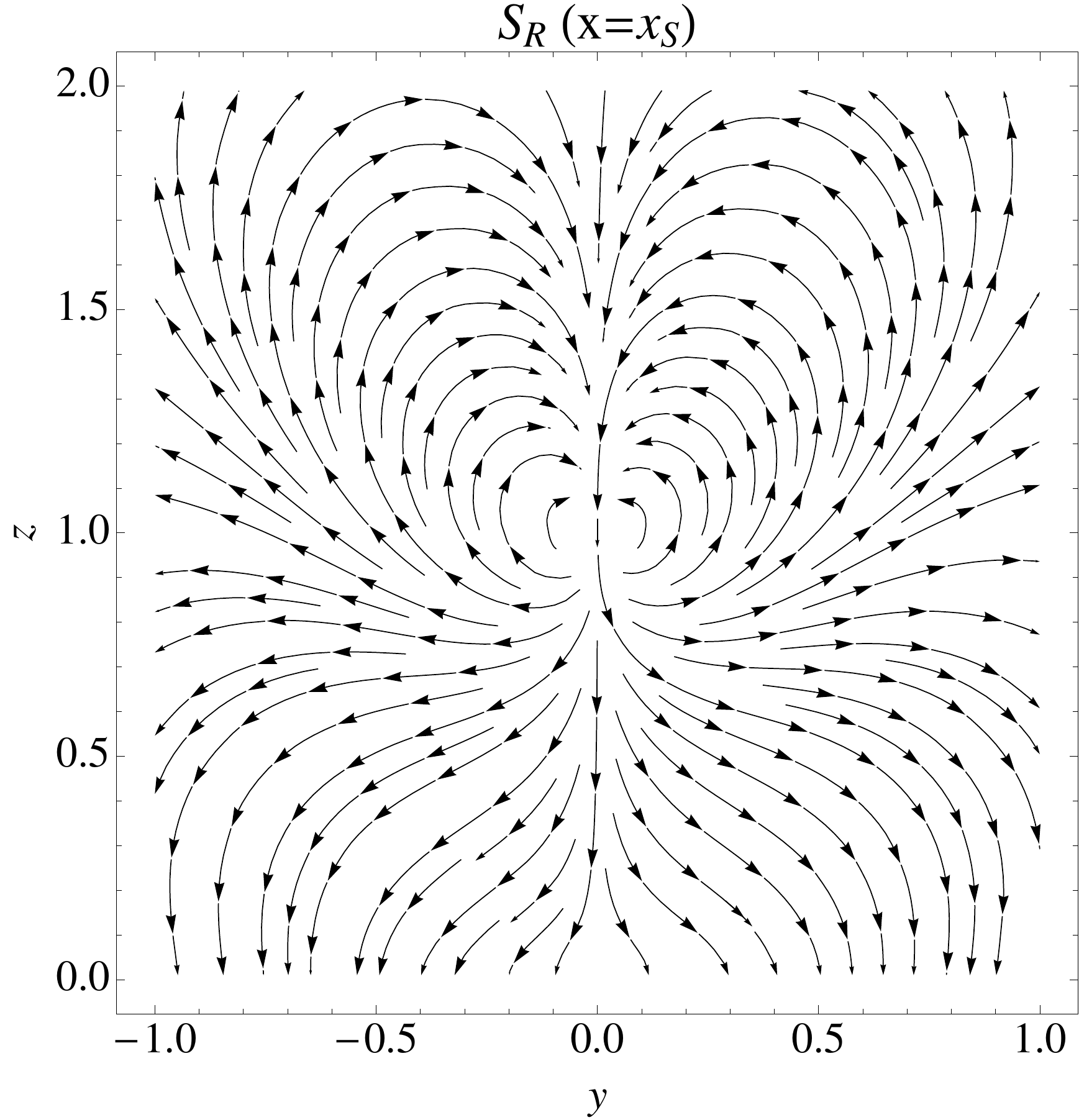} \\
\includegraphics[scale=0.38]{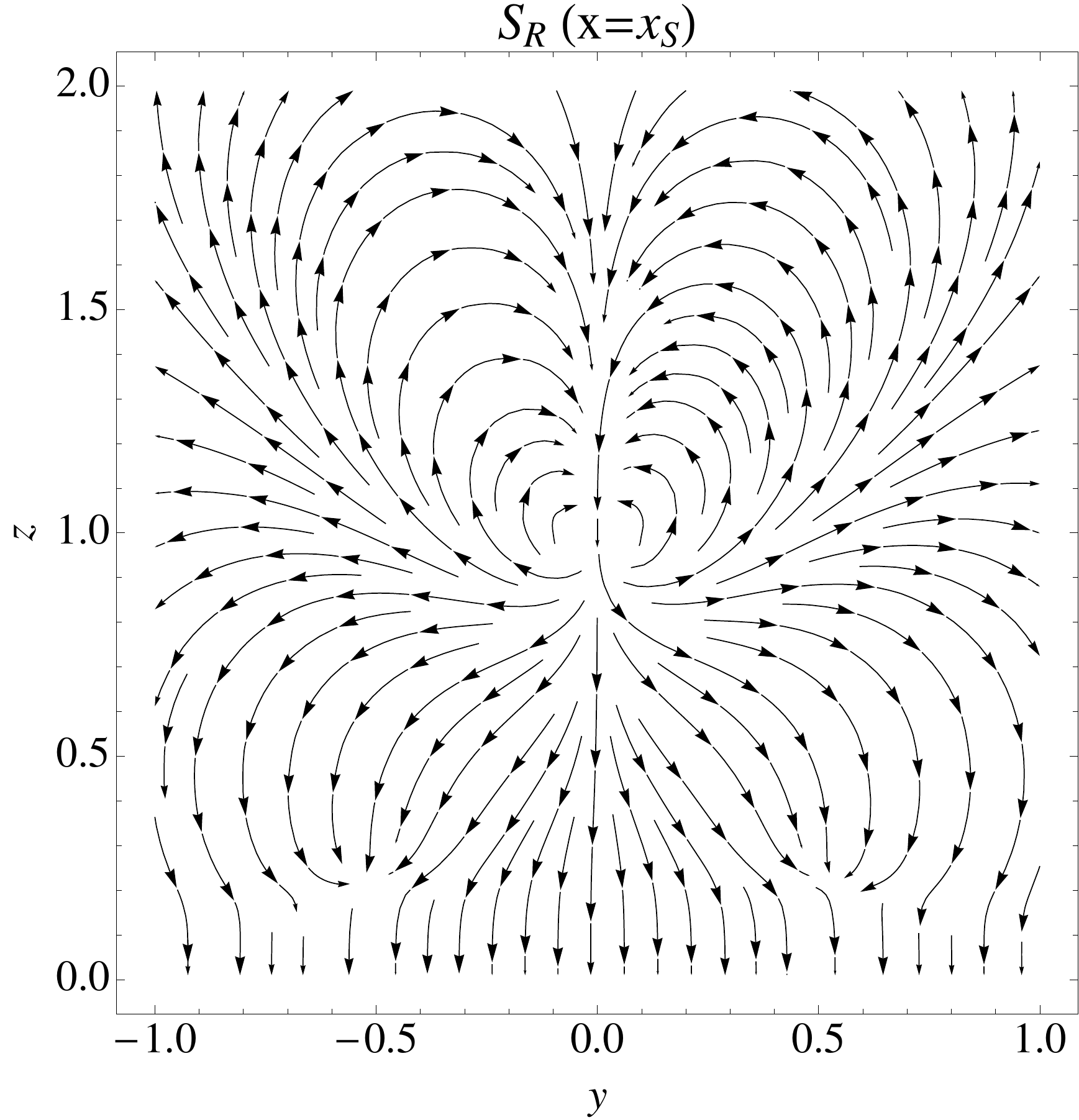}  \\
\includegraphics[scale=0.38]{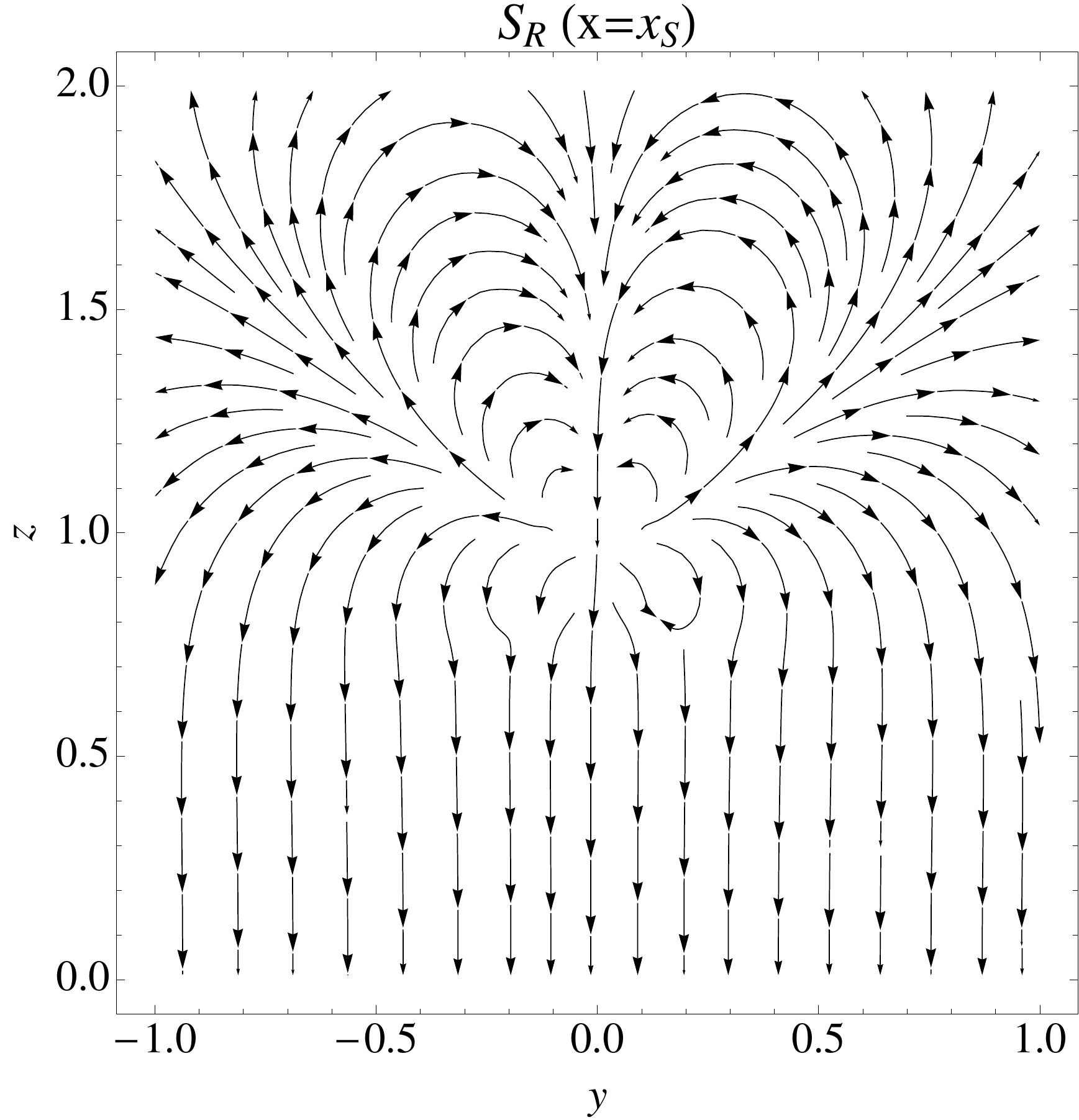} 
\end{array}$
\end{center}
\caption{An $x=x_S=0$ slice through the Poynting flux. The plane
  contains the source as viewed by Rindler observers for three
  different boost magnitudes in the $x$-direction, $\vb=0.1, 0.5,
  0.9$. The Poynting flux is 0 at infinity despite outward components of the field in the region plotted here. The axes are in units of $z_S$.}
\label{Poynt_Boost}
\end{figure}

 We also see this effect in Figure \ref{Poynt_Boost} which is a slice of the Poynting-flux vector field 
 in the $y-z$ plane containing the source. For large $\vb$, observers at small $z$ 
 see fields from when the dipole is relatively far away and thus do not see the 
 dipole structure of the field energy flowing past them, only nearly uniform $z$ and $y$-components.

\begin{figure*}
\begin{center}$
\begin{array}{cc}
\includegraphics[scale=0.34]{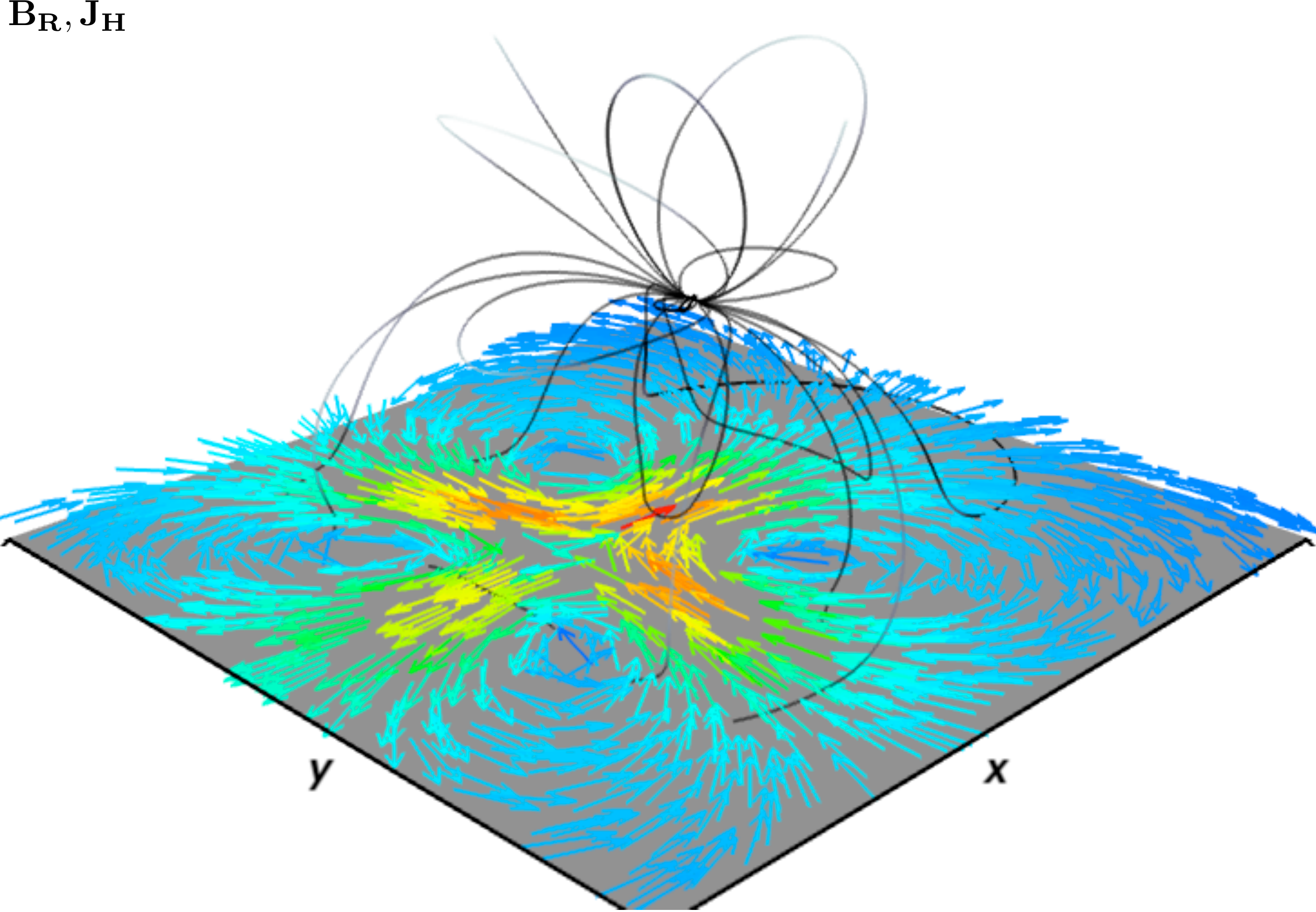} &
\includegraphics[scale=0.34]{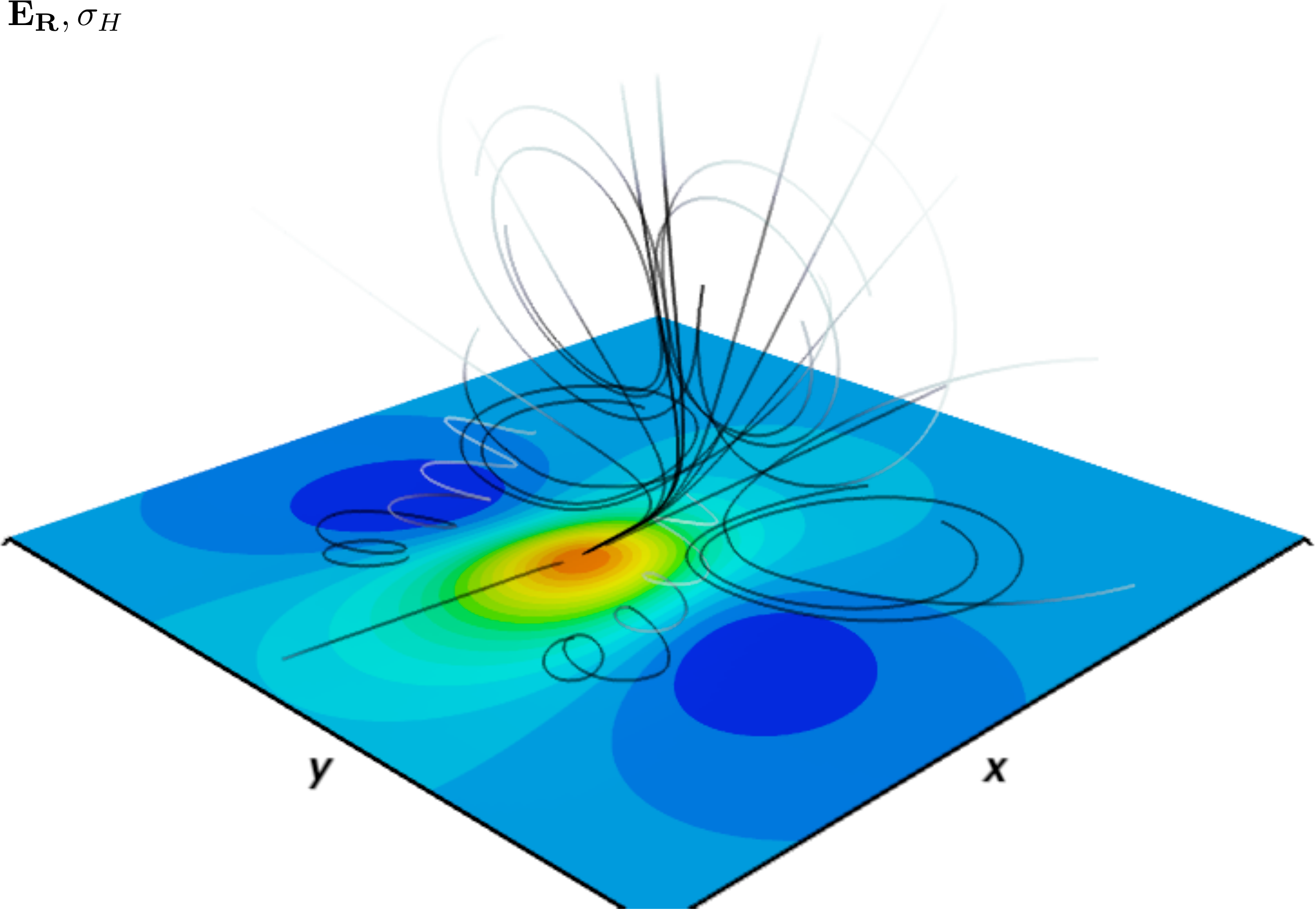}
\end{array}$
\end{center}
\caption{A 3D visualization of the magnetic fields lines and corresponding horizon currents $\mathbf{J_{\mathcal{H}}}$ (left) and electric field lines with the corresponding horizon charges $\sigma_\mathcal{H}$ (right) for the boosted Rindler dipole. The case shown is for $\vb=0.2$.}
\label{Fig:3DBoost}
\end{figure*}

\begin{figure}
\begin{center}$
\begin{array}{c}
\includegraphics[scale=0.35]{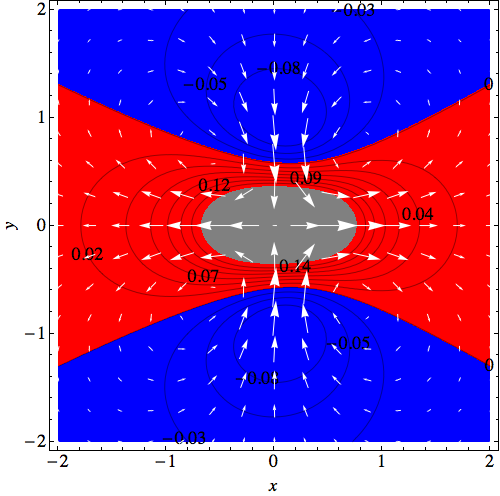} \\
\includegraphics[scale=0.35]{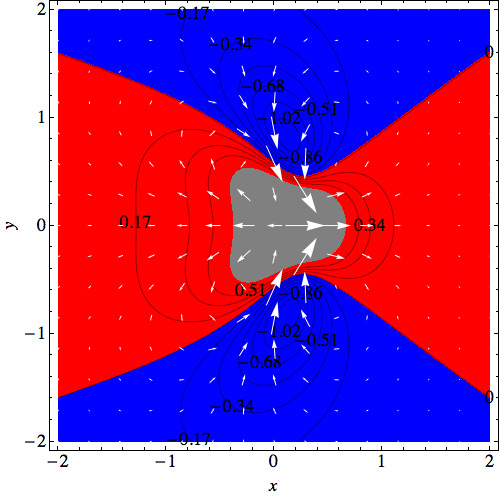}  \\
\includegraphics[scale=0.35]{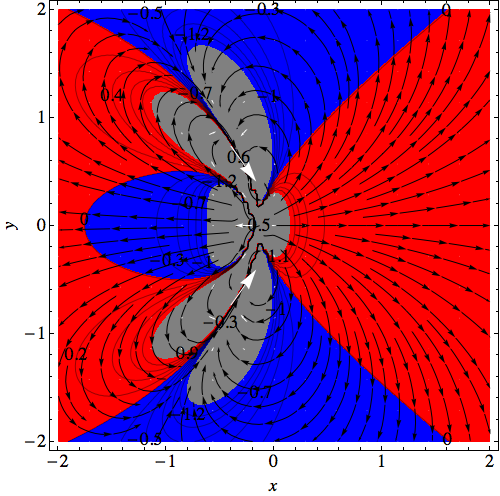} 
\end{array}$
\end{center}
\caption{An identical plot to Figure \ref{BInf_SigvBet} but for the boosted case. Current density vectors (white) are overlaid on contours of charge density on the stretched horizon ($\alpha_{\mathcal{H}}=10^{-4}$) of the boosted dipole with magnetic moment in the $y$-direction. The bottom panel also plots streamlines of the currents. From top to bottom, the magnitude of the boost in the $x$-direction increases from $\vb=0.1, 0.5, 0.9$. Each snapshot is taken at $g_H t=10$. The configuration drags along the stretched horizon keeping a constant lag distance behind the moving source.  The induced currents can be thought of as redistributing charge in order to slide the charge distribution along behind the boosted dipole. The gray regions are regions of steeply increasing $\sigma_\mathcal{H}$ which have been removed to more clearly view the contour structure. The axes are in units of $z_S$. }
\label{Fig:BoostJS_vs_Bet}
\end{figure}

For small $\vb$, the fields resemble those in the stationary case,
threading the horizon nearly perpendicularly. As $\vb$ is increased,
the dragging effect causes the fields near the horizon
to lay down tangentially to the horizon as the source moves
along. Figure \ref{Fig:3DBoost} shows a 3D representation of the dragging effect for the $\vb=0.2$ case. In the left panel of Figure \ref{Fig:3DBoost} we see that the tangential magnetic fields source horizon currents. Via horizon charge conservation, these currents build up horizon charge density which we observe in the right panel of Figure \ref{Fig:3DBoost} and interpret as the normal components of the induced electric fields. Figure \ref{Fig:BoostJS_vs_Bet} shows the
horizon charge and current densities for three different $\vb$, all at
$g_H t=10$. As time progresses, these same charge and current
distributions are dragged behind the dipole on the stretched horizon
at a lag distance which increases as $\vb$ increases, and also as the distance between the stretched
and true horizons decreases.

An interpretation of this behavior follows similarly to that of 
\citep{MP3_MS:1985} for the case of an electric point-charge boosted
parallel to the Rindler horizon. In the electric point-charge case,
the charge distribution induced on the horizon is also dragged behind
the boosted source. Via charge conservation, this necessitates horizon
currents to redistribute charges. Such horizon currents can be thought
of as due to tangential components of magnetic fields induced by the
moving point-charge.

The key result from this example is the explicit charge separation 
and therefore voltage drop across the event horizon.  We have 
established an event-horizon battery, a power source for a BH-NS 
electromagnetic circuit. The boosted Rindler dipole provides a proxy 
for a NS in orbit around a big BH. We will use this case to estimate 
some astrophysically relevant scales in the following section.

\section{Consequences for the BH-NS Binary}
\label{Consequences for the BH-NS Binary}
\subsection{Voltage, Luminosity, and Energy}

We utilize the electromagnetic field solutions of the boosted, Rindler
dipole of \S\ref{c} to estimate the power output and
maximum energy of radiation a BH battery can supply. 
We treat the BH-NS system as a series circuit containing
resistors and a battery with voltage given via (\ref{VH}) from the
electromagnetic field solutions. For simplicity, we imagine sticking
one wire of the circuit into the point of maximum horizon potential,
and the other wire, a distance of $2M$ away in the
y-direction. From (\ref{RindtoSch}), this separation can be compared to to a circuit connecting the pole and equator of a Schwarzschild BH. (Although this seems arbitrary, there is little dependence on the distance. We could have stuck the other wire
at infinity with little difference in results.) For
the boosted Rindler dipole solutions with $\mathbf{m_S} = m \mathbf{e_y}$, the y-component of the vector
potential vanishes. Then from the potential of Eq.\ (\ref{VH}) and
the electric field Eq.\ (\ref{Eq:EMBM}) in a Rindler coordinate frame,
we have $ \nabla_y V_{\mathcal{H}} = \alpha E^R_y =
\alpha^2 \nabla_y A^0_R $, and therefore $V_{\mathcal{H}}=\alpha^2 A_R^0$
across regions of charge separation estimates the voltage 
drop on the horizon. In terms of the Rindler retarded time, and with
physical constants restored, the horizon voltage is,
\begin{widetext}
\begin{align}
\label{BoostVolt}
V_{\mathcal{H}} &= \frac{c^2 \vb z_H m  }{8 G M}  \left(  v^2_{S, x} z_S
\frac{g_H}{c} t_* - \vb x + z_H \Sh \right)^{-3} \times   \\ 
& \left [    z_H \left( 1+ 2 v^2_{S, x} + \ChT  \right)   + 2  \vb
  \Sh \left( x - \vb z_S \frac{g_H}{c}  t_* \right) - 2 \Ch
  \left(\frac{x^2 + v^2_{S, x} z^2_S + z^2_H}{z_S}  + \vb \frac{g_H}{c}  t_*
  \left(  \vb z_S \frac{g_H}{c}  t_* - 2 x \right)   \right)
  \right]   \nonumber \\
{\rm with } & \nonumber \\
\Ch &= \cosh{\left[ \frac{g_H}{c} (t - t_*)\right] }, \qquad
\Sh = \sinh{\left[ \frac{g_H}{c} (t - t_*)\right] }, \qquad
\ChT = \cosh{\left[ 2 \frac{g_H}{c} (t - t_*)\right] }  \nonumber
\end{align}
\end{widetext}
where $m$ is the NS rest frame dipole moment (See (\ref{mnumbers}))
and the retarded time $t_*$ is a function of the (Rindler) observer coordinates. 

We compute power radiated by such a circuit from
(\ref{CircuitPower}). To do so, we estimate the physically relevant
values of the various parameters.
Very near the Schwarzschild horizon, in physical units, the gravitational acceleration is
\begin{align}
g_H &= \frac{c^4}{4GM}  \simeq  1.5 \times 10^{14} \left(\frac{ 10 M_{\odot} }{M}\right)  \frac{\rm{cm}}{\rm{s}^2}  \nonumber
\end{align}
about 100 billion times that on Earth for a $10 M_\odot$ black hole.
The magnitude of the NS's magnetic dipole moment written in terms
of the magnetic field strength at the NS's poles $B_p$ and
the radius of the NS $R_{NS}$ is of order, 
\begin{align}
\label{mnumbers}
m &= \frac{B_p R^3_{NS}}{2}\simeq 5 \times 10^{29}
\left( \frac{B_p}{10^{12} G}\right) \left(\frac{R_{NS}}{10^6  cm}\right)^3 \rm{G} \ \rm{cm}^3  \ . 
\end{align}
We must also approximate the resistances in our astrophysical circuit diagramed in Figure \ref{CircuitDiagram}. 
We have three resistors to consider: the horizon with resistance
$\Res_{H}$,\footnote{A material with resistivity $\rho$ has resistance $\Res = \rho \frac{L}{A}$ where $L$ and $A$ are the length and cross sectional-area of the material as seen by the current. In the case of a black hole horizon, A and L can be taken to both be of order $\pi 2M$.} the NS crust with resistance $\Res_{NS}$ and the plasma of the NS magnetosphere denoted by
$\Res_{\rm{plasma}}$. The horizon resistance is known from the membrane paradigm to be \cite{MPbook}
\begin{equation}
\Res_H \simeq \frac{4 \pi}{c} = 4.2 \times 10^{-10} \rm{s}
\ \rm{cm}^{-1}  =  377 \Omega \ \ .
\label{RH}
\end{equation}
The resistivity of the NS crust is likely very small compared to $\Res_H$, on the order of $10^{-24} s^{-1}$ (see \textit{e.g.} \citep{Piro:2012}); thus we set $\Res_{NS} =0$. 

The value of $\Res_{\rm{plasma}}$ is an interesting unknown and
requires numerical exploration beyond the scope of this article. 
For our present purposes, as a rough guide, we choose an effective value of 
$\Res_{\rm{plasma}} = \Res_H/2$, because it gives maximum power output through $\Res_{\rm{plasma}}$. Ref.
\cite{McL:2011} choose $\Res_{\rm{plasma}}$ based on equating the power dissipated due to curvature radiation with the power dissipated due to ohmic dissipation $I^2 \Res_{\rm{plasma}}$. Upon solving for $\Res_{\rm{plasma}}$, they find $\Res_{\rm{plasma}} = \Res_H$ when the plasma velocity is $\sim 0.7c$. However, the estimate is sensitive to the plasma particle velocity.
The dependence of power output on
$\Res_{\rm{plasma}}$ for a similar NS-NS circuit with non-zero
$\Res_{NS}$ is explored in \cite{Piro:2012}, however the NS-BH case is simpler since the denominator of the power formula (\ref{CircuitPower}) is dominated by $\Res_{H}$.

Since we have set the NS resistance to 0, we focus on the power
radiated in the space between the NS and BH, \textit{i.e.} $\Res_i =
\Res_{\rm{plasma}}$ in (\ref{CircuitPower}).  Since the horizon
potential is symmetric around the line $y=y_S$, which contains the
maximum of the potential and thus one of the circuit wires, we
multiply the above luminosity by a factor of 2. The combination of our
choices for $\Res_H$ and $V_{\mathcal{H}}$ will correspond to maximum
achievable bolometric luminosities when $\Res_{NS}$ is ignored. 

The circuit is connected if the BH is within the light
cylinder of the NS:
 \begin{align}
R_{lc} & =\frac{c}{\Omega_{NS}} = 5 \times 10^9 \left(\frac{P}{ 1
  s}\right) \ \rm{cm} \nonumber \\
& \sim  3 \times 10^3 \left(\frac{P}{ 1 s}\right) \left (\frac{10
  M_\odot}{M}\right ) \frac{GM}{c^2}
 \label{RLC}
 \end{align}
where $P$ is the period of the NS spin and in the last line we quote
the radius in units of $M$. 
We choose a fiducial horizon distance of $z_S = 3GM/c^2$ where 
the Rindler limit is valid and the pair has approached extremely
close prior to merger. 
For BH's with $M\gtrsim 10^4 M_{\odot}$ 
our fiducial value of $z_S$ is larger than the light cylinder of the
NS and the pair is unplugged.
We address the case for these larger black holes in the next section.
For lighter BHs, the circuit will connect when the NS is a distance
above the horizon
$\sim R_{lc}$ and the power supplied will grow until it reaches a maximum around
our fiducial distance $z_s=3GM/c^2$ just prior to merger. Realistically, the compact objects will plunge
extremely rapidly at such close separations so we only use these values to get a sense of
the maximum blast of luminosity. For a
$10M_\odot$ BH, $R_{lc}\gg z_S$ and so the circuit is connected
for many orbits before maximum is reached.

\begin{figure}
\begin{center}$
\begin{array}{c}
\includegraphics[scale=0.3]{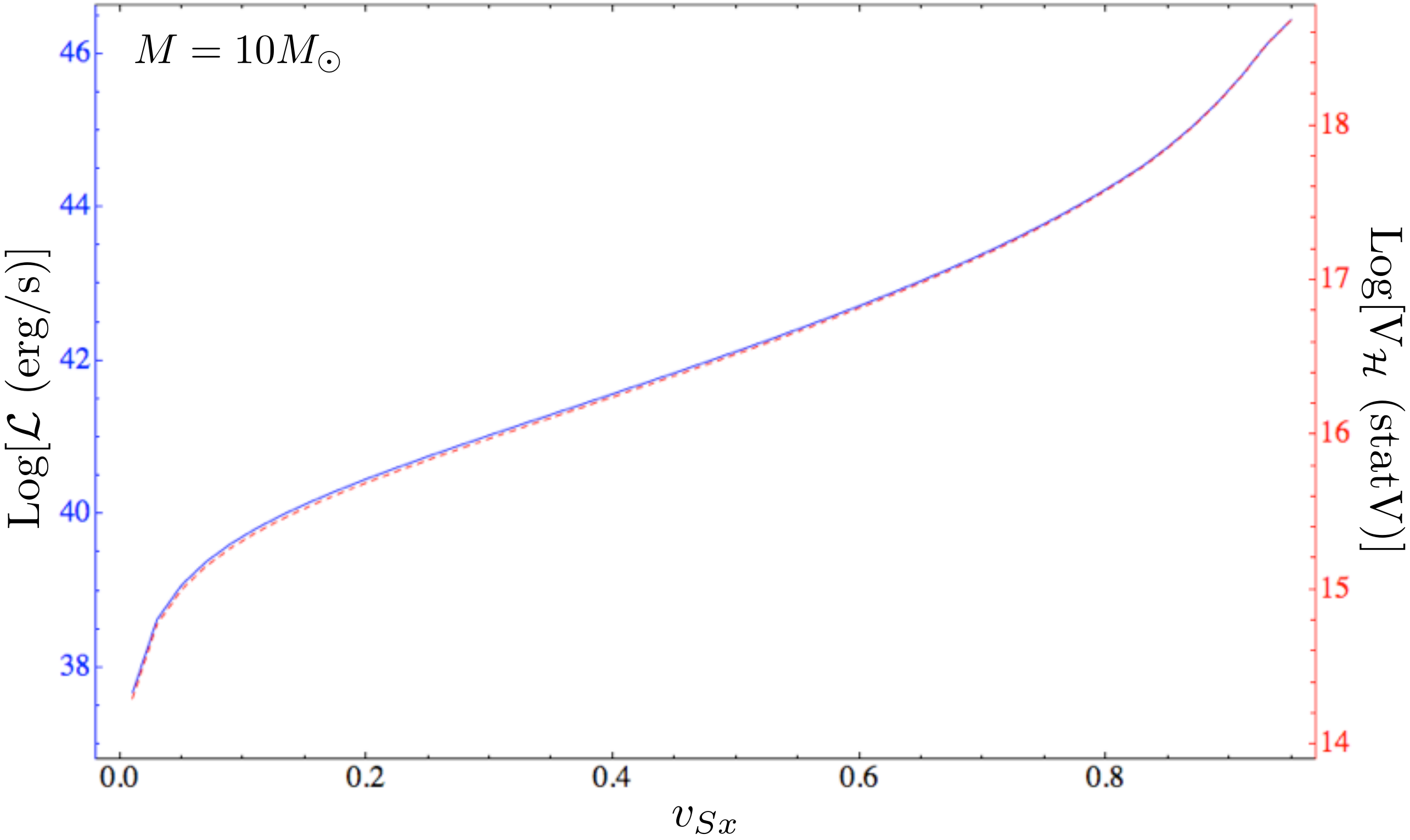} \\
\includegraphics[scale=0.3]{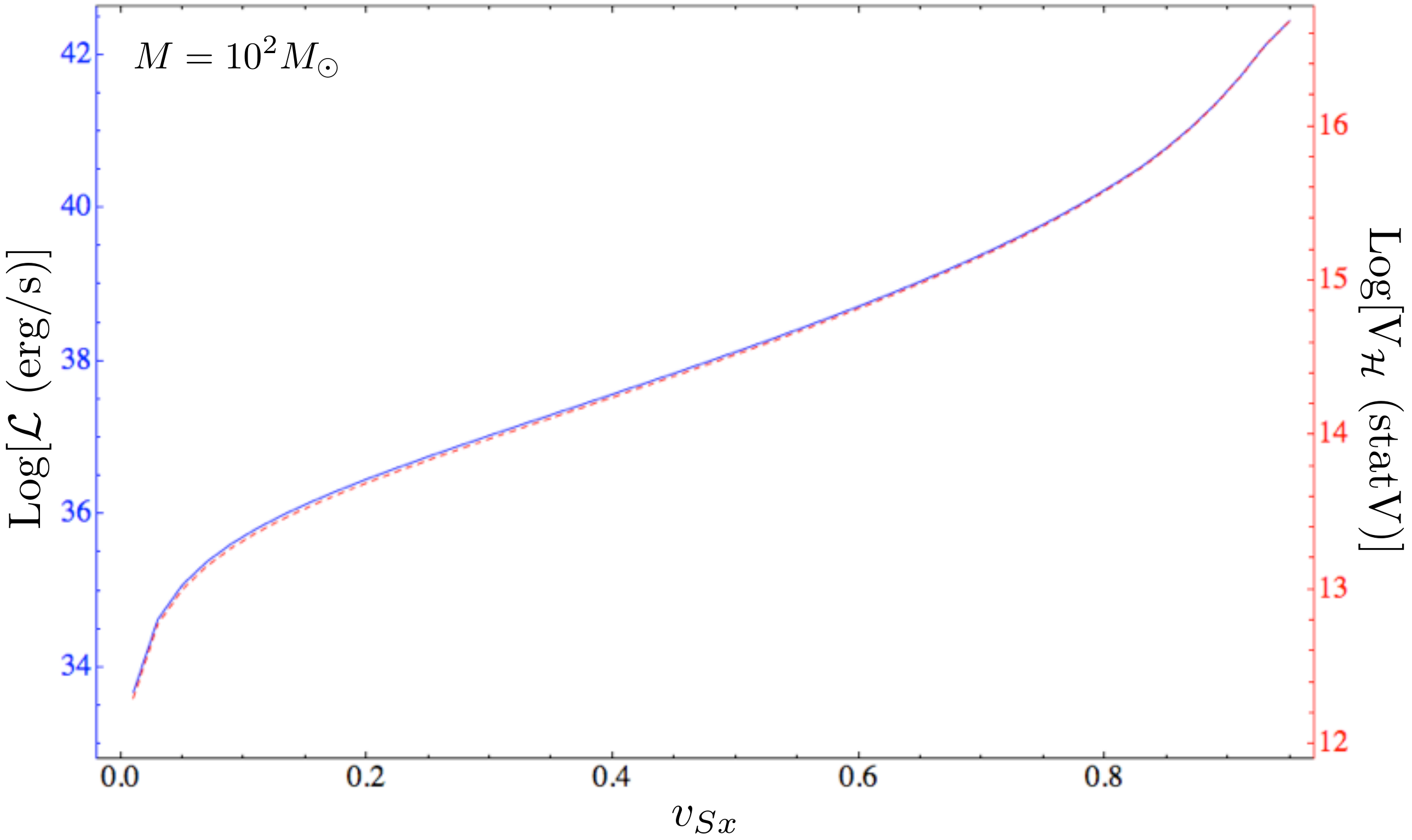} \\
\includegraphics[scale=0.3]{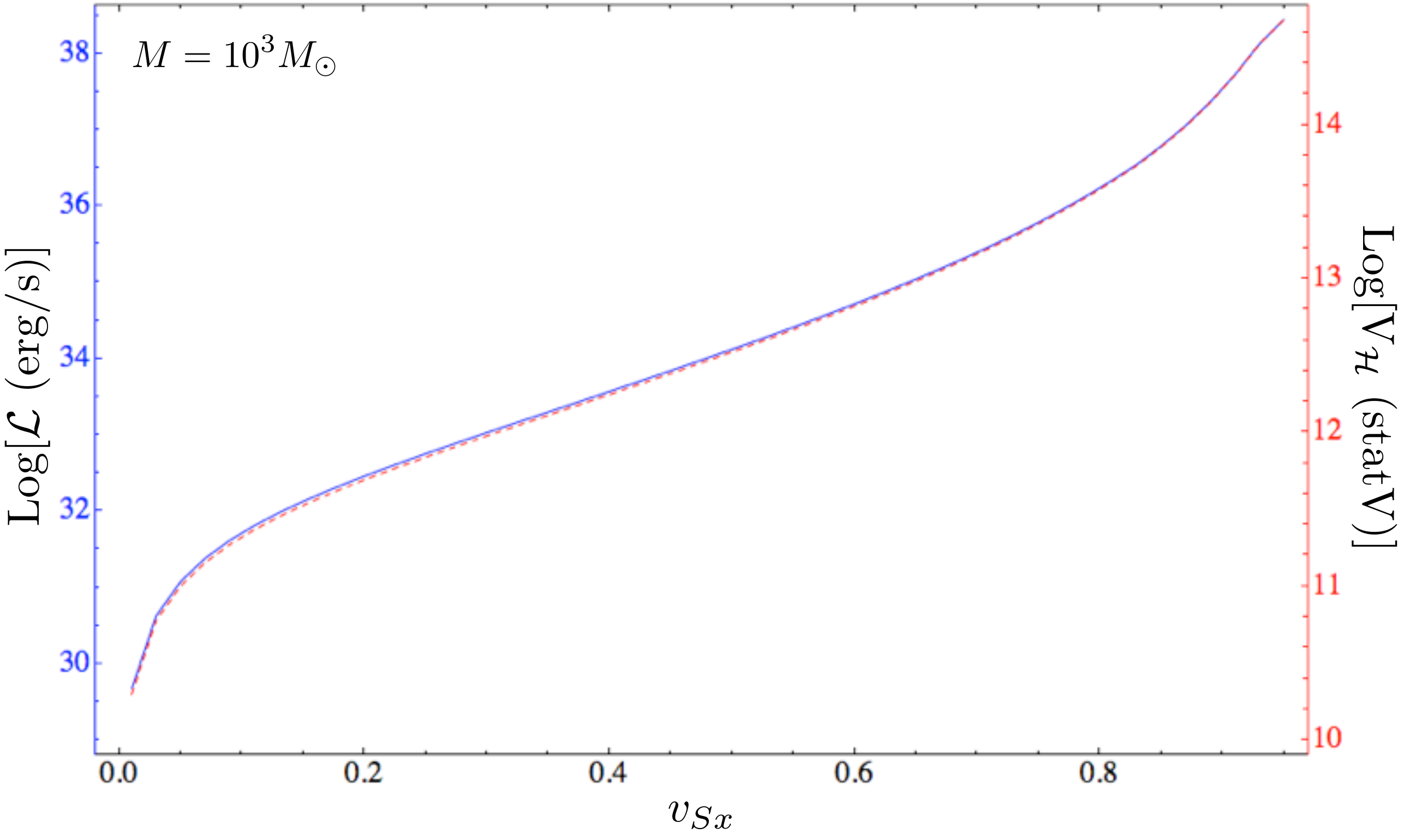} 
\end{array}$
\end{center}
\caption{Log-luminosity computed from Eq.\ (\ref{CircuitPower}) (blue, solid line and leftmost y-axis labels) and representative log-voltage drop on the horizon (red, dashed line and rightmost y-axis labels). Luminosities and voltages are computed for $M=10 M_{\odot}, 10^2 M_{\odot}$, and $10^3 M_{\odot}$ with the dipole at Rindler height $z_S = 3M$ as a function of $\vb$ varying from $0.01$ to $0.95$. The last stable circular orbit in the Schwarzschild spacetime would have $\vb = 0.5$.}
\label{LumV}
\end{figure}

The maximum horizon voltage and corresponding luminosities are plotted for 
$10  M_{\odot}$, $10^2  M_{\odot}$, and $10^3 M_{\odot}$ BH's in Figure \ref{LumV} 
for $\vb$ varying from $0.01$ to $0.95$.  For comparison, $v_{S,x}\sim 0.5$
at the last stable circular Schwarzschild orbit.

Finally then we have our answer. We estimate the voltage of our
BH battery to be $\sim 10^{16}$ statvolts for a $10M_\odot$
BH,  $\sim 10^{14}$ statvolts for a $10^2 M_\odot$ BH, and 
$\sim 10^{12}$ statvolts for a $10^3 M_\odot$ BH when $v_{S,x}\sim 0.5$.
The luminosities in this limited approximation are $\sim
10^{42}$ erg/s, $10^{38}$ erg/s, and $10^{34}$ erg/s respectively.

As is suggested by the pre-factor in (\ref{BoostVolt}), the horizon
voltage and the luminosity decrease with increasing BH mass. 
For $z_S \rightarrow 0$, the horizon voltage scales as $M^{-1}$ and the luminosity as $M^{-2}$.
Otherwise, terms proportional to $g_H$ inside the brackets in
Eq.\  (\ref{BoostVolt}) dominate and thus the voltage goes as $M^{-2}$
causing the luminosity to scale as $M^{-4}$. Comparison of the three
panels of Figure \ref{LumV} confirms this scaling with BH mass. 

This scaling also agrees
with \citep{McL:2011} where the BHNS battery was first proposed. In \citep{McL:2011} the physical mechanism 
was sketched out in a non-relativistic calculation. The analytic, relativistic solutions obtained here are in agreement with the findings of \citep{McL:2011}. Specifically, \citep{McL:2011}'s Eq.\  (6) for $\mathcal{L}$ exhibits the same scaling with black hole mass and NS magnetic field strength as does our Eq.\  (\ref{Eq:conc}). Also, \citep{McL:2011}'s Eq.\  (6) is calculated for similar parameters that we use in our calculation. They choose $\Res_{\rm plasma} = \Res_{\rm H}, \Res_{NS}=0$, $B_p=10^{12}$,  $M=10 M_{\odot}$, and a separation corresponding to the NS orbiting at the light ring of a Schwarzschild BH. Hence we may also compare magnitudes of the computed luminosities in each study, and we find that they agree in order of magnitude. Note that the inverse BH mass dependence arises because we are comparing luminosities for different BH masses while holding the distance of the dipole source from the horizon at a fixed number of gravitational radii (which scales with $M$).




We can also compute the maximum energy given to magnetosphere
particles by the horizon battery. To be clear, we are not calculating
the spectrum -- which promises to be complicated -- just the maximum energy 
scale. The magnitude of horizon voltages
plotted in Figure \ref{LumV} makes evident that the highest energy
particles accelerated via the horizon battery will radiate their
energy via curvature radiation.  A Rindler observer at the
instantaneous location of an accelerating plasma particle will measure
a local energy given by the characteristic energy of curvature
radiation,
\begin{align}
\epsilon_{R} = \frac{3hc}{4 \pi} \frac{\gamma^3_p}{\eta R_{LC}}
\label{Ecurv}
\end{align}
where $h$ is Planck's constant, $\gamma_p$ is the Lorentz
factor of the plasma particle (electron or positron) measured by a
Rindler observer,\footnote{The Lorentz factor in units of Rindler 
proper time are related to the Lorentz factor in units of universal time by
$\gamma_p(\tau_R) = \gamma_p(t)/\alpha$.} and we have 
parameterized the radius of curvature of a magnetic field line by a 
constant $\eta$ times the NS light cylinder radius. We choose $\eta=0.1$ throughout.
The energy measured by an observer at infinity is found by multiplying
by a factor of $\alpha = (g_H/c^2)  z = (4GM/c^2)^{-1} z$, which
accounts for the gravitational redshift, which is to be evaluated at the
Rindler $z$ coordinate of emission.
\begin{align}
\epsilon_{\infty} = \alpha \frac{3hc}{4 \pi} \frac{\gamma^3_p}{\eta R_{LC}}
\label{EcurvInfty}
\end{align}
Keep in mind however, that far enough from the horizon, where the
Rindler limit to the Schwarzschild spacetime breaks down ($\alpha
\gtrsim 1$), Rindler-$\alpha$ does not predict the correct gravitational
redshift. So, we can only use Eq.\ (\ref{EcurvInfty}) for emission
that originates close to the horizon, consistent with the regime in
which we are working.

We next solve for the values of $\gamma_p$. For a given BH mass and
horizon distance $z_S$ as a function of dipole boost $\vb$, we
estimate the maximum $\gamma_p$ in the radiation reaction limit, in
which the rate of energy gain from the horizon battery is balanced by
the rate of energy loss due to curvature radiation. 

A Rindler observer at the instantaneous location of an accelerating plasma particle will measure the following energy per unit proper time being radiated from the particle due to dipole radiation,
\begin{align}
\mathcal{P} = \frac{d \epsilon_R}{d\tau_R} = \frac{2}{3} e^2 c
\frac{\gamma^4_p}{(\eta R_{LC})^2}. 
\label{Eq:Larm}
\end{align} 
Eq.\  (\ref{Eq:Larm}) is the standard relativistic Larmor formula for
the power.

Then for a plasma particle moving on the path $\mathbf{s}(t)$, the radiation reaction limited $\gamma_p$ is given by,
\begin{align}
e \frac{d V}{ds^i} \frac{ds^i}{d \tau_R} =  \frac{2}{3} e^2 c \frac{\gamma^4_p}{(\eta R_{LC})^2},
\end{align}
where use of the  locally observed potential, $V=V_{\mathcal H}/\alpha$, is justified since we are only 
considering an infinitesimal potential difference, not a global value.

Upon inspection of the currents in Figure \ref{Fig:BoostJS_vs_Bet}, it is apparent that representatively large horizon electric fields exist at $y=0$ in the $ \pm x$ direction. Thus we choose $\mathbf{ds} = dx \mathbf{e_{\hat{x}}}$ which allows us to write
\begin{align}
| E^x_R  |   \left( 1 - \frac{1}{\gamma^2_p} \right)^{1/2}  =  \frac{2}{3} e  \frac{\gamma^4_p}{(\eta R_{LC})^2} 
 \label{MaxGamEq}
\end{align}
where we have written the 3-velocity of the particle (assumed to be only in the x direction) in terms of $\gamma_p$. Note that it is $E^x_R$ and not $\alpha E^x_R$ which should be on the LHS of (\ref{MaxGamEq}) because $qE^x_R$ is the rate of change of momentum as viewed by Rindler observers and we are asking the Rindler observer to locally balance the competing sources of momentum loss and gain.  

\begin{figure}
\begin{center}$
\begin{array}{c}
\includegraphics[scale=0.3]{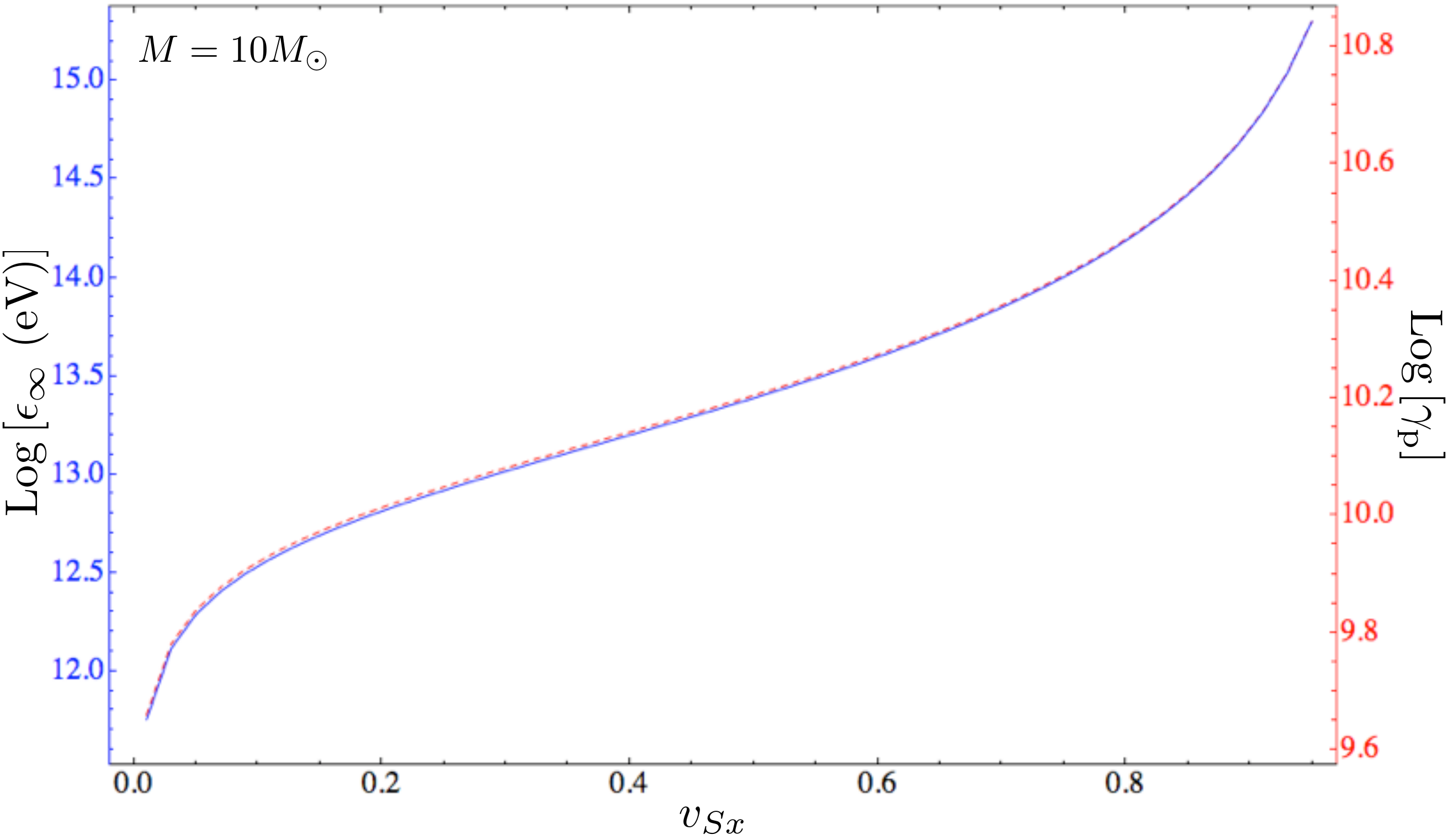} \\
\includegraphics[scale=0.3]{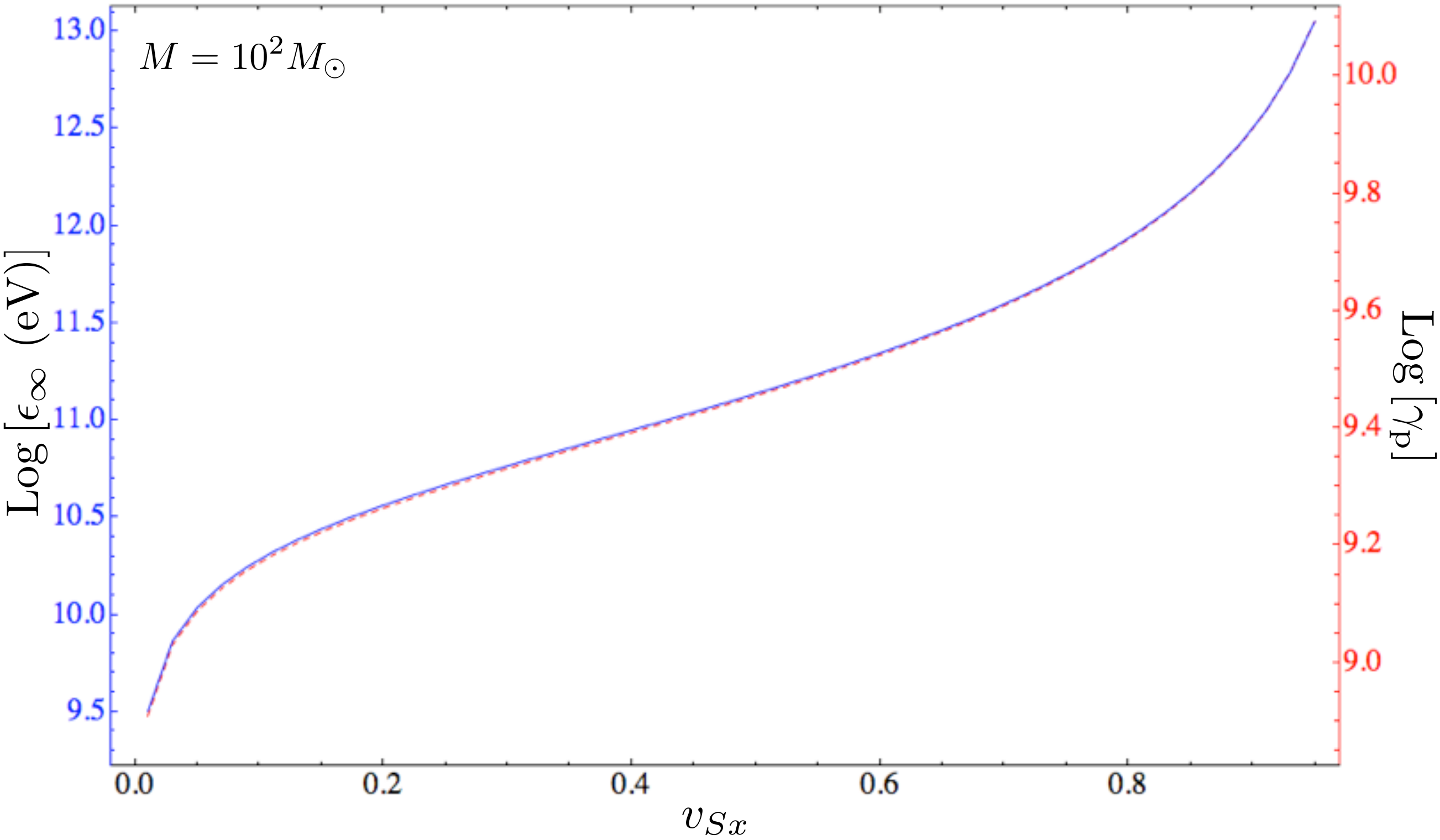}  \\
\includegraphics[scale=0.3]{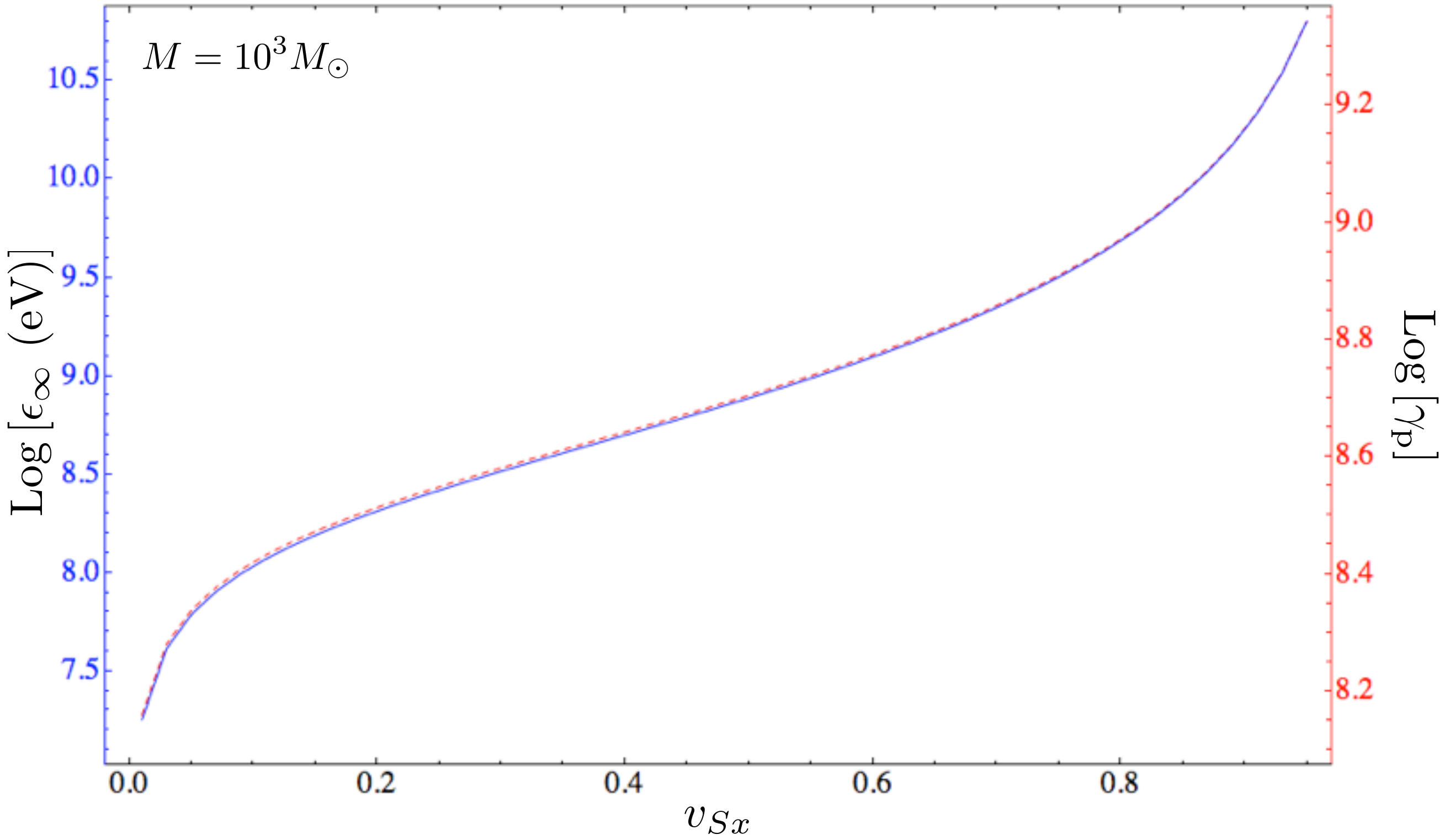} 
\end{array}$
\end{center}
\caption{Maximum curvature radiation energies computed from Eq.\ (\ref{EcurvInfty}) (blue, solid line and leftmost y-axis labels) and corresponding maximum $\gamma_p$ (Eq.\ \ref{MaxGamEq}) to which electrons/positrons can be accelerated (red, dashed line and rightmost y-axis labels). Both are computed for $M=10 M_{\odot}, 10^2 M_{\odot}$, and $10^3 M_{\odot}$ with the dipole at Rindler height $z_S = 3M$ as a function of $\vb$ varying from $0.01$ to $0.95$.}
\label{EGam}
\end{figure}

Solving the above equation for $\gamma_p$ then gives the radiation reaction limited Lorentz factor, as observed by Rindler observers, as a function of time. Since, however, the fields are stationary in the frame which drags along with the horizon charges, we need only find the maximum $\gamma_p$ at any time and choose that as our fiducial maximum $\gamma_p$. Substituting this into (\ref{EcurvInfty}) and evaluating $\alpha$ at the same $z$ position as we evaluated (\ref{MaxGamEq}), gives the maximum energy due to curvature radiation that the horizon battery can produce at a given $z_S$, $\vb$, and $M$, according to observers at infinity. In practice we find the largest values of $\epsilon_{\infty}$ when evaluating (\ref{MaxGamEq}) and (\ref{EcurvInfty}) at the stretched horizon, although varying the point of evaluation from $z=z_H$ up to $z=4M$ changes the result for $\epsilon_{\infty}$ by less than an order of magnitude. Figure \ref{EGam} plots the maximum $\gamma_p$ and $\epsilon_{\infty}$ as a function of $\vb$ at dipole height of $z_S = 3GM/c^2$ for BH masses $M=10 M_{\odot}, 10^2 M_{\odot}$, and $10^3 M_{\odot}$. 

We estimate maximum $\gamma_p$'s of our
BH battery to be $\sim 10^{10.2}$ for a $10M_\odot$
BH,  $\sim 10^{9.5}$ for a $10^2 M_\odot$ BH, and 
$\sim 10^{8.7}$ for a $10^3 M_\odot$ BH when $v_{S,x}\sim 0.5$. However, recall that these
are the Lorentz factors measured by the Rindler observers at the stretched horizon and do not correspond
to the tremendous energies which (\ref{Ecurv}) would imply. It is the energy measured at infinity given by (\ref{EcurvInfty}) which 
carries the only physical relevance here. 
These $\g_p$'s correspond to maximum curvature radiation energies at infinity of approximately $
30$ TeV, $100$ GeV, and $1$ GeV respectively at $\vb=0.5$. Because of the decrease in horizon voltage 
for larger mass BH's, as for the luminosity, the curvature radiation energies are smaller for 
larger mass BH's. 

We reiterate that the radiation energies plotted here represent the
highest energies of radiation that could be emitted by the NS-BH
circuit. Firstly this is because we are not accounting for any plasma 
effects which may act to screen the maximum fields quoted here. 
In addition to this, we are using the radiation reaction limited 
$\gamma_p$ computed for plasma particles with velocities aligned 
with the largest values of the electric field across the horizon. 
There will also be a spectrum of lower energy synchro-curvature 
radiation not calculated here.

Although we have not computed timescales or detailed spectra of
emission we note that with luminosities reaching up to $10^{42}$
erg/s ($10^{48}$ erg/s for magnetars) and with the capability of
producing photons with energies reaching into the TeV range, the
mechanism discussed here for a BH mass of  $10 M_{\odot}$ could be
capable of producing bursts of gamma-rays.
Further investigation of this mechanism and the
timescale, as well as any variability, of emission is needed in order
to say whether the BH-NS circuit is responsible for previously detected 
high-energy bursts, or rather, if it is responsible for an as of yet 
unobserved phenomenon.

We now look closer at the larger BH case, where the Rindler limit is an even better proxy for the physical situation.

\subsection{NS plummet into a SMBH}
Although motivated by an interest in stellar mass BHs, the solutions
we have found in the Rindler limit well approximate the end of a NS's
plummet into an intermediate mass or super-massive black hole (IMBH,
SMBH). We have included analysis for IMBH's in the previous section,
here we consider a SMBH. Recall that for the mechanism to operate,
the BH horizon must be within the magnetosphere of the NS; the
distance of the NS from the horizon must be less than the light cylinder radius
(\ref{RLC}) of the NS. 
In the previous subsection we always had that $R_{lc} \geq z_S = 3M$.
For the SMBH case however, the NS  light cylinder is smaller than $3M$. Thus we locate the dipole at $z_s=R_{lc}$ so that the
circuit is connected. In this case the Rindler approximation is good for the entire time that luminosity
can be generated by the BH battery. 

\begin{figure}
\begin{center}$
\begin{array}{c}
\includegraphics[scale=0.3]{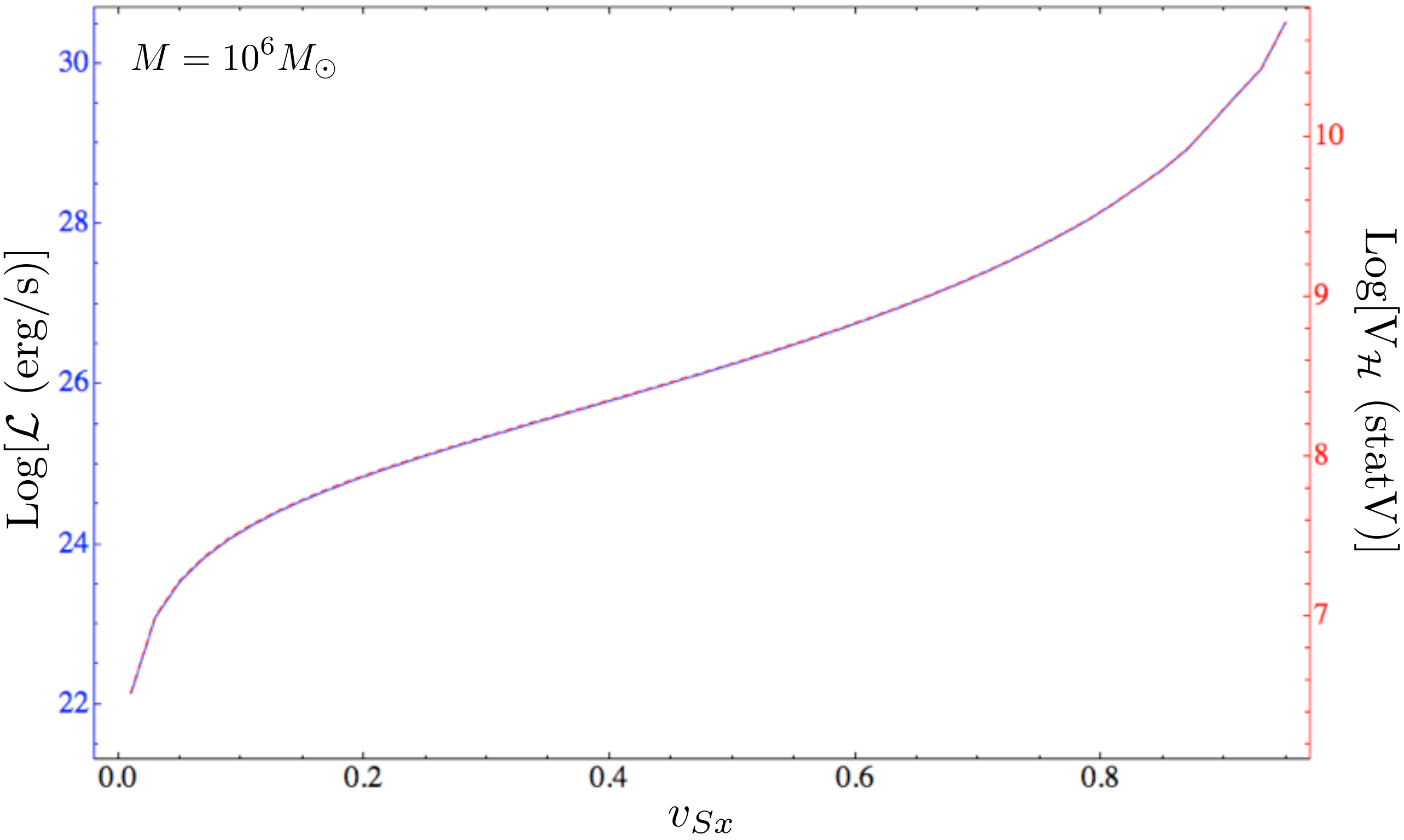} \\
\includegraphics[scale=0.3]{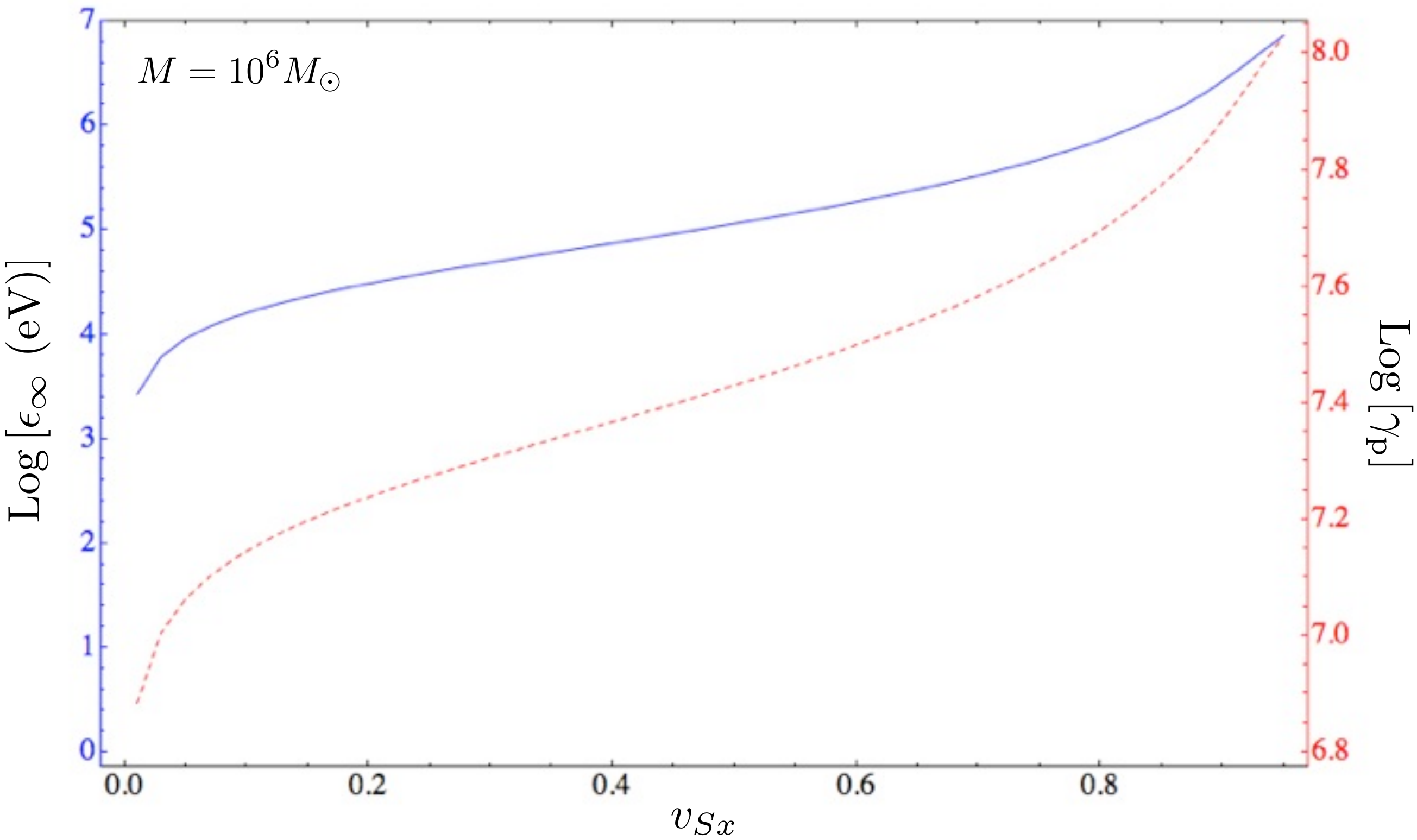} 
\end{array}$
\end{center}
\caption{Luminosity, voltage (top) and energy, Lorentz-factor (bottom)
  plots identical to those portrayed in Figures \ref{LumV} and
  \ref{EGam} respectively. Here we have plotted both panels for a
  $10^{6} M_{\odot}$ BH and at a much smaller horizon distance than in
  the previous Figures: $z_S=R_{lc}$, corresponding to the maximum
  separation where the BH-NS circuit remains connected. }
\label{M6BH}
\end{figure}

In Figure \ref{M6BH}, we plot the luminosities and energy of curvature
radiation that could be generated by the dipole at a distance $R_{lc}$
from a $10^{6} M_{\odot}$ BH horizon. We find that for $\vb = 0.5$,
luminosities of order $10^{26}$ erg/s can be achieved by the SMBH-NS
circuit. The maximum $\gamma_p$'s are still rather large reaching
values of $\sim 10^{7.4}$ at $\vb = 0.5$. 
However, recall that these are 
the maximum $\gamma_p$'s as measured by Rindler observers at the 
stretched horizon and so observed energies at infinity are reduced by a 
factor of  $\alpha_{\mathcal{H}} = 10^{-4}$ from what would be inferred from
$\gamma_p$ alone. 
For $\vb =0.5$ the SMBH-NS circuit could generate
energies of curvature radiation peaking in the X-ray at $\sim 100$ keV.

At luminosities of $\sim 10^{26}$ erg/s and peak radiation energies of
$\sim 100$ keV, even if the SMBH were in our own Galactic Center, this
signal would be difficult to detect, as it emanates from
 a noisy galactic nucleus and could be beamed in a direction not guaranteed to intersect Earth.
Note however, that in the optimal case of a magnetar with $B_p \simeq
10^{15} G$, and a slower spin period of $\sim 10 s$, the circuit would
be connected at a $10 \times$ greater distance from the horizon and
emit at a peak luminosity of $\sim 10^{32}$ erg/s. 
Such events, if beamed in our direction may produce a short, if faint, X-ray burst
coming from the Galactic-Center.
We can put a type of upper limit on
the length of such a magnetar-SMBH X-ray burst by noting that the
infall time observed at infinity for the NS falling from $R_{lc}$ to a
$R_{NS}$ at the speed of light is of order a minute. However, the
energetics of the magnetosphere could limit any emission to a much
shorter interval. For comparison,
X-ray flares at the Galactic Center are observed with durations of order an hour and X-ray
luminosities of $\sim 10^{35}$ erg/s in the energy range $2-10$ keV
\citep{Degenaar:2012:GCXrayFlares}.

NS-BH systems with BH mass in the range $10^3-10^5 M_{\odot}$ can
also be accurately described by the Rindler limit and could reside
nearby within globular clusters in the halo of our galaxy (see
\textit{e.g.} \citep{Lutzgendorf:2012}). The bottom panels of Figures
\ref{LumV} and \ref{EGam} show that for a $10^3 M_{\odot}$ BH the
BH-NS system could generate luminosities of order $10^{34}$ erg/s
peaking at a maximum achievable energy of radiation at a few GeV. 

To determine whether such a signal from a NS-IMBH binary would be detectable with currently operating instruments we consider flux sensitivities of the SWIFT Burst Alert Telescope (BAT) \citep{SWIFTBAT:2005} and the FERMI Gamma-ray Burst Monitor (GBM) \citep{FERMIGBM:2009} which are well suited for observing such transient high-energy events. From the BAT flux sensitivity, $\sim 10^{-8} \ \rm{ergs} \ \rm{cm}^{-2}  \rm{s}^{-1}$ and the GBM trigger rate $0.6 \ \rm{photons}\ \rm{cm}^{-2} \rm{s}^{-1}$ respectively, we may compute the minimum flux over the instrument energy range needed to detect a NS plunge event with luminosity computed from our model in the previous section. With this we can calculate a maximum observable distance for which our NS-BH circuit signal would be detectable. Assuming that the radiation is beamed into a solid angle $\Delta \Omega = 100 \ \rm{deg}^2$, taking a photon index of $5/3$ for curvature radiation, and integrating over the energy range of the instrument (15 to 150 KeV for the BAT and 150 KeV to 40 MeV for the GBM) we find 
\begin{align}
&D_{\rm{max}} \simeq  3.8 \ \rm{Kpc}  \sqrt{ \left(\frac{\mathcal{L}}{1.3 \times 10^{34} \rm{erg/s}} \right) \left( \frac{100 \rm{deg}^2}{\Delta \Omega} \right)  } \nonumber \\
& \rm{SWIFT} \ \ \rm{BAT}  \ \ 15\ \rightarrow 150\ \rm{KeV} \nonumber 
\end{align}
or
\begin{align}
&D_{\rm{max}} \simeq  0.4 \  \rm{Kpc} \sqrt{ \left(\frac{\mathcal{L}}{1.3 \times 10^{34} \rm{erg/s}} \right) \left( \frac{100 \rm{deg}^2}{\Delta \Omega} \right)  }\nonumber \\
& \rm{FERMI} \ \ \rm{GBM}  \ \ 150\ \rm{KeV} \rightarrow 40\ \rm{MeV}\nonumber 
\end{align}
where for the luminosity we have used the $\vb=0.5$ value for a $10^3 M_{\odot}$ mass black hole system (See Figure \ref{LumV}). Note that the above distances scale directly with the NS magnetic field strength. For a magnetar, maximum observable distances are on order a Mpc. Since Galactic globular clusters exist within a few Kpc of Earth such NS-IMBH inspirals could be observationally interesting events if the mechanism for EM radiation discussed here operates and if IMBH's exist in globular clusters.

\section{Conclusions}
\label{Conclusions}

When a magnetized NS and a BH 
approach within the NS light cylinder, an electromagnetic circuit is established.
In the Rindler limit, this corresponds to a magnetic dipole boosted parallel to the flat-wall horizon.
The power supplied to the
circuit will increase as the pair draws closer, reaching a maximum just
before merger. 
The maximum voltage 
the battery attains
and the maximum luminosities powered at this final stage scale roughly as
\begin{align}
V^{\rm{max}}_\mathcal{H} &\simeq 3.3 \times 10^{16} \left( \frac{B_p}{10^{12}\ \rm{G}} \right) \left( \frac{M}{10 M_{\odot} } \right)^{-2} {\rm{statvolts}} \\ \nonumber
 \mathcal{L}^{\rm{max}} &\simeq 1.3 \times 10^{42} \left( \frac{B_p}{10^{12}\ \rm{G}} \right)^2 \left( \frac{M}{10 M_{\odot} } \right)^{-4} \frac{\rm{erg}}{\rm{s}} \\ 
 ( z_S = &3M, \quad \vb=0.5, \quad \Res_{\rm{plasma}} =
 \Res_{\mathcal{H}}, \quad \Res_{NS} = 0   ) \nonumber \ \ .
\end{align}
The scaling changes if the NS does not maintain the fixed height of
$3M$ above the horizon and depends as well on the unknowns $\Res_{\rm{plasma}}$ and
$\Res_{\rm{NS}}$.
The estimated maximum could be higher when BH and NS spins are included. NS spin
can be thought of as increasing the effective $\vb$. BH spin adds
extra power from the analogue of the BZ effect. 

There are many caveats to consider when formulating
observational features of an event-horizon battery, such as potential short
circuits in the system.
Charges from the NS and its surrounding magnetosphere can act to 
screen the induced electric fields. In addition to these charges, 
if both the horizon voltage and the magnetic field strength are large 
enough, pair production could become an important source of screening charges.
Hence, the structure of the NSBH magnetosphere needs to be investigated 
further in order to determine the viability of an event-horizon battery powered electromagnetic signal.
Another concern is that at such high voltages 
the current generated along the magnetic field lines would 
be so great that the magnetic fields induced exceed those of 
the original dipole.\footnote{Short circuits could turn the mechanism off
temporarily until the current builds up again. A possible signature of
this short circuit transient might be repeated spikes in the emissions }
However, \citep{DLai:2012} has shown 
that this effect should not be large enough to short out the circuit for
a NS-BH system due to the large resistance of the horizon.

It would be essential to pin down the
timescales of the various emission mechanisms associated with this phenomenon, although the
solutions presented here give us no special advantage in doing so. 
Numerical results are needed to carefully characterize this EM signal in greater
detail --  although we can conjecture that 
there are potentially several 
distinct channels: 1) a brief jet, 2) beamed synchrotron and curvature radiation that sweeps
across the sky, 3) a faint hot
spot as charged particles hit the NS pole. 

These caveats aside, in light of this analysis we can say that BH-NS binaries with
BH's of order 10's of $M_{\odot}$ could conceivably produce luminosities of order 
$10^{42}$ erg/s ($10^{48}$ erg/s if the NS is a magnetar) and emit
high-energy gamma rays, possibly consistent with a sub-class of gamma ray bursts.
Therefore, stellar mass BH-NS binaries detectable
by AdLIGO, could power high-energy electromagnetic radiation, possibly into the TeV range, detectable moments prior to the gravitational radiation burst at merger.
Discovery of these important pairs could
probe NS properties as well as population rates in the pre-AdLIGO era.
Also intriguing is the possibility of an
IMBH in a binary with a highly magnetized NS. Although less energetic, their emissions
may nonetheless be detectible.

\bigskip

{\bf{Acknowledgements}}

\medskip

We would like to thank Jules Halpern and Sean McWilliams as well as participants
of the KITP ``Rattle and Shine'' conference (July 2012) for useful discussions. We also 
thank the anonymous referee for useful suggestions on improving the manuscript.
3D visualizations of field lines were computed using Mayavi \citep{mayavi}.
This research was supported by an NSF Graduate Research Fellowship Grant No. DGE1144155
(DJD), NSF grant AST-0908365 (JL), 
a KITP Scholarship under Grant no. NSF PHY05-51164 (JL), and a Guggenheim
Fellowship (JL).

\clearpage

\appendix
\section{Detailed Solution to the Field Equations}
\label{AppendixA}
In Minkowski spacetime, Maxwell's equations for the 4-potential $A^{\alpha}$ are,
\begin{equation}
\Box A^{\alpha}(\nobf{x}) - \partial^{\alpha} \left( \partial_{\beta} A^{\beta} \right)= \frac{4 \pi}{c}  J^{\alpha}(\nobf{x})
\label{MxPotMink1}
\end{equation}
where $J^{\alpha}(\nobf{x})$ is the 4-current as a function of the coordinates and we retain factors of c in the appendix. Working in the Lorentz  Gauge $  \partial_{\beta} A^{\beta}  = 0$,  Maxwell's equations become sourced wave-equations,
\begin{equation}
\Box A^{\alpha}(\nobf{x}) = \frac{4 \pi}{c} J^{\alpha}(\nobf{x}).
\label{MxPotMinkL}
\end{equation}
The solution for $A^{\alpha}$ 
can be written in terms of the retarded (or advanced) Green's function given by,
\begin{equation}
\Box_{\nobf{x}} G(\nobf{x}, \nobf{\bar{x}}) = \delta^{(4)}\left[ \nobf{x} -\nobf{\bar{x}} \right]
\label{Gdef}
\end{equation}
Where $\nobf{x}$ is the observer spacetime coordinates, and $\nobf{\bar{x}}$ is the spacetime position 4-vector to be integrated over. The above equation shows that $G(\nobf{x}, \nobf{ \bar{x}})$ must depend on $\nobf{x}$ and $\nobf{\bar{x}}$ only via the 4-vector $\nobf{x} - \nobf{\bar{x}}$, so we write the retarded Green's function as $G(\nobf{x} - \nobf{\bar{x}})$ and solve Eq.\ (\ref{Gdef}) to find (\cite{Jackson}), 
\begin{equation}
G(\nobf{x} - \nobf{\bar{x}}) = \frac{1}{2 \pi} H \left( x^0  -  \bar{x}^0 \right)  \delta \left[ \left(\nobf{x} -\nobf{\bar{x}}\right)^2 \right]
\label{Gr}
\end{equation}
where the Heaviside function H picks out the retarded as opposed to the advanced Green's function.We may then write the solution to Eq.\ (\ref{MxPotMinkL}),
\begin{equation}
A^{\alpha} = \frac{4 \pi}{c} \int{ G(\nobf{x}-\nobf{\bar{x}}) J^{\alpha}(\nobf{\bar{x}}) \sqrt{-\bar{g}}  \hbox{  } d^4\bar{\nobf{x}}}
\label{GreenSoln}
\end{equation}
where, in Cartesian coordinates, the metric determinant $g=-1$. A
choice of source distribution for the 4-current in (\ref{GreenSoln})
gives the 4-potential from which the fields may be computed.

\subsection{Point Charge}
Although the derivation of an electric point charge can be found in a standard text on electrodynamics (we follow \cite{Jackson} below), we include the derivation here to better elucidate, and put in context, the derivation for the dipole to follow.

The 4-current in terms of the 4-position $x^{\mu}_S(\tau)$ of a point charge $q$ with arbitrary 4-velocity $V^{\alpha}$ is
\begin{equation}
J^{\alpha}(\nobf{x}) = c \int {q V^{\alpha}(\tau) \delta^{(4)}\left[ \nobf{x} -\nobf{x_S}(\tau) \right]  d \tau}
\label{PC4J}
\end{equation}
where $\tau$ is the proper time of the source charge. Substituting Eq.\  (\ref{PC4J}) and (\ref{Gr}) into (\ref{GreenSoln}) yields,
\begin{widetext}
\begin{align}
A^{\alpha}  = 2q \int{ H \left( x^0  -  \bar{x}^0 \right)  \delta \left[ \left( \nobf{x} -\nobf{\bar{x}}\right)^2 \right]  V^{\alpha}(\tau) \delta^{(4)}\left[ \nobf{\bar{x}} -\nobf{x_S}(\tau) \right]  d^4\bar{\nobf{x}} d \tau}
\label{PCA1}
\end{align}
\end{widetext}
Integrating over the volume,
\begin{align}
A^{\alpha} =2q \int{ H \left( x^0  -  x^0_S(\tau) \right) \delta \left[ \left(\nobf{x} -\nobf{x_S}(\tau)\right)^2 \right]  V^{\alpha}(\tau)    d \tau}.
\label{PCA2}
\end{align}
To evaluate this we use the rule,
\begin{equation}
\delta \left[ f(x) \right] = \sum_i{\frac{\delta(x-x_i)}{| \left( \partial f / \partial x \right)_{x=x_i}|}}
\label{DelFofx}
\end{equation}
where the sum is over the $i^{\hbox{th}}$ root of $f(x)$. This allows us to write, 
\begin{equation}
\delta \left[  \left(\nobf{x} -\nobf{x_S}(\tau) \right)^2 \right] =  \frac{ \delta(\tau-\tau_*) }{ | -2 \ r_{\mu}(\tau) V^{\mu}(\tau)|_{*} }
\label{delf2}
\end{equation}
where, as in the text, $r^{\mu}  = x^{\mu} - x^{\mu}_S(\tau)$ and $\tau_*$ is the proper time of the point charge given by the light cone condition,
\begin{equation}
r_{\mu} r^{\mu}|_* =0
\end{equation}

Using (\ref{delf2}) to simplify (\ref{PCA2}), we find,
\begin{equation}
A_{LW}^{\alpha} = \frac{  q V^{\alpha} (\tau) }{  V_{\mu}(\tau) r^{\mu}(\tau)  } \Bigg{|}_{\tau_*}.
\label{LWMink}
\end{equation}
which are the Lienard-Wiechert Potentials for a moving point charge. Note that all quantities are evaluated at the retarded point $\tau_*(x)$, the proper time at which the source coordinates are coincident with the past light cone of the observer at $\nobf{x}$. Transforming these to Rindler space (lower case or primed), we find
\begin{align}
A^{\alpha'} = L^{\alpha'}_{ {\ \alpha}} A_{LW}^{\alpha} &=  L^{\alpha'}_{ {\ \alpha}} \left[ \frac{  q  L^{\alpha}_{{\ \beta'}} V^{\beta'}  }{  r^{\beta}   L^{\mu'}_{{\ \beta}} V_{\mu'}  } \right]_{\tau_*} \nonumber \\
&=  \frac{  q  L^{\alpha'}_{ {\ \alpha}}  L^{\alpha}_{* { \beta'}} V_*^{\beta'}  }{  V_{* \mu'} L^{\mu'}_{* {\beta}}  \left( {x}^{\beta} - {x^{\beta}}_* \right)   } .
\label{LWRind}
\end{align}
as found in \cite{MP3_MS:1985}. Here $L^{\alpha'}_{{\alpha}}$ denotes the transformation matrix $\frac{\partial x^{\alpha'}}{\partial x^{\alpha}}$. The potentials are written in terms of the vectors at the retarded point, (subscript $*$ above) and the observer point coordinates (no subscript). To write them in terms of only the observer coordinates, and thus obtain the full solution, we use the light cone condition to solve for the intersection of the past light cone of the observer, and the trajectory of the source.  In Minkowski coordinates,
\begin{equation}
(X^i-X^i_*)^2= (T-T_*)^2,
\end{equation}
and in Rindler coordinates
\begin{equation}
(x-x_*)^2 + (y-y_*)^2 + z^2 + z^2_* -2z z_* \hbox{cosh}\left[g_H(t-t_*)\right] = 0
\label{LC}
\end{equation}

\subsection{A General Dipole Solution}
We now derive the analog of the Lienard-Wiechert potential for a pure dipole source with arbitrary 4-velocity. We start again from Eq.\ (\ref{GreenSoln}) but write the 4-current for a point dipole source
\begin{align}
J^{\alpha}(\nobf{x}) =  \hbox{  } c \nabla_{\mu} \int { Q^{\alpha \mu}(\tau) \delta^{(4)}\left[ \nobf{x} -\nobf{x_S}(\tau) \right]   d \tau}
\label{eq:4JDipole}
\end{align}
where the antisymmetric dipole tensor, 
\begin{align}
Q^{\alpha \mu}(\tau) &= V^{\alpha}p^{\mu} - p^{\alpha}V^{\mu} + \epsilon^{\alpha \mu}_{ \hbox{  } \hbox{  } \hbox{  }\rho \sigma} V^{\rho} m^{\sigma},
\label{DipoleTensor}
\end{align}
is the antisymmetric decomposition of electric and magnetic
parts given by
\cite{RS:1995}. We discuss this decommposition further in the next
appendix. Eq.\ (\ref{DipoleTensor}) is a general decomposition
of any antisymmetric rank two tensor given vectors $V$, $p$ and $m$
such that $p \cdot V = m \cdot V =0$. Such vectors $p$ and $m$ can be
chosen in terms of the dipole moments as measured in the instantaneous
rest frame of the source, $\bf{p_S}$ and $\bf{m_S}$,
\begin{align}
p^{\alpha}&= \left(\gamma_S \beta_j {p^j_S}, \hbox{ } {p^i_S} + \frac{\gamma_S-1}{\beta^j \beta_j} \beta_j {p^j_S} \beta^i \right)  \nonumber \\
m^{\alpha}&= \left( \gamma_S \beta_j {m^j_S}, \hbox{ } {m^i_S} + \frac{\gamma_S-1}{\beta^j \beta_j} \beta_j {m^j_S}  \beta^i \right)
\label{App:4Moments}
\end{align}

In Minkowski coordinates we set $\sqrt{-g}=1$ and substitute (\ref{eq:4JDipole}) into (\ref{GreenSoln}) to get,
\begin{widetext}
\begin{align}
A^{\alpha} &= 2 \int \left \{
 H \left( x^0  -  \bar{x}^0 \right)  \delta \left[ \left(\nobf{x}
   -\nobf{\bar{x}}\right)^2 \right]
\bar{\nabla}_\mu\left (
\int  Q^{\alpha \mu}(\tau) \delta^{(4)}\left[ \nobf{\bar{x}}
  -\nobf{x_S}(\tau) \right]   d \tau
\right )   \right \}
\hbox{  }
d^4\bar{\nobf{x}} .
\end{align}

Since $Q^{\alpha \mu}$ depend on source coordinates while $\bar{\nabla}_\mu$ is taken with respect to observer coordinates, we may take $Q^{\alpha \mu}$ out of the derivative,

\begin{align}
A^{\alpha} &= 2 \int \int \left \{
 H \left( x^0  -  \bar{x}^0 \right)  \delta \left[ \left(\nobf{x}
   -\nobf{\bar{x}}\right)^2 \right]
  Q^{\alpha \mu}(\tau) \bar{\nabla}_\mu\left ( \delta^{(4)}\left[ \nobf{\bar{x}}
  -\nobf{x_S}(\tau) \right] 
\right )   \right \}
\hbox{  }
  d \tau d^4\bar{\nobf{x}} .
\end{align}

To evaluate the above integrals we use the notion of a generalized derivative and employ integration by parts to write,
\begin{align}
&\int^a_b{  \partial_{\bar{x}} \delta(x-\bar{x}) f(\bar{x}) d\bar{x} } = 
 \delta(x-\bar{x}) f(\bar{x}) \bigg{|}^a_b  -  \int^a_b{  \delta(x-\bar{x}) \partial_{\bar{x}}  f(\bar{x}) d\bar{x} }  =  -\left[ \partial_{\bar{x}} f(\bar{x}) \right]_{\bar{x}=x}
\end{align}
for a continuous, once differentiable function $f(x)$. We assume $x \subset (a,b)$ so that the boundary terms disappear.
Generalizing to multiple dimensions, 
\begin{align}
\int^a_b{\int^a_b{ \partial_{\bar{x}} \left[ \delta(x-\bar{x})  \delta(y-\bar{y}) \right]f(\bar{x}, \bar{y}) d\bar{x} d\bar{y}}} &=  -\left[\partial_{\bar{x}} f(\bar{x},y) \right]_{\bar{x}=x},
\label{KronD2}
\end{align}
and to the case at hand,
\begin{align}
\int{ \partial_{\alpha} \left[ \delta^{n}(\nobf{x}-\nobf{\bar{x}}) \right] F^{\alpha \beta ...}(\bar{x}^0, \bar{x}^1...\bar{x}^n) \sqrt{-\bar{g}}  \ d^{n}\bar{x}} =   -\left[\partial_{\alpha} F^{\alpha \beta ...}(\bar{x}^0, \bar{x}^1 ... \bar{x}^n) \sqrt{-\bar{g}} \right]_{\bar{x}^{\alpha}=x}.
\label{KronD}
\end{align}

Integrating by parts (\textit{i.e.} using \ref{KronD})
\begin{align}
A^{\alpha} &= -2 \int 
 \left[\bar{\nabla}_{\mu} H \left( x^0  -  \bar{x}^0 \right)  \delta \left[ \left(\nobf{x}
   -\nobf{\bar{x}}\right)^2 \right]  
  Q^{\alpha \mu}(\tau)  \right ]_{\bar{x}=x_S(\tau)}  \hbox{  }  d \tau \nonumber \\
  &= -2 \int 
\nabla^S_{\mu} \left( H \left( x^0  -  x^0_S(\tau) \right)  \delta \left[ \left(\nobf{x}
   -\nobf{x_S(\tau)}\right)^2 \right]  
   \right )  Q^{\alpha \mu}(\tau) \   d \tau 
   \label{AEvalDeriv}
\end{align}
where $\nabla^S_\mu = \frac{\partial}{\partial x^{\mu}_S(\tau)}$ and the second line follows because $Q^{\alpha \mu}(\tau)$ does not depend on $\bar{x}$, and the function being differentiated and then evaluated at $x_S(\tau)$ does not depend anywhere on $x_S(\tau)$ before being evaluated. Applying the product rule to the derivative, we obtain two terms to evaluate
\begin{align}
A_1^{\alpha} &= -2 \int 
\nabla_\mu^S \left[ H \left( x^0  -  x^0_S(\tau) \right) \right] \delta \left[ \left(\nobf{x}
   -\nobf{x^0_S(\tau)}\right)^2 \right]  
  Q^{\alpha \mu}(\tau)  \hbox{  } d \tau  \\
 A_2^{\alpha} &= -2 \int 
 H \left( x^0  -  x^0_S(\tau) \right) \nabla^S_{\mu} \left[ \delta \left[ \left(\nobf{x}
   -\nobf{x^0_S(\tau)}\right)^2 \right]   \right]
  Q^{\alpha \mu}(\tau)  \hbox{ } d \tau  
\end{align}
Using 
\begin{align}
\left [\bar{\nabla}_{\mu} H(x^0-{\bar{x}}^0) \right ]_{\bar x=x_S}
&=-\delta^0_\mu \delta(x^0-x^0_S) 
\label{GradH}
\end{align}
and
\begin{align}
\delta \left[  \left(\nobf{x} -\nobf{x_S}(\tau) \right)^2 \right] =  \frac{ \delta(\tau-\tau_*) }{ | -2 \ r_{\nu}(\tau) V^{\nu}(\tau)|_{*} }
\label{eq:delf2}
\end{align}
we find
\begin{align}
A_1^\alpha(x) & = \int d\tau
 Q^{\alpha 0} \
\delta(x^0-x_S^0)
\frac{\delta(\tau-\tau_*)}{\left[ r \cdot V \right]_{*}} \nonumber \\
&=\frac{Q^{\alpha 0} \
}{\left( r \cdot V \right)} \delta(x^0-x_S^0) \bigg{|}_*
\label{eq:firstterm}
\end{align}
which is only non-zero at the retarded point, so for observers at the location of the dipole. 
The second term can be written,
\begin{align}
 A_2^{\alpha} &= -2 \int 
  Q^{\alpha \mu}(\tau)  H \left( x^0  -  x^0_S(\tau) \right) \nabla^S_{\mu} \tau \frac{d}{d \tau} \left[  \frac{ \delta(\tau-\tau_*) }{ | -2( r \cdot V)|_{*} } \right]
 \hbox{ } d \tau 
\end{align}
where we have used the chain rule to rewrite $\nabla^S_{\mu}$. Integrating by parts we find,
\begin{align}
 A_2^{\alpha} &=  \int 
 \frac{d}{d \tau} \left[   Q^{\alpha \mu}(\tau)  H \left( x^0  -  x^0_S(\tau) \right) \nabla^S_{\mu} \tau  \right]   \frac{ \delta(\tau-\tau_*) }{ (r \cdot V)_* }
 \hbox{ } d \tau 
\end{align}
which, upon integration over $\tau$, we may write as 
\begin{align}
 A_2^{\alpha} &=  
 \frac{d}{d \tau_*} \left[   Q^{\alpha \mu}(\tau_*)  H \left( x^0  -  x^0_S(\tau_*) \right) \nabla^S_{\mu} \tau_*  \right]   \frac{ 1 }{ (r \cdot V)_* }
\end{align}
where we have again exploited the fact that the function being
differentiated in square brackets does not depend anywhere on
$\tau_*$. This allows us to first evaluate the function at $\tau_*$
and then take the derivative wrt $\tau_*$, instead of differentiating
first and then evaluating.

Now, we can use
\begin{align}
\nabla_\mu^{S} \tau  \big{|}_* = \nabla_\mu \tau  \big{|}_* = \frac{r_{\mu}}{ r \cdot V }  \bigg{|}_*
\label{AppA:GradTau*}
\end{align}
which follows from writing out the gradient of the null condition with respect to source coordinates,
\begin{align}
r_{\nu}\nabla_\mu^{S} r^{\nu}  \big{|}_* &= r_{\nu}  \left(  \nabla_\mu^{S} x^{\nu}  - \delta^{\nu}_{\mu} \right)\big{|}_*  = 0 \nonumber \\
 r_{\nu}\nabla_\mu^{S} x^{\nu}  \big{|}_* &= r_{\mu} \big{|}_* \ ,
\end{align}
so that we may write
\begin{align}
\nabla_\mu^{S} \tau\big{|}_* = \nabla_\mu^{S} x^{\gamma} \nabla_{\gamma} \tau  \big{|}_* = \frac{r_{\mu}}{ r \cdot V }  \bigg{|}_*.
\end{align}
Using this, 
we can combine terms
and include $A_1^{\alpha} $ to obtain,
\begin{align}
A^{\alpha}(x) = \nabla_{\mu} \left[ \frac{Q^{\alpha \mu} }{(r \cdot
    V)} \right]_* -  \left[ \gamma_S \frac{Q^{\alpha \mu} r_{\mu}}{(r
    \cdot V)^2}  - \frac{Q^{\alpha 0} }{\left( r \cdot V \right)}
  \right]_*  \delta(\hbox{r})
\label{Eq:it}
\end{align}
The first term is our desired solution for observers off of the source worldline.  The terms which turn on at the position of the dipole fall off one factor of $-(r \cdot V)$ more slowly than the worldline terms. Still however, the off-worldline terms and their curl blow up at the position of the dipole. 
A very different means to the first term in Eq.\  (\ref{Eq:it}) can be found
in \cite{Rowe2:1986}.

For completeness we write out the transformed Rindler potential (for observers not at the position of the dipole) in terms of positions, velocities, accelerations, and dipole moments in the Rindler frame analogous to Eq.\ (\ref{LWRind}),
\begin{align}
A^{\alpha'} (x') &=  L^{\alpha'}_{\ \alpha} \epsilon^{\alpha \mu}_{\quad \rho
    \sigma} \left[ \frac{ L^{\rho}_{\ \rho'} a^{\rho'} L^{\sigma}_{\ \sigma'} m^{\sigma'} L^{\mu'}_{\ \mu}  r_{\mu'} + L^{\rho}_{\ \rho'} V^{\rho'} 
  L^{\sigma}_{\ \sigma'} \dot{m}^{\sigma'} L^{\mu'}_{\ \mu}  r_{\mu'} }{(r\cdot V)^2}  -  \frac{ L^{\rho}_{\ \rho'} V^{\rho'} L^{\sigma}_{\ \sigma'} m^{\sigma'} L^{\mu'}_{\ \mu} r_{\mu'}}{(r\cdot V)^3} (1+r\cdot a)  \right]_{*} .
  \label{DpoleSolnRind}
\end{align}

\end{widetext}

\section{Dipole Moments}
\label{DipoleMoments}

To clarify the interpretation of the dipole moment 4- vectors, we will
work in direct analogy to the electric and magnetic field vectors for
which the antisymmetric Maxwell tensor is
\begin{equation}
F^{\alpha \mu}=u^\alpha E^\mu - E^\alpha u^\mu +\epsilon^{\alpha
  \mu}_{\ \ \ \rho \sigma} B^\rho u^\sigma
\end{equation}
The fields as measured by an observer with 4-velocity $u^\mu$ are then
given by
\begin{align}
E^\alpha & = F^{\alpha \mu} u_\mu \nonumber \\
B^\alpha & = \frac{1}{2} \epsilon^{\alpha \mu \gamma \delta} F_{\gamma
  \delta} u_{\mu}
\end{align}
The electric field of one observer is {\it not} a coordinate transformation
of the electric field of another observer, although the two electric
fields can be related.

To relate the fields of two different observers, con- sider first a
Minkowski observer. Having 4-velocity $u_M^\mu = (1, 0, 0, 0)$. She sees
\begin{align}
\EM & = -F^{i0} \nonumber \\
\BM & = \frac{1}{2}\epsilon^{ijk}F_{jk}
\end{align}
so that we can build the Maxwell tensor in Minkowski
coordinates:
\begin{equation}
F^{\alpha  \mu}=\begin{pmatrix}
0 & -\EM \\
\EM & \epsilon^{ij}_{\ k}\BM^k
\end{pmatrix} \ . 
\end{equation}
On the other hand, the EM fields measured by an ob- server that moves
with generic 4-velocity $u^\mu = \gamma (1,\bb)$ according to our Minkowski
observer can be related to the EM fields measured by our Minkowski
observer through:
\begin{align}
E_o^{\alpha'} &= \frac{\partial x^{\alpha'}}{\partial x^{\alpha}}
F^{\alpha \mu} u_{\mu } \nonumber \\
B_o^{\alpha'} &= \frac{1}{2} \frac{\partial x^{\alpha'}}{\partial x^{\alpha}} \epsilon^{\alpha \mu \gamma \delta}F_{\gamma \delta} u_{\mu}.
\end{align}
Expanding gives the relation
\begin{align}
{\mathbf E}_o' &= \hb (\hb \cdot \EM)(1-\gamma)+\gamma \EM +\gamma(\bb
\times \BM) \nonumber \\
{\mathbf B}_o' &= \hb (\hb \cdot \BM)(1-\gamma)+\gamma \BM -\gamma(\bb
\times \EM) 
\end{align}
(equivalent to Eqs.\ (\ref{Eq:trans}).) For emphasis, $\EM$ are the components of
the electric field measured by a Minkwoski
observer in her inertial frame basis while by contrast ${\bf E}_o'$ are the components
of the electric field measured by an observer boosted (relative to the
Minkowski observer) expressed in the (boosted) observer’s coordinate
basis. The fields $\EM$ and ${\bf E}_o'$ are not related solely by a coordinate
transformation. 

To construct the relevant objects and interpretations for the EM
dipole moments we note that
\begin{align}
Q & \leftrightarrow F \nonumber \\
p & \leftrightarrow E \nonumber \\
m & \leftrightarrow B \nonumber \\
V & \leftrightarrow u 
\end{align}
Working in analogy with the above, we begin with the
antisymmetric dipole tensor $Q^{\alpha\mu}$ . Any antisymmetric rank two tensor can be
decomposed given a timelike unit vector $V$ and two vectors $p$ and $m$
which are orthogonal to $V$
\begin{equation}
Q^{\alpha \mu}=V^\alpha p^\mu-p^\alpha V^\mu +\epsilon^{\alpha
  \mu}_{\ \ \rho \sigma}V^\rho m^\sigma \quad ,
\end{equation}
The identitites
\begin{align}
p^\alpha & = Q^{\alpha \mu} V_\mu \nonumber \\
m^\alpha & = \frac{1}{2} \epsilon^{\alpha \mu \gamma \delta} F_{\gamma
  \delta} V_{\mu}
\label{Eq:pm}
\end{align}
provide definitions for the covariant dipole
moments. We will use $V^\alpha = V^\alpha_S = \gamma_S(1,\bb_S)$, the
$4$-velocity of the source.
An observer at rest in the coordinate basis in which $Q$ is expressed
measures dipole moments (\ref{Eq:pm}) for a source moving with velocity
$V_S^\mu$ with respect to that basis.

In direct analogy to the EM field tensor, the dipole moment tensor
$Q^{\alpha \mu}$
is the physically significant entity while, in direct analogy to
the electric and magnetic fields, $p$ and $m$ are observer dependent. 

In the rest frame of the source, $V_S^{\mu'} = (1, 0, 0, 0)$, and we can
define the rest-frame moments:
\begin{align}
p^{\alpha '} &\equiv (0,{\bf p}_S) = -Q^{i'0'} \nonumber \\
m^{\alpha '}& \equiv (0, {\bf m}_S) = \frac{1}{2}\epsilon^{i'j'k'} Q_{j'k'}
\end{align}
so that we can build the dipole tensor in Minkowski coordinates:
\begin{equation}
Q^{\alpha'}_{\ \mu'}=\begin{pmatrix}
0 & -{\bf p}_S \\
{\bf p}_S & \epsilon^{i'j'}_{\ k'}{\bf m}^{k'}
\end{pmatrix} \ . 
\end{equation}
In general, the source may be moving with respect to the natural basis.
If we want to express the components of $Q$ in a basis with respect to
which the source is moving, we
Lorentz transform
\begin{align}
Q^{\alpha\mu} & =\frac{\partial x^{\alpha }}{\partial x^{\alpha '}}
\frac{\partial x^{\mu }}{\partial x^{\mu '}} Q^{\alpha' \mu'} \ .
\end{align}
The Lorentz boosted source velocity is
$V_S^{\mu}=\gamma_S(1,\bb_S)$
and the Lorentz boosted dipole moments are
\begin{align}
p^{\alpha } & =\left ( \gamma_S \bb_S \cdot {\bf p}_S, {\bf p}_S
+(\gamma_S-1)(\hb_S \cdot {\bf p}_S )\hb_S \right ) \nonumber \\
m^{\alpha } & = \left (\gamma_S \bb_S \cdot {\bf m}_S, {\bf m}_S
+(\gamma_S-1)(\hb_S \cdot {\bf m}_S) \hb_S  \right )
\label{Eq:transform}
\end{align}
which is identical to Eqs.\ (\ref{Eq:onlydipole}) with 
${\bf p}_S\ne 0$.
Expressions
(\ref{Eq:transform}) 
are the moments as measured by an
observer that sees the source boosted.

There is a subtlety to be noted. $V$
determines the basis in which you are expressing $Q$, unlike EM fields
where $u$ does not determine the basis in which you are expressing $F$. In other words,
you can choose a basis and write the components of $F$ in that
basis. However, there are no restrictions on which observer you
consult in that basis and therefore no restrictions on which $u$
to contract with in the definitions of $E,B$. By contrast, once you
choose a basis and write out the components of $Q$ in that basis, you
have fixed the source velocity $V_S$. There
is one and only one $V_S$ with respect to a given basis, and so one
and only one $V_S$ to contract with in the definitions in $p$ and $m$.

\bibliography{NSBH}

\end{document}